\documentclass[12pt,a4paper ]{report}
\usepackage[utf8x]{inputenc}
\usepackage[T1]{fontenc}
\usepackage[francais]{babel}
\usepackage{amssymb}
\usepackage[left=2cm,right=2cm,top=2cm,bottom=2cm]{geometry}
\usepackage{graphicx,float}
\usepackage{lmodern}
\usepackage{color}
\usepackage{subfig}
\usepackage{url}
\usepackage{amsmath}
\usepackage{amsfonts}
\usepackage{verbatim}
\usepackage{caption}
\newcommand{\HRule}{\rule{\linewidth}{0.5mm}}

\usepackage{array}
\newcolumntype{L}[1]{>{\raggedright\let\newline\\\arraybackslash\hspace{0pt}}m{#1}}
\newcolumntype{C}[1]{>{\centering\let\newline\\\arraybackslash\hspace{0pt}}m{#1}}
\newcolumntype{R}[1]{>{\raggedleft\let\newline\\\arraybackslash\hspace{0pt}}m{#1}}

\begin{document}

\begin{titlepage}
  \begin{sffamily}
  \begin{center}

    \includegraphics[scale=0.5]{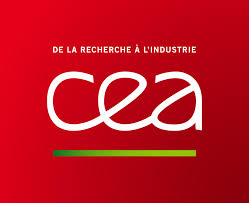}
    \\[2cm]

    \medskip 
    \medskip
    \textsc{\Large Rapport de stage }\\[1.5cm]

    \HRule \\[0.4cm]
    { \huge \bfseries Modélisation Magnétohydrodynamique\\[0.4cm] }

    \HRule \\[2cm]

    \begin{minipage}{0.4\textwidth}
      \begin{flushleft} \large
        Quentin \textsc{Cauvet}\\
      \end{flushleft}
    \end{minipage}
    \begin{minipage}{0.4\textwidth}
      \begin{flushright} \large
        \emph{Tuteur :} M.  \textsc{Bernecker}\\
        \emph{Chef de laboratoire : } \\
        M. \textsc{Pichon}
      \end{flushright}
    \end{minipage}

    \vfill

    {\large 1\ier{} Avril 2019 — 6 Septembre 2019}

  \end{center}
  \end{sffamily}
\end{titlepage}

\tableofcontents

\newpage
\section*{Notations}

\begin{tabular}{lC{5cm}L{8.cm}}
	
	& Expression & Description (Unité) \\

	&$\mathbf{A}$& Vecteur ou tenseur  \\
	& $A$  avec $\mathbf{A}\in \mathbb{R}^m$& Norme du vecteur $\mathbf{A}$ \\
	& $\mathbf{A}\cdot \mathbf{B}$ avec $\mathbf{A},\mathbf{B} \in \mathbb{R}^m$ & Produit scalaire usuel entre deux vecteurs $\mathbf{A}$ et $\mathbf{B}$ \\  
	& $\mathbf{A}\times \mathbf{B}$ avec $\mathbf{A},\mathbf{B} \in \mathbb{R}^m$ & Produit vectoriel usuel entre deux vecteurs  $\mathbf{A}$ et $\mathbf{B}$ \\ 
	& $\mathbf{A}\otimes \mathbf{B} \in \mathbb{R}^m\times\mathbb{R}^m$  avec  $\mathbf{A},\mathbf{B} \in \mathbb{R}^m$ & Produit tensoriel usuel entre les deux vecteurs  $\mathbf{A}$ et $\mathbf{B}$ tel que $(\mathbf{A}\otimes\mathbf{B})_{ij}=\mathbf{A}_i \mathbf{B}_j$ avec $i,j \in [1,m]$ \\ 
	& $\partial_t$ (resp. $\partial_x$)  & Opérateur de la dérivée temporelle: $\frac{\partial}{\partial t}$ (resp. spatialle: $\frac{\partial}{\partial x}$) \\ 
	& $\nabla$ & Opérateur nabla: $\nabla = (\frac{\partial}{\partial x},\frac{\partial}{\partial y},\frac{\partial}{\partial z})$ \\ 
	& $\nabla A$ & Gradient du scalaire A \\ 
	&  $\nabla . \mathbf{A}$ & Divergence du vecteur $\mathbf{A}$\\
	& $\nabla \times \mathbf{A}$ & Rotationnel du vecteur $\mathbf{A}$ \\
	& $\rho$ & Masse volumique du plasma\\ 
	& $\rho_s$ & Masse volumique du fluide $s$ ($e$ pour les électrons, $i$ pour les ions et $n$ pour les neutres) avec $\rho_s = m_s n_s$ \\ 
	& $\mathbf{V} \in \mathbb{R}^3$ & Vecteur vitesse du plasma \\ 
	& $\mathbf{V}_s \in \mathbb{R}^3$ & Vecteur vitesse du fluide $s$ ($e$ pour les électrons, $i$ pour les ions et $n$ pour les neutres) \\ 
	& $p$ & Pression cinétique du plasma \\ 
	& $p_t$ & Pression totale du plasma (cinétique + magnétique) \\ 
	& $p_s$ & Pression cinétique du fluide $s$ ($e$ pour les électrons, $i$ pour les ions et $n$ pour les neutres) \\ 
	& $\mathcal{E}$ & Énergie mécanique du plasma \\ 
	& $\mathcal{E}_{EM}$ & Énergie électromagnétique \\
	& $\mathcal{E}_t$ & Énergie totale du plasma (mécanique + électromagnétique) \\ 
 	& $\mathcal{E}_s$ & Énergie mécanique du fluide $s$ ($e$ pour les électrons, $i$ pour les ions et $n$ pour les neutres) \\ 
	& $n$ & Concentration du plasma \\
	& $n_s$ & Concentration du fluide $s$, ($e$ pour les électrons et $i$ pour les ions) \\
	& $\mathbf{E}$ & Vecteur du champ électrique \\ 
	& $\mathbf{B}$ & Vecteur du champ magnétique \\ 
	& $\mathbf{J}$ & Vecteur de la densité de courant électrique \\
	& $Q$ & Densité de charge électrique \\
	& $\eta$ & Résistivité électrique \\ 
	& $\sigma$ & Conductivité électrique \\ 
	& $m_e$ (resp. $m_i$) & Masse de l'électron (resp. de l'ion) \\ 
	
\end{tabular}

\newpage

\begin{tabular}{lC{5cm}L{8.cm}}
	& $\omega_{pe}$ (resp. $\omega_{pi}$) & Pulsation plasma des électrons (resp. des ions) \\

	& $\Omega_{ce}$ (resp. $\omega_{ci}$) & Pulsation cyclotronique des électrons (resp. des ions) \\ 
	& $\nu_{\alpha \beta}$ & Fréquence efficace d'échange de quantité de mouvement entre les particule $\alpha$ et $\beta$ \\ 
	& $\mathbf{F}_{\alpha \beta}$ &  Force de friction due aux collisions entre les espèces $\alpha$ et $\beta$ \\ 
	&  $\gamma$ & Indice adiabatique \\ 
	& $e$ & Charge électrique élémentaire ($1.6 \times 10^{-19} \  \mathrm{ C}$) \\
	& $\epsilon_0$ & Permittivité du vide ($\approx \frac{1}{36 \pi}\times 10^{-9} \ \mathrm{ F.m^{-1}}$) \\
	&$\mu_0$ & Perméabilité du vide ($\approx 4 \pi \times 10^{-7}\ \mathrm{ H.m^{-1}} $) \\
	& $c$ & Célérité de la lumière dans le vide \\
	& $\mathbf{U}$ & Vecteur des variables conservatives \\ 
	& $\mathbf{F}(\mathbf{U})$ & Fonction de flux \\ 
	&$\Delta t$ (resp. $\Delta x$) & Pas de temps du schéma numérique (resp. pas en espace)\\

\end{tabular}

\newpage

\chapter{Introduction}

\section{Contexte}

L'ionosphère terrestre est un gaz partiellement ionisé qui enveloppe la Terre et qui, dans une certaine mesure, joue le rôle d'interface entre l'atmosphère et l'espace. Puisque le gaz est ionisé, il ne peut être entièrement décrit par les équations de la dynamique des fluides pour les neutres. En réalité, la compréhension de la physique de l'ionosphère nécessite la prise en compte de la physique des plasmas. Néanmoins, la densité des particules neutres dépasse celle du plasma et ne doit donc pas être ignorée. Ainsi, pour étudier l'ionosphère, il est nécessaire d'être familier avec la mécanique des fluides \og classique \fg \,  et la physique des plasmas. 
Cependant, même ces deux branches de la physique ne sont pas toujours suffisantes, puisque l'ionosphère fait office d'interface entre deux milieux très différents. Il est également conseillé d'avoir des bases en dynamique atmosphérique (vent d'altitude...) et interaction Soleil-Terre (rayonnement solaire, vent solaire, éjection de masse coronale...) pour comprendre les caractéristiques et l'évolution de l'ionosphère. 
A tout cela, s'ajoute la connaissance de la chimie et photochimie qui pilotent la production et les pertes des espèces (ionisation, recombinaison, photo-ionisation...) \cite{kelley2009earth}. 

\subsection{Intérêt}

Bien que l'ionosphère puisse nous paraître lointaine, car elle débute à environ 100 km d'altitude, certains phénomènes peuvent quand même avoir des impacts négatifs sur les activités humaines, aussi bien civiles que militaires.
L'ionosphère permet de communiquer sur de grandes distances puisqu'elle réfléchit les ondes radio haute-fréquence qui reviennent ensuite sur Terre, sur un rayon pouvant atteindre plusieurs centaines de km. Néanmoins, ce mode de communication longue distance est soumis aux aléas de l'ionosphère, notamment avec le cycle Jour/Nuit qui peut modifier cette distance. 
Des  recherches sont en cours pour permettre aux satellites  d'utiliser l'ionosphère comme moyen de générer de l'électricité ou de se propulser grâce à un long câble électrodynamique \cite{cosmo1997tethers}. 

\subsection{Phénomènes naturels}

L'ionosphère, tout comme l'atmosphère, est sensible aux différentes fluctuations qui peuvent avoir lieu. Ainsi, nous allons présenter deux perturbations: l'orage magnétique qui peut impacter notre vie quotidienne et les jets équatoriaux qui altèrent plus spécifiquement les communications radio. 

\paragraph{Orage magnétique (\og geomagnetic storm \fg)}

Cette perturbation est la plus connue du grand public, puisque c'est lors de ces orages magnétiques qu'apparaissent les aurores boréales.
Provoqués par des éjections coronales provenant du Soleil, ces orages perturbent le champ magnétique terrestre et apportent également une quantité non négligeable de particules ionisées à haute énergie dans l'ionosphère.
Certaines de ces tempêtes sont restées célèbres pour leur intensité, notamment l'orage géomagnétique de mars 1989 qui a provoqué une coupure d'électricité générale dans une partie du Québec \cite{quebec}, et a aussi perturbé les communications d'une opération militaire australienne en Namibie pendant presqu'une semaine \cite{horner2011australia}. 
Outre les conséquences sur nos systèmes électriques et nos communications radio, d'autres effets moins directs sont observés comme par exemple l'augmentation de la vitesse de corrosion des pipelines due aux courants induits \cite{osella1998currents}. 

\paragraph{Equatorial Spread F (ESF)}

Les \og Equatorial Spread F\fg  (ESF), également appelés \og Convective Equatorial Ionospheric Storm \fg, ont été observés pour la première fois dans les années 30. En certaines occasions, l'écho des ondes réfléchies par l'ionosphère était étalé ( \og spread \fg) en espace et en fréquence. 

On sait maintenant que c'est un phénomène fréquent apparaissant au niveau de l'équateur juste après le coucher du Soleil et qu'il provient d'une perturbation de la région F de l'ionosphère. 
Cet étalement spectral est provoqué par une bulle de plasma peu dense, de l'ordre de $n_i=10^6 \ \mathrm{ cm^{-3}}$, qui s'élève au travers d'une région plus dense, de l'ordre de $n_i=10^4\ \mathrm{  cm^{-3}}$. Le mécanisme majeur de ce phénomène est l'instabilité de Rayleigh-Taylor magnétisée. D'autres phénomènes, comme les ondes de gravité ou les instabilités de cisaillement entrent également en jeu pour initier la perturbation et ainsi déterminer l'ordre de grandeur de l'instabilité de Rayleigh-Taylor \cite{kelley2009earth}.

\subsection{Ordres de grandeur}

Dans le tableau \ref{ordrephy}, sont répertoriées les grandeurs caractéristiques physiques de l'ionosphère. 

\begin{table}[!h]
    \centering
 \begin{tabular}{|l|l|l|}
     \hline
      Grandeur physique & Notation & Ordre de grandeur  \\
     \hline
     Temps caractéristique & $t$ & $10^3 \ \mathrm{ s} $\\
     Longueur caractéristique & $L$ & $10^2\ \mathrm{ km}$ \\ 
     Vitesse du plasma & $V$ & $10^2 \ \mathrm{ m}$ \\ 
     Concentration du plasma & $n$ & $10^6\ \mathrm{ cm^{-3}}$ \\ 
     Pression cinétique & $p$ & $10^{-5} \sim 10^{-8}\ \mathrm{ Pa}$ \\
     Champ magnétique & $B$ & $10^{-5}\ \mathrm{ T}$ \\ 
     Résistivité & $\eta$ & $10^{-1} \ \mathrm{\Omega}$ \\
     \hline
\end{tabular}
    \caption{Valeur dans l'ionosphère des grandeurs physiques importantes}
    \label{ordrephy}
\end{table}

Le plasma ionosphérique est un plasma de faibles concentration et température; on peut le comparer à d'autres types de plasma grâce au tableau \ref{plasmatype}. 

\begin{table}[!h]
    \centering
 \begin{tabular}{|l|l|l|}
     \hline
     Type de plasma & Concentration ($\mathrm{ cm^{-3}}$) & Température ($ \mathrm{ K}$) \\
     \hline
     Ionosphère & $10^4 \sim 10^6$ & $10^2 \sim 10^3$ \\
     Centre du Soleil & $10^{25} \sim 10^{26}$ & $ 10^7 $ \\ 
     Tokamak & $10^{13} \sim 10^{14}$ & $10^8$ \\
     FCI & $10^{23} \sim 10^{26}$ & $10^8 $\\
     \hline  
\end{tabular}
    \caption{Concentration et température pour différents types de plasma}
    \label{plasmatype}
\end{table}

\newpage

\section{Objectif du stage}

Pour comprendre l'apparition et l'évolution de ces phénomènes, le CEA développe un code de physique qui décrit la dynamique des particules neutres et ionisées dans le champ magnétique terrestre. 
Il utilise la méthode des Volumes Finis de type Godunov avec des solveurs de Riemann de Roe ou de type HLL pour résoudre les équations de l'hydrodynamique (équations d'Euler), aussi bien que pour résoudre celles de la magnétohydrodynamique (équations d'Euler-Maxwell).\\
L'étape de validation, qui finalise la phase de développement de chaque solveur, permet de comparer les résultats du code à des cas-tests dont les solutions sont connues, soit de manière analytique, soit de manière expérimentale.
A défaut, on compare, si possible quantitativement mais souvent qualitativement, le résultat du code à ceux d'autres codes.\\

L'objet de ce stage est de constituer une série de cas-tests représentatifs de divers phénomènes physiques pouvant apparaitre dans l'ionosphère : propagation d'ondes, émergence d'instabilités, etc.\\


Dans un premier temps, on décrira le CEA et plus précisément la Direction des applications militaires.
Dans la deuxième partie, on rappellera les équations de conservation, et les équations de Maxwell pour finalement aboutir au système d'équations de la MHD idéale.
La troisième partie traitera des solveurs de Riemann utilisés dans le code avec notamment le solveur de Roe et les solveurs de type HLL.
La quatrième partie listera les cas-tests implémentés avec leurs conditions initiales et des exemples de simulations.
Finalement dans une cinquième partie, une étude plus approfondie de l'instabilité de Rayleigh-Taylor sera réalisée afin de proposer une approche plus quantitative.
Et enfin une conclusion générale sur le stage terminera ce document.
     
\clearpage

\chapter{Présentation de l'organisme d'accueil}

\section{Le CEA}

Acteur majeur de la recherche, du développement et de l'innovation, le Commissariat à l’énergie atomique et aux énergies alternatives intervient dans quatre domaines : 

\begin{itemize}
 \item la défense et la sécurité ;
 \item les énergies bas-carbone (nucléaires et renouvelables) ;
\item la recherche technologique pour l’industrie ;
\item la recherche fondamentale (sciences de la matière et sciences de la vie).
\end{itemize}

S’appuyant sur une capacité d’expertise reconnue, le CEA participe à la mise en place de projets de collaboration avec de nombreux partenaires académiques et industriels.
Le CEA est implanté sur neuf centres répartis dans toute la France. Il développe de nombreux partenariats avec les autres organismes de recherche, les collectivités locales et les universités. A ce titre, le CEA est partie prenante des alliances nationales coordonnant la recherche française dans les domaines de l’énergie (ANCRE), des sciences de la vie et de la santé (AVIESAN), des sciences et technologies du numérique (ALLISTENE), des sciences de l’environnement (AlIEnvi) et des sciences humaines et sociales (ATHENA).
Reconnu comme un expert dans ses domaines de compétence, le CEA est pleinement inséré dans l'espace européen de la recherche et exerce une présence croissante au niveau international.
Le CEA compte 15 942 techniciens, ingénieurs, chercheurs et collaborateurs pour un budget de 5 milliards d'euros (chiffres publiés fin 2017).
\begin{figure}[!h]
    \centering
    \includegraphics[scale=0.35]{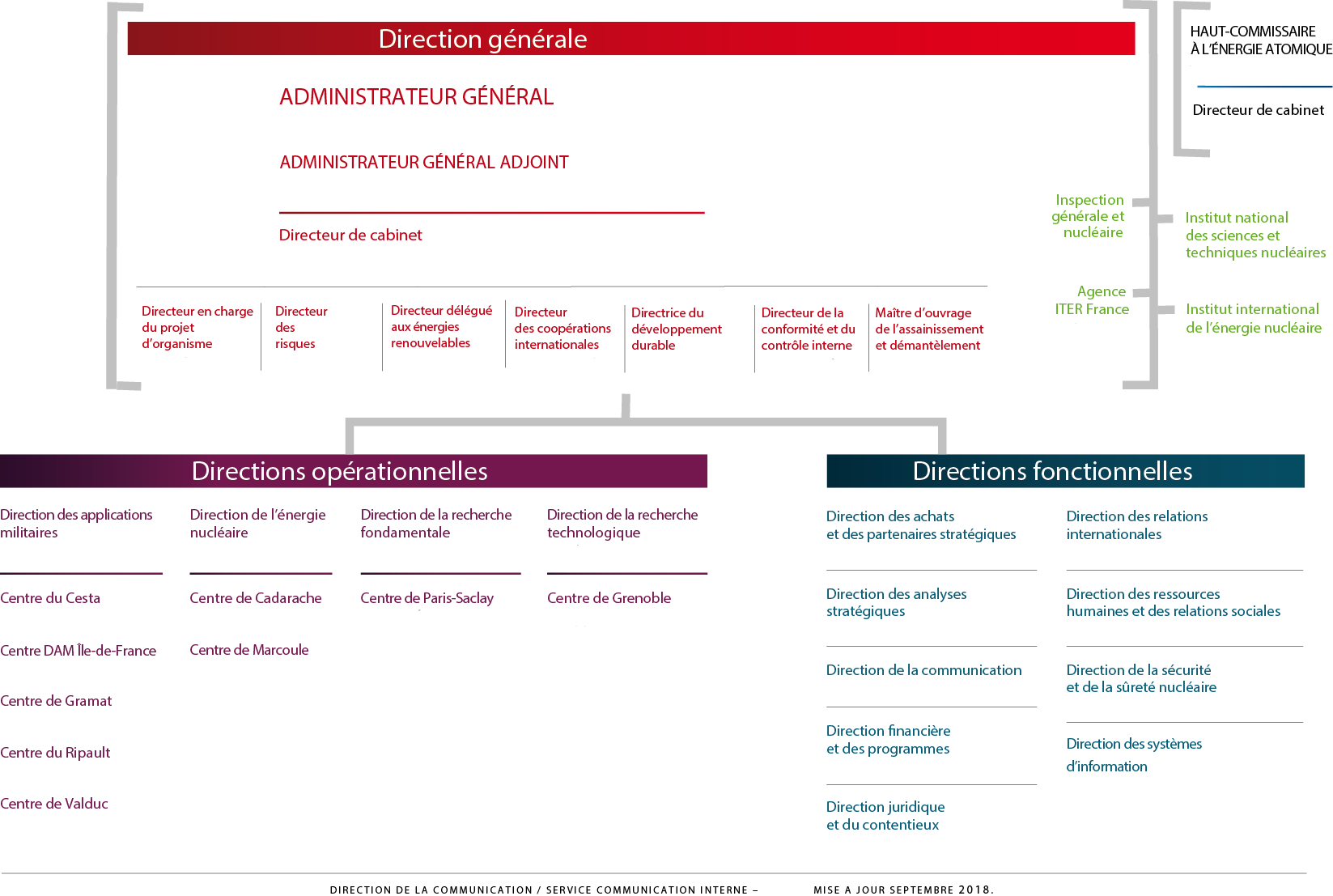}
    \caption{L’organigramme du CEA}
\end{figure}

\section{La Direction des applications militaires}

\subsection{Une direction au service de la dissuasion}

La Direction des applications militaires (DAM) du CEA, a pour mission de concevoir, fabriquer, maintenir en condition opérationnelle, puis démanteler les têtes nucléaires qui équipent les forces nucléaires aéroportées et océaniques françaises.
La DAM est chargée de la conception et de la réalisation des réacteurs et de cœurs nucléaires équipant les bâtiments de la Marine nationale, sous-marins et porte-avions. Elle apporte son soutien à la Marine nationale pour le suivi en service et le maintien en condition opérationnelle de ses réacteurs.
La DAM est également responsable de l'approvisionnement des matières nucléaires stratégiques pour les besoins de la dissuasion.
Dans un monde en profonde mutation, la DAM contribue aussi à la sécurité nationale et internationale à travers l’appui technique qu’elle apporte aux autorités, pour les questions de lutte contre la prolifération nucléaire et le terrorisme et de désarmement.
Depuis le transfert du centre de Gramat en 2010 de la Direction générale de l’armement au CEA, la DAM apporte son expertise à la Défense dans le domaine de l’armement conventionnel.

\subsection{Une direction ouverte à la recherche}

Le partage national et international des connaissances (lorsqu’il est possible), la confrontation à l’évaluation scientifique extérieure, l’intégration à des réseaux de compétences constituent des gages de crédibilité scientifique.
Les équipes de la DAM réalisent chaque année environ 2000 publications et communications scientifiques. Cette ouverture de la DAM passe également par la mise à la disposition de la communauté des chercheurs de ses moyens expérimentaux et par la contribution de ses équipes à d’autres programmes de recherche.

\subsection{Une direction actrice de la politique industrielle française}

La DAM partage très largement son activité avec l’industrie française : c’est ainsi que le montant des achats, auprès de celle-ci, représente plus des deux tiers de son budget ; le dernier tiers se répartit entre les salaires des personnels (un cinquième) et les taxes.

La politique industrielle de la DAM est originale à plus d’un titre :

\begin{itemize}
\item d’abord parce que la DAM conserve la maîtrise d’œuvre d’ensemble de la grande majorité des systèmes dont elle a la responsabilité : elle veille ainsi au juste équilibre entre les grands groupes industriels de la Défense et les PME souvent innovantes, en contractualisant directement avec ces dernières, leur permettant ainsi de recevoir la juste rémunération de leur production;
\item ensuite, parce que la répartition de son budget est sous-tendue par une répartition des travaux : la DAM conduit la recherche dans ses laboratoires grâce à son personnel de haut niveau scientifique et technologique. Une fois la définition d’un produit acquise, la DAM transfère la définition et les procédés vers les industriels qui en réalisent le développement, puis la production.

\end{itemize}

La DAM a également pour objectif que ses centres participent à la vie économique locale par leur implication dans les pôles de compétitivité. Hors de son propre champ d’utilisation, elle valorise ses recherches par le transfert de technologies vers l’industrie et le dépôt de nombreux brevets.

\subsection{Le format}

La DAM comprend cinq centres aux missions homogènes, dont les activités se répartissent entre la recherche de base, le développement et la fabrication :

\begin{itemize}

\item \textbf{DAM Ile-de-France (DIF)}, à Bruyères-le-Châtel, où sont menés les travaux de physique des armes, les activités de simulation numérique et de lutte contre la prolifération nucléaire ; DIF est aussi le centre responsable de l’ingénierie à la DAM ; enfin, au centre DIF est rattachée l’INBS-Propulsion Nucléaire du centre CEA/Cadarache, en région Provence Alpes-Côte d’Azur, où sont implantées les installations d'essais à terre et une partie des fabrications de la propulsion nucléaire ;

\item   \textbf{Le Cesta, en Aquitaine}, consacré à l’architecture des armes, aux tests de tenue à l’environnement. Il met en œuvre le Laser Mégajoule, équipement majeur de la Simulation ;

\item \textbf{Valduc}, en Bourgogne, dédié aux matériaux nucléaires et à l’installation expérimentale Epure du programme Simulation ;

\item \textbf{ Le Ripault}, en région Centre, dédié aux matériaux non nucléaires (explosifs, chimiques…);

\item \textbf{Gramat}, (ex-DGA) en Midi-Pyrénées, qui conduit au profit de la Défense des activités en vulnérabilité des systèmes et efficacité des armements.
\end{itemize}

\begin{figure}[!h]
    \centering
    \includegraphics[scale=0.6]{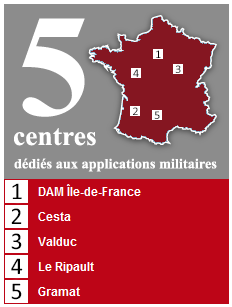}
    \caption{Carte des centres de la direction des applications militaires}
\end{figure}

\section{Le centre DAM Ile de France}

Le CEA/DAM - Île de France (DIF) est l'une des directions opérationnelles de la DAM.
Le site de la DIF compte environ 2000 salariés CEA et accueille quotidiennement environ 600 salariés d'entreprises extérieures. Il est situé à Bruyères-le-Châtel à environ 40 km au sud de Paris, dans l'Essonne.
Les missions de la DIF comprennent :

\begin{itemize}
\item la conception et garantie des armes nucléaires, grâce au programme Simulation. L'enjeu consiste à reproduire par le calcul les différentes phases du fonctionnement d'une arme nucléaire et à confronter ces résultats aux mesures des tirs nucléaires passés et aux résultats expérimentaux obtenus sur les installations actuelles (machine radiographique, lasers de puissance, accélérateurs de particules) ;
\item la lutte contre la prolifération et le terrorisme, en contribuant notamment au programme de garantie du Traité de Non-Prolifération et en assurant l'expertise technique française pour la mise en œuvre du Traité d'Interdiction Complète des Essais Nucléaires (TICE);
\item l'expertise scientifique et technique, dans le cadre de la construction et du démantèlement d'ouvrages complexes ainsi que pour la surveillance de l'environnement et les sciences de la terre ;
\item l'alerte des autorités, mission opérationnelle assurée 24h sur 24, 365 jours par an, en cas d'essai nucléaire, de séisme en France ou à l'étranger, et de tsunami dans la zone Euro-méditerranéenne. La DIF fournit aux autorités les analyses et synthèses techniques associées.
\end{itemize}

\begin{figure}[!h]
    \centering
    \includegraphics[scale=0.3]{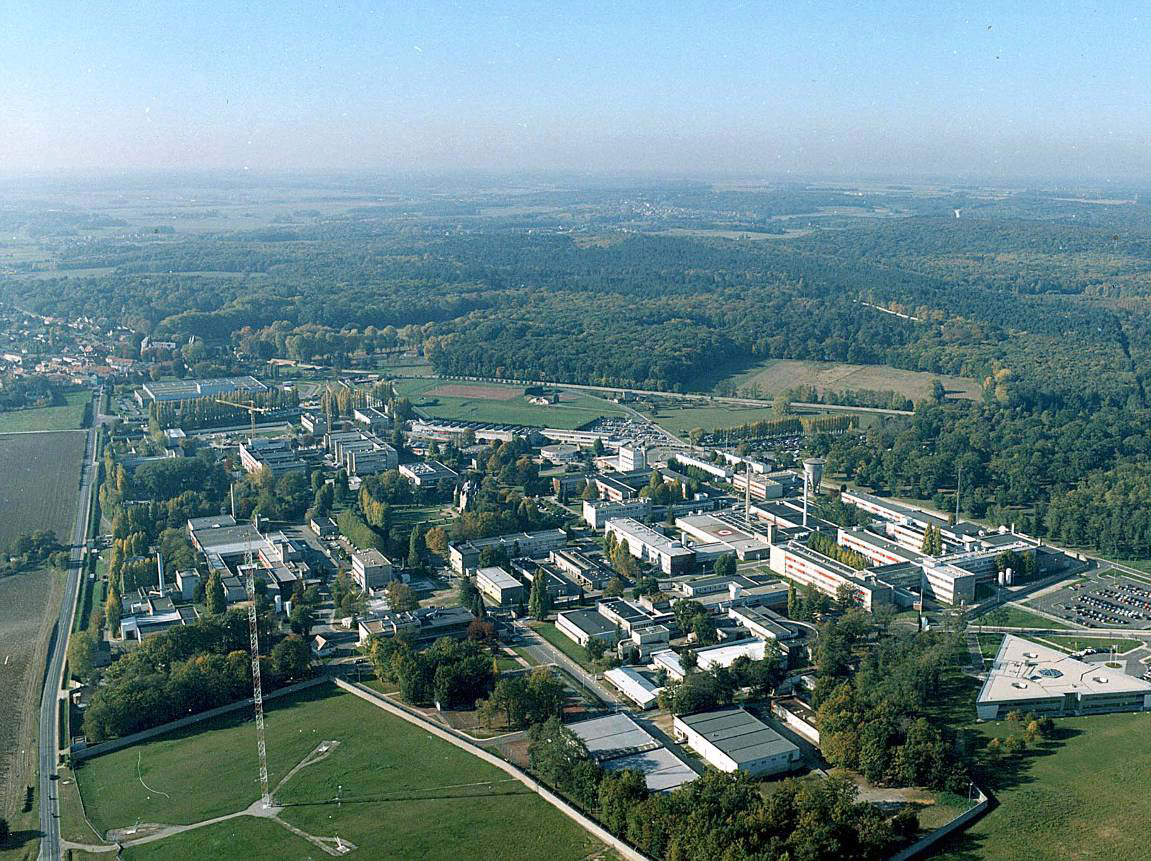}
    \caption{Le centre DAM Île-de-France}
\end{figure}

Depuis 2003, le centre DAM-Île-de-France héberge le complexe de calcul scientifique du CEA, qui regroupe l’ensemble des supercalculateurs du CEA, et qui comprend à ce jour:

\begin{itemize}

\item le supercalculateur Tera1000-1 pour les besoins du programme Simulation du CEA/DAM, mis en service en 2016, qui dispose d’une puissance de calcul de $1{,}25$ petaflops, c’est à dire capable d’effectuer $2{,}5$ millions de milliards d’opérations par seconde.
Il est complété en 2018 par Tera1000-2, autre composante du projet Tera1000, qui préfigure les architectures et technologies du futur supercalculateur qui sera installé à l’horizon 2020. Sa puissance de calcul est de 25 petaflops.

\item le supercalculateur Cobalt du Centre de Calcul pour la Recherche et la Technologie (CCRT), ouvert à la communauté civile de la recherche et de l’industrie, pour une puissance globale de $1{,}5$ petaflops ;

\item le supercalculateur IRENE, d’une puissance de 9 petaflops, deuxième élément d’un réseau de supercalculateurs de classe petaflopique destiné aux chercheurs de la communauté scientifique européenne. Ce supercalculateur est hébergé au TGCC (Très Grand Centre de Calcul) et exploité par les équipes du CEA, qui apporte ainsi sa contribution à la participation de la France au projet PRACE (Partnership for Advanced Computing in Europe).

\end{itemize}

\clearpage

\chapter{Dynamique d'un plasma}

\section{Équations fondamentales}

Un plasma est composé d'ions et d'électrons soumis tous deux aux forces électromagnétiques déterminées par le champ électrique $\mathbf{E}$ et le champ magnétique $\mathbf{B}$. Pour étudier la dynamique de ce milieu, il faut donc d'abord considérer les équations hydrodynamiques pour les deux fluides. Soit l'équation de conservation de la masse, avec $\rho_s$ la masse volumique de du fluide $s$ et $\mathbf{V}_s$ la vitesse du fluide  $s$,

\begin{equation}
           \partial_t \rho_{e,i} + \nabla \cdot (\rho_{e,i} \mathbf{V}_{e,i}) = 0 \quad (+S-L)
    \label{masses_ei}
\end{equation}

l'équation de conservation de la quantité de mouvement, 

\begin{equation}
        \partial_t (\rho_{e,i} \mathbf{V}_{e,i}) + \nabla \cdot (\rho_{e,i} \mathbf{V}_{e,i} \otimes \mathbf{V}_{e,i} ) = -\nabla p_{e,i} + q_{e,i} n_{e,i} (\mathbf{E} + \mathbf{V}_{e,i} \times \mathbf{B}) +\mathbf{F}_{ei,ie} \quad (+\mathbf{F}_{en,in}+\rho_{e,i} \mathbf{g} + \mathbf{F}_{e,i}^{SL}) 
    \label{mouvement_ei}
\end{equation}

et l'équation de conservation de l'énergie hydrodynamique $\mathcal{E}_s$ définie par $\mathcal{E}_s = \frac{p_s}{\gamma_s -1} + \frac{1}{2} \rho_s \mathbf{V}_s^2$ avec $\gamma_s$ l'indice adiabatique du fluide $s$ (pour un gaz mono-atomique $\gamma_e=\gamma_i=\frac{5}{3} = \gamma$). 

\begin{equation}
    \partial_t    \mathcal{E}_{e,i}   + \nabla \cdot [(\mathcal{E}_{e,i} + p_{e,i})\mathbf{V}_{e,i}] = q_{e,i} n_{e,i} \mathbf{E} \cdot \mathbf{V}_{e,i} + Q_{ei,ie} \quad (+ Q_{en,in}+ \rho_{e,i} \mathbf{V}_{e,i} \cdot \mathbf{g} + Q_{e,i}^{SL})
    \label{energie_ei}
\end{equation}

où $e$, $i$ et $n$ font référence aux électrons, aux ions et aux neutres respectivement, $p_s$ est la pression cinétique du fluide $s$, $q_s$ est la charge de la particule $s$ (ici $q_i=-q_e= e$; on ne prend pas en compte les ionisations multiples), $n_s$ est la concentration du fluide $s$, $\mathbf{F}_{\alpha \beta}$ est la force de friction que subissent les particules $\alpha$ entrant en collision avec les particules $\beta$ définie par $\mathbf{F}_{\alpha \beta}= - \nu_{\alpha \beta} \rho_\alpha (\mathbf{V}_\alpha - \mathbf{V}_\beta)$, $Q_{\alpha \beta}$ est le terme de variation d'énergie associée à la friction $\mathbf{F}_{\alpha \beta}$, $\mathbf{g}$ est la force de gravité, $S$ est un terme source provenant de l'ionisation de neutres par divers phénomènes, $L$ est un terme de pertes dues aux recombinaisons entre les ions et les électrons pour reformer un atome neutre et $\mathbf{F}_s^{SL}$ et $Q_s^{SL}$ représentent les variations de quantité de mouvement et d'énergie respectivement dues aux termes source et de pertes pour le fluide $s$. 

Les termes entre parenthèses à droite représentent les effets de la gravité et la présence de neutres dans un plasma partiellement ionisé qui sont spécifiques à l'étude de l'ionosphère et qui ne feront que quelques apparitions dans la suite de notre étude. Nous avons ici négligé le terme de viscosité $\nabla \mathbf{\sigma}_{vs}$, où $\mathbf{\sigma}_{vs}$ est le tenseur des contraintes visqueuses du  fluide $s$, et le flux de chaleur.  

Pour compléter ce jeu d'équations, nous avons besoin des équations de Maxwell qui décrivent l'évolution spatiale et temporelle  du champ électromagnétique: 

\begin{equation}
    \begin{array}{lr}
         \nabla \times \mathbf{E} = - \partial_t \mathbf{B}  &\textrm{(Maxwell-Faraday)} \\
        \nabla \times \mathbf{B} =  \mu_0 \mathbf{J} + \frac{1}{c^2} \partial_t \mathbf{E}&\textrm{(Maxwell-Ampère)}  \\ 
        \nabla \cdot \mathbf{E} = \frac{Q}{\epsilon_0} &\textrm{(Maxwell-Gauss)}\\
        \nabla \cdot \mathbf{B} = 0&\textrm{(Maxwell-Flux)}
    \end{array}
\end{equation}

où $\mathbf{J}$ est la densité de courant définie par $\mathbf{J} = \sum \limits_{e,i} q_s n_s \mathbf{V_s}$, $Q$ est la densité de charges électriques définie par $Q = \sum \limits_{e,i} q_s n_s$, $\mu_0$ est la perméabilité du vide, $\epsilon_0$ est la permittivité du vide et $c$ est la célérité de la lumière dans le vide avec $c^2= \frac{1}{\epsilon_0 \mu_0}$. 
On peut aussi écrire le théorème de Poynting qui décrit la variation de la conservation de l'énergie électromagnétique: 

\begin{equation}
    \partial_t \mathcal{E}_{EM} + \nabla\cdot \left(\frac{\mathbf{E}\times \mathbf{B}}{\mu_0}\right) = - \mathbf{E} \cdot \mathbf{J}     
\end{equation}
 
où $\mathcal{E}_{EM}= \frac{\epsilon_0 \mathbf{E}^2}{2} + \frac{\mathbf{B}^2}{2\mu_0 }$.

\section{Hypothèses de la Magnétohydrodynamique (MHD)}

\subsection{Généralités}

 Nous allons maintenant en déduire le système d'équations MHD usuel. Comme la masse des ions est très grande devant la masse des électrons ($m_i \gg m_e$ avec $m_i$ la masse d'un ion et $m_e$ la masse d'un électron), nous allons assimiler la masse volumique et la quantité de mouvement du plasma à celle des ions ($\rho = \rho_i + \rho_e \approx  \rho_i$ et $\rho \mathbf{V} = \rho_i \mathbf{V}_i + \rho_e \mathbf{V}_e \approx \rho_i \mathbf{V}_i$), cette dernière approximation n'étant valable que si la vitesse des électrons n'est pas très grande devant celle des ions ($V_i \approx V_e$ ou $V_i  \gg V_e$). Ainsi nous obtenons à partir de (\ref{masses_ei}), l'équation de conservation de la masse pour notre plasma: 
\begin{equation}
\partial_t \rho + \nabla \cdot (\rho  \mathbf{V}) = 0 
\end{equation}

Nous travaillons également dans l'hypothèse de quasi-neutralité $n_e=n_i = n $; le courant devient  $\mathbf{J}  = en(\mathbf{V}_i - \mathbf{V}_e)$. En additionnant la quantité de mouvement des ions et des électrons, on obtient alors pour la quantité de mouvement du plasma:

\begin{equation}
\partial_t ( \rho \mathbf{V}) + \nabla \cdot (\rho \mathbf{V} \otimes \mathbf{V}) =  \mathbf{J} \times \mathbf{B}-\nabla p
\end{equation}

avec $p=p_e+p_i$. 

De même, nous obtenons l'équation de la conservation d'énergie hydrodynamique à partir de la somme des équations (\ref{energie_ei}),

\begin{equation}
      \partial_t    \mathcal{E} + \nabla \cdot [(\mathcal{E} + p)\mathbf{V}] = \mathbf{E} \cdot \mathbf{J}
      \label{energiemeca}
\end{equation}

avec $\mathcal{E} =  \mathcal{E}_i + \mathcal{E}_e = \frac{p_e+p_i}{\gamma - 1 }+\frac{1}{2} \rho_i \mathbf{V}_i^2 +\frac{1}{2} \rho_e \mathbf{V}_e^2 $.

Nous allons maintenant introduire une nouvelle hypothèse, celle d'un plasma non relativiste. Nous sommes en effet intéressés par des écoulements fluides avec des vitesses caractéristiques d'écoulement très petites devant la vitesse de la lumière: $V \ll c$. Il est important de noter que ceci n'est valable que dans le cas de la MHD; pour des interactions rayonnement/matière ayant lieu dans le plasma cela n'est pas correct (exemple: absorption d'onde cyclotronique). Ici, nous nous plaçons dans un cas neutre et donc nous aurons des vitesses caractéristiques inférieures ou de l'ordre de grandeur de la vitesse des ions qui est rarement proche de la célérité de la lumière ($V_i \ll c$). Ainsi le terme $\frac{1}{c^2} \partial_t \mathbf{E}$ peut être négligé. Les équations de Maxwell deviennent: 

\begin{equation}
    \begin{array}{lr}
         \nabla \times \mathbf{E} = - \partial_t \mathbf{B}  &\textrm{(Maxwell-Faraday)} \\
        \nabla \times \mathbf{B} =  \mu_0 \mathbf{J} &\textrm{(Maxwell-Gauss)}  \\ 
        \nabla \cdot \mathbf{E} = 0 &\textrm{(Maxwell-Faraday)}\\
        \nabla \cdot \mathbf{B} = 0&\textrm{(Maxwell-Flux)}
    \end{array}
    \label{Maxwelllaw}
\end{equation}
Sous cette même hypothèse, l'énergie magnétique devient $\mathcal{E}_{EM} = \frac{\mathbf{B}^2}{2\mu_0  }$.

\subsection{Loi d'Ohm}

Nous allons maintenant dériver la loi d'Ohm, qui établit une relation entre le champ électrique, le champ magnétique et le courant. Suivant les approximations faites à ce niveau, le système d'équations MHD peut devenir un système MHD idéale, résistif ou encore prendre d'autres formes. 

Nous obtenons facilement sous les mêmes hypothèses: quasi-neutralité $n=n_e=n_i$ et $ m_i \| \mathbf{V}_i\|  \gg m_e \| \mathbf{V}_e \|$ ($ \rho_s = m_s n_s$), que $\mathbf{V}_i = \mathbf{V}$ et $\mathbf{V}_e  = \mathbf{V}-\frac{1}{ne}\mathbf{J}$. Ainsi les équations de conservation de la quantité de mouvement deviennent:

\begin{equation}
\begin{split}
\partial_t \mathbf{V} -\frac{1}{ne} \partial_t \mathbf{J} + \nabla \cdot [ (\mathbf{V} - \frac{1}{ne} \mathbf{J} ) \otimes (\mathbf{V} - \frac{1}{ne} \mathbf{J} )] = - \frac{1}{nm_e} \nabla p_e -\frac{e}{m_e} \mathbf{E}  \\
-\frac{e}{m_e} (\mathbf{V}-\frac{1}{ne}\mathbf{J}) \times \mathbf{B}  + \frac{\nu_{ei}}{ne} \mathbf{J} 
\end{split}
\label{ebe}
\end{equation}

\begin{equation}
\partial_t \mathbf{V} + \nabla \cdot  (\mathbf{V}  \otimes \mathbf{V}  ) = - \frac{1}{nm_i} \nabla p_i +\frac{e}{m_i} \mathbf{E} + \frac{e}{m_i} \mathbf{V} \times \mathbf{B} - \frac{\nu_{ie}}{ne} \mathbf{J} 
\label{ibi}
\end{equation}

En multipliant par $e$ et en soustrayant (\ref{ebe}) à (\ref{ibi}), nous obtenons la loi d'Ohm généralisée: 

\begin{equation}
\begin{split}
\frac{1}{ne} [ \partial_t \mathbf{J} + \nabla \cdot ( \mathbf{V}\otimes \mathbf{J} + \mathbf{V} \otimes \mathbf{J}) - \frac{1}{ne} \nabla \cdot \mathbf{J} \otimes \mathbf{J}] = \frac{1}{nm_e} \nabla p_e - \frac{1}{nm_i} \nabla p_i + \left(\frac{1}{m_e} + \frac{1}{m_i} \right)e \mathbf{E} \\
+ \left(\frac{1}{m_e} + \frac{1}{m_i} \right) e\mathbf{V} \times \mathbf{B} - \frac{1}{m_ene} \mathbf{J} \times \mathbf{B}-\frac{\nu_{ei} + \nu_{ie}}{ne} \mathbf{J}
\end{split}
 \end{equation}

soit:
\begin{equation}
\begin{split}
\frac{m_em_i}{m_e+m_i} \left[ \partial_t \mathbf{J} + \nabla \cdot ( \mathbf{V}\otimes \mathbf{J} + \mathbf{V} \otimes \mathbf{J}) - \frac{1}{ne} \nabla \cdot \mathbf{J} \otimes \mathbf{J} \right] =  ec_i \nabla p_e - ec_e \nabla p_i \\ + e^2n( \mathbf{E} + \mathbf{V} \times \mathbf{B}-\eta \mathbf{J}) - c_i\mathbf{J} \times \mathbf{B}
\end{split}
\end{equation}

avec $c_e= \frac{m_e}{m_e+m_i}$, $c_i= \frac{m_i}{m_e+m_i}$ et $\eta = \frac{\nu_{ei}+\nu_{ie}}{ne^2} \frac{m_em_i}{m_e+m_i}$. 

Comme nous travaillons avec l'approximation $m_e \ll m_i$, $c_e \rightarrow 0 $ et $ c_i \approx 1$. En ajoutant l'hypothèse: $p_i = p_e=\frac{1}{2}p $ (ou de façon équivalente $T_e=T_i$), nous pouvons faire une première approximation de notre loi d'Ohm: 

\begin{equation}
m_e \left[ \partial_t \mathbf{J} + \nabla \cdot ( \mathbf{V}\otimes \mathbf{J} + \mathbf{V} \otimes \mathbf{J}) - \frac{1}{ne} \nabla \cdot \mathbf{J} \otimes \mathbf{J} \right] =  \frac{1}{2} \nabla p + e^2n( \mathbf{E} + \mathbf{V} \times \mathbf{B}-\eta \mathbf{J}) - \mathbf{J} \times \mathbf{B}
\label{omh1}
\end{equation}

\subsubsection{Adimensionnement}

Pour déterminer l'importance de chaque terme, nous allons maintenant adimensionner notre équation (\ref{omh1}) (ceci de manière similaire à la thèse d'Estibals \cite{estibals2017mhd}). Nous définissons la température de référence $T_0$, la longueur de référence $L_0$, la densité de référence $n_0$ et le champ magnétique de référence $B_0$. 

Nous définissons la vitesse de référence comme: 

\begin{equation}
V_0 = \sqrt{\frac{2k_B T_0}{m_i}}
\end{equation}

Nous définissons le temps de référence comme: 

\begin{equation}
    t_0 = \frac{L_0}{V_0}
\end{equation}

D'après l'équation d'état des gaz parfaits, nous définissons la pression de référence comme:

\begin{equation}
    p_0 = n_0 k_B T_0 
\end{equation}

De la loi de Maxwell-Ampère (\ref{Maxwelllaw}), nous obtenons le courant de référence: 

\begin{equation}
    J_0 = \frac{B_0}{L_0 \mu_0}
\end{equation}

De la loi d'Ohm en MHD idéale (voir la suite (\ref{idealohm})), nous obtenons le champ électrique de référence: 

\begin{equation}
    E_0 = V_0 B_0
\end{equation}

Nous définissons maintenant nos paramètres adimensionnés, premièrement, le beta magnétique, très utilisé en science des plasma puisqu'il décrit le rapport entre la pression cinétique et la pression magnétique: 

\begin{equation}
    \beta = \frac{m_i n_0 V_0^2}{B_0^2/\mu_0} = \frac{n_0 k_B 2 T_0 }{B_0^2 \mu_0}
\end{equation}

Deuxièmement les paramètres: 

\begin{equation}
    \begin{array}{c|c}
        \delta_e^* = \frac{\delta_e}{L_0} & \delta_i^* = \frac{\delta_i}{L_0} \\
    \end{array}
\end{equation}

avec 

\begin{equation}
    \begin{array}{c|c}
        \delta_e^2 = \frac{c^2}{\omega_{pe}^2}=\frac{m_e}{n_0 e^2 \mu_0} &\delta_i^2 = \frac{c^2}{\omega_{pi}^2}=\frac{m_i}{n_0 e^2 \mu_0}  \\
    \end{array}
\end{equation}

où $\omega_{pe}$ et $\omega_{pi}$ sont les pulsations plasma des ions et des électrons et $\delta_e$ et $\delta_i$ sont les longueurs d'inertie des ions et des électrons. A noter que dans notre approximation $\delta_e \ll \delta_i$ puisque $m_e \ll m_i$. 

Et enfin le nombre de Reynolds magnétique : 

\begin{equation}
    R_m = \frac{\mu_0 L_0 V_0}{\eta_0}
\end{equation}
qui est en réalité le rapport entre la vitesse d'écoulement et la dissipation magnétique. Dans le cas d'un nombre de Reynolds élevé nous pouvons alors négliger la dissipation magnétique devant les vitesses d'écoulement. 
Ainsi  notre loi d'Ohm généralisée s'écrit: 

\begin{equation}
\begin{split}
\delta_e^{2*} \frac{1}{n}\left[ \partial_t \mathbf{J} + \nabla \cdot ( \mathbf{V}\otimes \mathbf{J} + \mathbf{V} \otimes \mathbf{J})\right] - \frac{\delta_i^* \delta_e^{2*}}{\sqrt{\beta}} \frac{1}{n^2}\nabla \cdot \mathbf{J} \otimes \mathbf{J} = \\
 \sqrt{\beta}\delta_i^* \frac{1}{2n} \nabla p +  \mathbf{E} + \mathbf{V} \times \mathbf{B}-\frac{1}{R_m}\eta \mathbf{J} - \frac{\delta_i^*}{\sqrt{\beta}}\frac{\mathbf{J} \times \mathbf{B}}{n}
\end{split}
\end{equation}

soit en négligeant le terme de gauche puisque $\delta_e^* \rightarrow 0$ ($\delta_e^* \ll \delta_i^*$), notre loi d'Ohm contenant le terme de Hall : 

\begin{equation}
  \mathbf{E} + \mathbf{V} \times \mathbf{B}= \frac{1}{R_m}\eta \mathbf{J} -\sqrt{\beta}\delta_i^* \frac{1}{2n} \nabla p + \frac{\delta_i^*}{\sqrt{\beta}}\frac{\mathbf{J} \times \mathbf{B}}{n}
\end{equation}

Dans notre rapport le terme de pression est négligeable car nous allons travailler dans un scénario avec un faible beta. 

Nous obtenons ainsi les trois types de lois d'Ohm majeures qui nous donnent trois dénominations courantes de la MHD. 

La MHD idéale en considérant $R_m \rightarrow \infty$ et $\delta_i^* \rightarrow 0$:

\begin{equation}
    \mathbf{E} + \mathbf{V} \times \mathbf{B} = 0 
    \label{idealohm}
\end{equation}

La MHD résistive en considérant $R_m$ fini et $\delta_i^* \rightarrow 0$: 
\begin{equation}
    \mathbf{E} + \mathbf{V} \times \mathbf{B} = \eta \mathbf{J}
\end{equation}

et la MHD résistive avec effet Hall en considérant $R_m$ et $\delta^*_i$ finis  : 

\begin{equation}
    \mathbf{E} + \mathbf{V} \times \mathbf{B} = \eta_0 \mathbf{J} +\eta_H \mathbf{J} \times \mathbf{B}
    \label{OhmHall}
\end{equation}

Un adimensionnement similaire peut être fait pour les équations de conservation, mais nous obtiendrons le même résultat, c'est à dire que le terme de l'effet Hall est prépondérant par rapport aux termes que nous avons déjà négligés (dans le cas d'un faible beta, $\beta \rightarrow 0$). Ces approximations sont justifiées par les ordres de grandeur des différents paramètres adimensionnés listés dans le tableau \ref{adimtab}.

\begin{table}[!h]
    \centering
 \begin{tabular}{|l|L{7cm}|l|}
     \hline
      Paramètre adimensionné & Signification   & Ordre de grandeur  \\
     \hline
     $ \varepsilon = \frac{m_e }{m_i}$ & Rapport entre la masse de l'électron et la masse de l'ion & $10^{-3}$ \\
     $ \beta = \frac{p_0 }{B_0^2 \mu_0}$ & Rapport entre la pression cinétique et la pression magnétique & $10^{-4} \sim 10^{-1}$ \\
    $R_m = \frac{\mu_0 L_0 V_0}{\eta_0}$ & Rapport entre la vitesse d'écoulement et la dissipation magnétique & $10^3$ \\
     $\delta_e^*$ & Rapport entre la longueur d'inertie des électrons et la longueur caractéristique  & $10^{-6} \sim 10^{-5}$ \\
     $\delta_i^*$ & Rapport entre la longueur d'inertie des ions et la longueur caractéristique  & $10^{-4} \sim 10^{-3}$ \\
    $\alpha = \frac{u}{c}$ & Rapport entre la vitesse du plasma et la célérité de la lumière & $10^{-6}$ \\
    \hline
\end{tabular}
    \caption{Ordre de  grandeur pour les paramètres adimensionnés.}
    \label{adimtab}
\end{table}

\newpage

Il est donc bon de préciser que ces trois classes de système MHD ne sont valables que dans le cadre de nos hypothèses qui, pour rappel, sont: 

\begin{itemize}
    \item Plasma quasi-neutre
     \item Plasma non relativiste 
    \item Viscosité et flux de chaleur négligeables
    \item Faible beta magnétique 
\end{itemize}

\medskip
Nous avons également effectué le calcul de la loi d'Ohm en faisant apparaître les termes supplémentaires résitif et de l'effet Hall pour une approche tri-fluides. Ce calcul est disponible dans l'annexe \ref{OHm3}. 

L'autre forme de la loi d'Ohm connue dans le domaine de l'étude de l'ionosphère et donnée en annexe \ref{notationprésicion}.

\subsection{MHD idéale}

Nous pouvons maintenant définir le jeu complet d'équations de la MHD idéale: 

\begin{equation}
    \left\{
    \begin{array}{l}
          \partial_t \rho + \nabla \cdot (\rho  \mathbf{V}) = 0 \\
          \partial_t  (\rho \mathbf{V}) + \nabla \cdot (\rho \mathbf{V} \otimes \mathbf{V}) =  \mathbf{J} \times \mathbf{B}-\nabla p\\
          \partial_t    \mathcal{E} + \nabla \cdot [(\mathcal{E} + p)\mathbf{V}] = \mathbf{E}  \cdot \mathbf{J}\\
        \partial_t \mathbf{B} =-  \nabla \times \mathbf{E} \\
           \mu_0 \mathbf{J} =\nabla \times \mathbf{B} \\
          \mathbf{E} =- \mathbf{V} \times \mathbf{B} \\
         \nabla \cdot \mathbf{B} = 0
    \end{array}
    \right.
\end{equation}

\newpage

\chapter{Solveurs de Riemann pour la MHD idéale}
\label{Riemansolveur}
Avec les progrès récents sur les solveurs de Riemann, comme le solveur HLLD \cite{miyoshi2005multi, miyoshi2010hlld} et sa variante (plus rapide) munie de l'approximation de Boris \cite{matsumoto2019new}, il est maintenant possible de modéliser des problèmes MHD avec une bonne précision et des temps de calcul acceptables. 
Des solveurs plus anciens comme le solveur de Roe \cite{roe1981approximate} restent toujours très utilisés que ce soit pour l'hydrodynamique ou la magnétohydrodynamique. D'autres solveurs, comme le prédécesseur du solveur HLLD, le solveur HLLC \cite{toro1994restoration} sont suffisants pour étudier le transport d'un fluide neutre. 

Les solveurs de Riemann utilisent les solutions exactes ou approchées d'un problème de Riemann associées à la méthode des Volumes Finis, par exemple de type Godunov, pour résoudre des systèmes d'équations de conservation hyperboliques.
Le problème de Riemann et les caractéristiques des différents solveurs de Riemann sont très bien décrits dans l'ouvrage de E.Toro \cite{Toro}.

\section{Problème de Riemann}

Le problème de Riemann consiste à étudier l'évolution temporelle de deux états initiaux séparés par une discontinuité et qui suivent un système d'équations hyperbolique. Le problème de Riemann pour un système hyperbolique d'ordre $n\times n $ s'écrit sous la forme: 

\begin{equation}
    \left.
    \begin{array}{ll}
        \text{PDEs:} & \partial_t \mathbf{U}  + \mathbf{A } \partial_x \mathbf{U} =0 \\
        \text{IC:}  & \mathbf{U}(x,0) = \mathbf{U}^{(0)}(x) = 
        \left\{
        \begin{array}{c}
              \mathbf{U}_L \quad x<0, \\
              \mathbf{U}_R \quad x>0
        \end{array}
        \right.
    \end{array}
    \right\}
\end{equation}

avec $ \mathbf{A}$ la matrice jacobienne définie par $ \mathbf{A} = \frac{\partial \mathbf{F}(\mathbf{U})}{\partial \mathbf{U}}$. 

\begin{figure}[!h]
    \centering
    \includegraphics[scale=0.5]{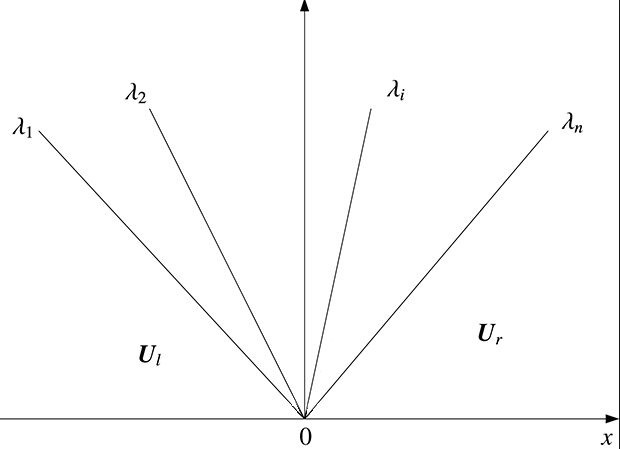}
    \caption{Structure de la solution du problème de Riemann pour un système hyperbolique $n\times n$ avec des coefficients constants}
    \label{fan}
\end{figure}

La structure de la solution au problème de Riemann est représentée sur la figure \ref{fan}. Elle est constituée de $n$ ondes provenant de l'origine, une pour chaque valeur propre $\lambda_i$ de $\mathbf{A}$ (nous avons classé les valeurs propres dans l'ordre croissant $\lambda_1<... <\lambda_i<...<\lambda_n$). Chaque onde porte un saut de discontinuité en $\mathbf{U}$ se propageant à la vitesse $\lambda_i$. Naturellement, la solution à gauche de $\lambda_1$ est l'état initial $\mathbf{U}_L$ et la solution à droite de $\lambda_n$ est l'état initial $\mathbf{U}_R$. Notre travail est de trouver les solutions dans les espaces entre les ondes $1$ et $n$.

\section{Application aux équations de la MHD}

Les équations de conservation s'écrivent sous la forme générique suivante:
\begin{equation}
\partial_t \mathbf{U} + \nabla \cdot \mathbf{F}(\mathbf{U})=0 
\end{equation}

où $\mathbf{U}$ est le vecteur des variables conservatives et $\mathbf{F}(\mathbf{U})$ est le vecteur de flux. 

Dans le cas de la MHD, les composantes de  $\mathbf{U}$ sont:

\begin{equation}
\mathbf{U}= 
\begin{pmatrix}
\rho \\
\rho u \\ 
\rho v \\ 
\rho w \\ 
B_y \\ 
B_z \\ 
\mathcal{E}_t \\
\end{pmatrix}
\end{equation}

avec $\mathbf{V}=(u,v,w)^T $ le vecteur vitesse et $\mathbf{B}=(B_x, B_y , B_z)^T$ le champ magnétique. Ici, à cause de la conservation du flux magnétique (Maxwell-Flux (\ref{Maxwelllaw})), $B_x$ est considéré constant dans le cas à une dimension. 
\medskip

Nous allons maintenant modifier un peu les équations de la MHD idéale pour déterminer $\mathbf{F}(\mathbf{U})$. 

Tout d'abord, l'équation de conservation de la masse devient:
\begin{equation}
\partial_t \rho + \partial_x (\rho u) =0
\end{equation} 

Maintenant, passons à l'équation de conservation de la quantité de mouvement. En utilisant l'équation de Maxwell-Ampère (\ref{Maxwelllaw}), nous pouvons exprimer $\mathbf{J} \times \mathbf{B}$ de la façon suivante: 

\begin{equation}
\mathbf{J} \times \mathbf{B} = \left(\frac{1}{\mu_0} \nabla \times \mathbf{B}\right) \times \mathbf{B} =\frac{1}{\mu_0}\left[ \nabla \cdot (\mathbf{B} \otimes \mathbf{B}) - \nabla\frac{B^2}{2}\right]
\end{equation}

où $\otimes$ est le produit tensoriel et $B^2 = \mathbf{B} \cdot \mathbf{B}$. 

L'équation de conservation de la quantité de mouvement devient alors: 

\begin{equation}
\partial_t \rho\mathbf{V} + \nabla \cdot (\rho \mathbf{V} \otimes \mathbf{V} + p_t \mathbf{I} - \frac{1}{\mu_0} \mathbf{B} \otimes \mathbf{B} ) =0 
\end{equation}

où $p_t$ est la pression totale définie par $p_t = p +\frac{B^2}{2\mu_0}
$. Soit en une dimension: 

\begin{equation}
    \left\{
    \begin{array}{l}
         \partial_t \rho u + \partial_x ( \rho u^2 +p_t - \frac{B_x^2}{\mu_0}   )  =0\\
          \partial_t \rho v + \partial_x ( \rho uv  - \frac{B_xB_y}{\mu_0}   )=0  \\
          \partial_t \rho w + \partial_x ( \rho uw  - \frac{B_xB_z}{\mu_0}   )  =0
    \end{array}
    \right.
\end{equation}

Puis nous combinons l'équation de Maxwell-Faraday (\ref{Maxwelllaw}) avec la loi d'Ohm en MHD idéale (\ref{idealohm}) pour obtenir: 

\begin{equation}
\partial_t \mathbf{B} = \nabla \times (\mathbf{V}\times \mathbf{B}) = \nabla \cdot ( \mathbf{B} \otimes \mathbf{V} - \mathbf{V} \otimes \mathbf{B}) 
\end{equation}

soit en une dimension: 

\begin{equation}
     \left\{
    \begin{array}{l}
         \partial_t B_y +\partial_x (B_y u-B_x v)=0\\
         \partial_t B_z +\partial_x (B_z u-B_x w)=0
    \end{array}
    \right.
\end{equation}

Et enfin l'équation de conservation de l'énergie s'obtient en sommant les équations d'énergie hydrodynamique $\mathcal{E} = \frac{p}{\gamma -1} +\frac{1}{2} \rho \mathbf{V}^2 $ et électromagnétique $\mathcal{E}_{EM}=\frac{B^2}{2\mu_0}$  qui sont respectivement: 

\begin{equation}
    \begin{array}{l}
      \partial_t \mathcal{E} +\nabla \cdot [ (\mathcal{E} +p)\mathbf{V}]=\mathbf{E}\cdot \mathbf{J}   \\
    \partial_t \mathcal{E}_{EM} + \nabla \cdot (\frac{\mathbf{E}\times \mathbf{B}}{\mu_0}) = - \mathbf{E}\cdot \mathbf{J}     
    \end{array}
\end{equation}

ce qui donne avec la loi de Maxwell-Ampère (\ref{Maxwelllaw}) et en une dimension:

\begin{equation}
\partial_t  \mathcal{E}_t +\partial_x \left[ (\mathcal{E}_t +p_t)u -(\mathbf{V}\cdot \mathbf{B})\frac{B_x}{\mu_0} \right]=0 
\end{equation}

où $\mathcal{E}_t$ est l'énergie totale définie par $\mathcal{E}_t = \frac{p}{\gamma -1} +\frac{1}{2} \rho \mathbf{V}^2 + \frac{B^2}{2 \mu_0}$. 
Pour alléger les équations nous allons maintenant inclure la perméabilité du vide dans le champ magnétique, soit $\mathbf{\Tilde{B}}= \mathbf{B}/\sqrt{\mu_0}$.

Ce qui donne donc (en omettant le tilde): 

\begin{equation}
\mathbf{F}(\mathbf{U}) = 
\begin{pmatrix}
\rho u \\ 
\rho u^2 +p_t -B_x^2 \\
\rho uv - B_xB_y \\
\rho uw - B_xB_z \\ 
B_y u -B_x v \\
B_z u - B_x w \\ 
(\mathcal{E}_t + p_t)u -(\mathbf{V}\cdot \mathbf{B})B_x
\end{pmatrix}
\end{equation}

Dans le cas MHD idéale, nous avons 7 valeurs propres distinctes qui correspondent aux vitesses de propagation des deux ondes d'Alfvén (aux vitesses $\lambda_2$ et $\lambda_6$), des quatre ondes magnéto-acoustiques (deux  rapides aux vitesses $\lambda_1$ et $ \lambda_7$ et deux lentes aux vitesses $\lambda_3$ et $\lambda_5$),  d'une onde d'entropie et d'une onde de divergence (à la vitesse $\lambda_4$): 

\begin{equation}
    \begin{array}{cccc}
        \lambda_{1,7} = u \pm c_f, & \lambda_{2,6} = u \pm c_a, & \lambda_{3,5} = u \pm c_s, & \lambda_4 = u  \\
         
    \end{array}
\end{equation}

où 

\begin{equation}
    \begin{array}{cc}
         c_a = \frac{\|B_x\|}{\sqrt{\rho}},  &c_{f,s} = \left( \frac{\gamma p + \|\mathbf{B}\|^2 \pm \sqrt{(\gamma p +\|\mathbf{B}\|^2 )^2 -4\gamma p B_x^2} }{2\rho} \right)^{\frac{1}{2}}  \\
         
    \end{array}
\end{equation}

les vitesses magnétosonores rapides et lentes peuvent également être mises sous la forme: 

\begin{equation}
    c_{f,s} = \sqrt{\frac{v_s^2 +v_a^2 \pm \sqrt{(v_s^2+v_a^2)^2-4v_s^2c_a^2}}{2}}
\end{equation}

où $v_s$ est la vitesse du son définie par $v_s = \sqrt{\gamma\frac{p}{\rho}}$  et $v_a=\frac{B}{\sqrt{\rho}}$. Trivialement, nous avons l'inégalité suivante: 

\begin{equation}
    \lambda_1 \leq \lambda_2 \leq \lambda_3 \leq \lambda_4 \leq \lambda_5 \leq \lambda_6 \leq \lambda_7.
\end{equation}

\section{Solveur de type HLL} 

\subsection{Solveur HLL}

Nous allons décrire un des solveurs de Riemann de type Godunov proposé par \textit{Harten, Lax et van Leer}, d'où le nom de HLL \cite{harten1983upstream}. 

Considérons la forme intégrale de notre loi de conservation hyperbolique pour un rectangle $(x_1,x_2) \times (t_1,t_2)$: 

\begin{equation}
    \int_{x_1}^{x_2} \mathbf{U}(x,t_2)dx - \int_{x_1}^{x_2} \mathbf{U}(x,t_1)dx  +\int_{t_1}^{t_2} \mathbf{F}(\mathbf{U}(x_1,t))dt -\int_{t_1}^{t_2} \mathbf{F}(\mathbf{U}(x_2,t))dt =0
    \label{Godunov-integral}
\end{equation}

Harten et al. \cite{harten1983upstream} ont montré que le schéma de type Godunov pouvait s'écrire sous la forme: 

\begin{equation}
    \mathbf{U}^{n+1}_i = \mathbf{U}^n_i  - \frac{\Delta t}{\Delta x} [\mathbf{F}(\mathbf{R}(0,\mathbf{U}^n_i,\mathbf{U}^n_{i+1}))-\mathbf{F}(\mathbf{R}(0,\mathbf{U}^n_{i-1},\mathbf{U}^n_i))]
     \label{Godunov-integral2}
\end{equation}

où $i$ indique la ième cellule ou ième volume, $n$ le nième pas de temps et  $\mathbf{R}(0,\mathbf{U}^n_i,\mathbf{U}^n_{i+1})$ est l'approximation de la solution de Riemann autour de l'interface $x_{i+1/2}$. Sous cette forme, le flux numérique approprié est obtenu grâce à la forme intégrale   (\ref{Godunov-integral}) de la loi de conservation sur le rectangle $(x_{i},x_{i+1/2} )\times (t^n,t^{n+1})$: 

\begin{equation}
    \mathbf{F}_{i+1/2} = \mathbf{F}_i -\frac{1}{\Delta t} \int_{x_i}^{x_{i+1/2}} \mathbf{R}\left(\frac{x-x_{i+1/2}}{\Delta t},\mathbf{U}^n_i,\mathbf{U}^n_{i+1}\right) dx + \frac{x_i-x_{i+1/2}}{\Delta t}\mathbf{U}^n_i
     \label{Flux_int}
\end{equation}

\begin{figure}[!h]
    \centering
    \includegraphics[scale=0.5]{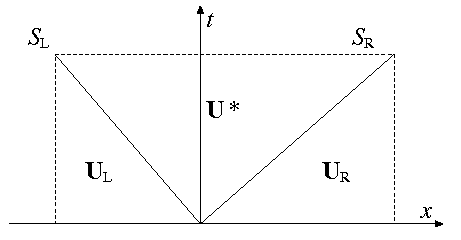}
    \caption{Structure de l'éventail de Riemann avec un état intermédiaire $\mathbf{U}^*$}
    \label{HLLfan}
\end{figure}

Le solveur HLL est construit en supposant un état seul intermédiaire moyen entre les 2 ondes les plus rapides, comme nous pouvons le voir sur la figure \ref{HLLfan}. On considère donc une solution d'un problème de Riemann avec un état intermédiaire, à l'interface entre un état gauche $\mathbf{U}_L$ et un état  droit $\mathbf{U}_R$, où la vitesse d'onde minimale $S_L$ et la vitesse d'onde maximale $S_R$ sont négative et positive respectivement.  En utilisant la forme intégrale (\ref{Godunov-integral2}) sur l'éventail de Riemann $(\Delta t S_L, \Delta t S_R) \times (0,\Delta t)$, on obtient l'état intermédiaire donné par :

\begin{equation}
    \mathbf{U}^* = \frac{S_R \mathbf{U}_R-S_L\mathbf{U}_L -\mathbf{F}_R+\mathbf{F}_L}{S_R-S_L}
\end{equation}

ce qui donne le flux en utilisant (\ref{Flux_int}): 

\begin{equation}
    \mathbf{F}^* = \frac{S_R\mathbf{F}_L-S_L\mathbf{F}_R +S_RS_L (\mathbf{U}_R-\mathbf{U}_L)}{S_R -S_L }
\end{equation}

Ainsi, le flux HLL est: 

\begin{equation}
    \mathbf{F}_{HLL}
    \left\{
    \begin{array}{lcc}
         F_L & si & 0 \leq S_L \\
         F^* & si & S_L \leq 0 \leq S_R \\
         F_R & si & S_R \leq 0 
    \end{array}
    \right.
\end{equation}

Il ne reste plus qu'à déterminer une approximation de $S_L$ et $S_R$; une des approximations utilisées est: 

\begin{equation}
    \begin{array}{l}
         S_L = \min (\lambda_1(\mathbf{U}_L),\lambda_1(\mathbf{U}_R))  \\
         S_R = \max (\lambda_7(\mathbf{U}_L),\lambda_7(\mathbf{U}_R)) 
         \label{SLSR}
    \end{array}
\end{equation}

où $\lambda_1$ et $\lambda_7$ sont la plus petite et la plus grande des valeurs propres de $\mathbf{A}$. La manière d'estimer $S_L$ et $S_R$ n'est pas unique. Celle présentée ici est utilisée par Davis \cite{davis1988simplified}, mais d'autres existent également comme la méthode utilisée par Einfeldt \cite{einfeldt1991godunov} basée sur les moyennes de Roe. Ainsi, par la suite, je ne redéfinirai plus les valeurs de $S_L$ et $S_R$ à cause de la grande variété d'approximations possibles.  

Bien que le solveur HLL soit très robuste, il reste néanmoins très diffusif, car il ne prend en compte qu'un seul état intermédiaire et deux vitesses sur les sept présentes en MHD. 

\subsection{Solveur HLLC}

Le C dans HLLC signifie \og contact \fg, car ce schéma permet de modéliser les discontinuités de contact en hydrodynamique. Nous allons décrire plus rapidement le fonctionnement de ce schéma, car il suit la même idée qu'un schéma HLL  \cite{toro1994restoration, Toro}.

\begin{figure}[!h]
    \centering
    \includegraphics[scale=0.5]{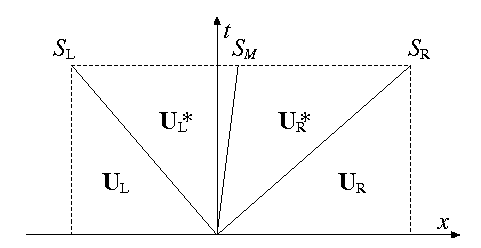}
    \caption{Structure de l'éventail de Riemann avec deux états intermédiaires $\mathbf{U}_L^*$ et $\mathbf{U}_R^*$}
    \label{HLLCfan}
\end{figure}

A la différence du schéma HLL, le schéma HLLC est constitué de deux états intermédiaires $\mathbf{U}_R^*$ et $\mathbf{U}_L^*$ séparés par une onde de contact $S_M$ comme on peut le voir sur la figure \ref{HLLCfan}. Nous supposons également que la vitesse normale et la pression sont constantes au passage d'une discontinuité de contact avec: 

\begin{equation}
    u_R^* = u_L^* = S_M
\end{equation}

et: 

\begin{equation}
    S_M = \frac{(S_R-u_R) \rho_R u_R -(S_L-u_L) \rho_L u_L -p_R+p_L }{(S_R-u_R) \rho_R  -(S_L-u_L) \rho_L }
\end{equation}

avec $S_L$ et $S_R$ définies en \ref{SLSR}. On peut ainsi déterminer le flux pour chaque état intermédiaire, ce qui donne comme flux HLLC: 

\begin{equation}
    \mathbf{F}_{HLLC}
    \left\{
    \begin{array}{lcc}
         F_L & si & 0 \leq S_L \\
         F^*_L & si & S_L \leq 0 \leq S_M\\
         F^*_R & si & S_M \leq 0 \leq S_R\\
         F_R & si & S_R \leq 0 
    \end{array}
    \right.
\end{equation}

Les flux $F^*_R$, $F^*_L$, $F_R$ et $F_L$ ne sont pas explicités ici par souci de concision, mais peuvent être déterminés grâce aux conditions de Rankine-Hugoniot et les invariants de Riemann à travers les différentes ondes. 
Bien que ce schéma soit idéal pour résoudre un problème HD, il n'est pas encore assez précis pour un problème MHD. 

\subsection{Solveur HLLD}

\begin{figure}[!h]
    \centering
    \includegraphics[scale=0.5]{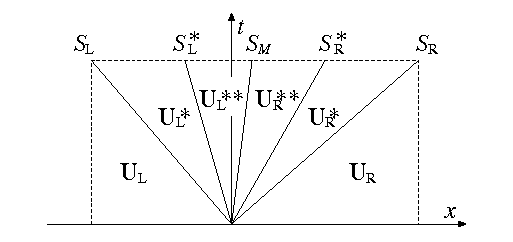}
    \caption{Structure de l'éventail de Riemann avec quatre états intermédiaires}
    \label{HLLDfan}
\end{figure}

HLLD est un schéma récent proposé par Miyoshi et al. \cite{miyoshi2005multi} spécialement pour résoudre des problèmes MHD avec précision. Il est composé de quatre états intermédiaires séparés par deux vitesses d'Alfvén $S_L^*$ et $S_R^*$ et vitesse de l'onde d'entropie $S_M$, comme on peut le voir sur la figure \ref{HLLDfan}. Les vitesses d'onde sont définies par: 

\begin{equation}
    S_M = \frac{(S_R-u_R) \rho_R u_R -(S_L-u_L) \rho_L u_L -p_{tR}+p_{tL} }{(S_R-u_R) \rho_R  -(S_L-u_L) \rho_L }
\end{equation}
 
 et 
 \begin{equation}
 \begin{array}{cc}
  S_L^* = S_M - \frac{\mid B_x\mid}{\sqrt{\rho^*_L}} ,   & S_R^* = S_M+ \frac{\mid B_x\mid }{\sqrt{\rho^*_R}}   
 \end{array}
  \end{equation}

 Le flux HLLD est donné par:
 
 \begin{equation}
    \mathbf{F}_{HLLD}
    \left\{
    \begin{array}{lcc}
         F_L & si & 0 \leq S_L \\
         F^*_L & si & S_L \leq 0 \leq S_L^*\\
         F^{**}_L & si & S_L^* \leq 0 \leq S_M\\
         F^{**}_R & si & S_M \leq 0 \leq S_R^*\\
         F^*_R & si & S_R^* \leq 0 \leq S_R\\
         F_R & si & S_R \leq 0 
    \end{array}
    \right.
\end{equation}

Pour plus de précision, on pourra se référer à Miyoshi et al. \cite{miyoshi2005multi}.

\section{Solveur de type Roe}

Le solveur de Roe \cite{roe1981approximate} est un autre solveur très répandu dans les codes consacrés à l'étude des problèmes MHD. Nous allons décrire l'approche utilisée par Roe \cite{roe1981approximate} dans les grandes lignes \cite{Toro}. Il s'agit d'un solveur de Riemann qui utilise la linéarisation de la matrice jacobienne. 

On rappelle la forme conservative explicite de notre problème: 

\begin{equation}
    \mathbf{U}_i^{n+1} = \mathbf{U}^n_i + \frac{\Delta t}{\Delta x} (\mathbf{F}_{i-1/2}-\mathbf{F}_{i+1/2})
\end{equation}

où $\mathbf{F}_{i+1/2}=\mathbf{F}(\mathbf{U}_{i+1/2}(0))$. L'approche de Roe utilise la loi de conservation sous une forme contenant la Jacobienne $\mathbf{A}$: 

\begin{equation}
     \partial_t\mathbf{U} + \mathbf{A} (\mathbf{U}) \partial_x \mathbf{U} = 0 
\end{equation}

En effet, l'élément clé dans l'approche de Roe est l'approximation de la matrice jacobienne par une matrice jacobienne constante sur $\Delta t$:

\begin{equation}
    \mathbf{\Tilde{A}} = \mathbf{\Tilde{A}} (\mathbf{U}_L,\mathbf{U}_R)
\end{equation}

ce qui donne: 

\begin{equation}
     \partial_t\mathbf{U} + \mathbf{\Tilde{A}} (\mathbf{U}) \partial_x \mathbf{U} = 0 
\end{equation}

Pour un système hyperbolique de $m$ lois de conservation, la matrice jacobienne de Roe  $\mathbf{\Tilde{A}}$ doit satisfaire  les propriétés suivantes: 

\subparagraph{Propriété (A):}
Pour avoir un système hyperbolique, $\mathbf{\Tilde{A}}$ doit être diagonalisable et avoir des valeurs propres $\Tilde{\lambda_j}= \Tilde{\lambda_j} (\mathbf{U}_L, \mathbf{U}_R)  $ réelles que nous ordonnons de telle sorte que: 

\begin{equation}
    \Tilde{\lambda_1} \leq \Tilde{\lambda_2} \leq \cdots \leq \Tilde{\lambda_m} 
\end{equation}

et une séquence complète de vecteurs propres indépendants:

\begin{equation}
    \mathbf{\Tilde{K}}^{(1)}, \quad \mathbf{\Tilde{K}}^{(2)}, \cdots , \mathbf{\Tilde{K}}^{(m)}.
\end{equation}

\subparagraph{Propriété (B):}

Cohérence avec la matrice Jacobienne réelle: 

\begin{equation}
    \mathbf{\Tilde{A}}(\mathbf{U},\mathbf{U}) = \mathbf{A}(\mathbf{U}). 
\end{equation}

\subparagraph{Propriété (C):}

Conservation à travers les discontinuités: 

\begin{equation}
    \mathbf{F}(\mathbf{U}_R) - \mathbf{F}(\mathbf{U}_L) = \mathbf{ \Tilde{A}} (\mathbf{U}_R-\mathbf{U}_L)  
\end{equation}

\medskip

On définit alors l'amplitude de l'onde $j$ comme étant $\Tilde{\alpha}_j = \Tilde{\alpha}(\mathbf{U}_L,\mathbf{U}_R)$ et vérifiant: 

\begin{equation}
    \Delta \mathbf{U} = \mathbf{U}_R - \mathbf{U}_L = \sum_{j=1}^m \Tilde{\alpha}_j \mathbf{\Tilde{K}}^{(j)}
\end{equation}

On peut ainsi monter que:

\begin{equation}
    \mathbf{F}_{i+1/2} = \frac{1}{2} (\mathbf{F}_L +\mathbf{F}_R) - \frac{1}{2}\sum_ {j=1}^m \Tilde{\alpha_j} \|\Tilde{\lambda}_j\| \mathbf{\Tilde{K}}^{(j)}
\end{equation}

\section{Vers la MHD non-idéale}

Les solveurs de Riemann présentés ici permettent de traiter des problèmes comprenant des discontinuités fortes comme dans la propagation d'ondes de choc mais dans le cadre de la MHD idéale.
Pour traiter la MHD non-idéale, c'est à dire munie d'une loi d'Ohm comprenant davantage de termes comme la résistivité ou encore l'effet Hall, l'utilisation des solveurs de Riemann n'est pas immédiate.
Deux manières sont possibles, la première consiste à séparer le système d'équations en un système hyperbolique, comprenant la MHD idéale, et un système parabolique comprenant les termes résistifs ou de Hall.
La seconde possibilité est d'inclure dans le flux une approximation des termes de Hall.
On peut retrouver une recherche bibliographique sur ces différentes méthodes en annexe \ref{MHDNONIDEAL}.

\newpage

\chapter{Les cas-tests MHD}
Les cas-tests qui ont été choisis permettent de vérifier de différentes manières la qualité des solveurs. Le cas de Brio et Wu \cite{brio1988upwind} est un cas 1D qui nous permet d'évaluer la capacité des solveurs à capturer les différentes ondes présentes dans notre système d'équations MHD. 

Les cas 2D d'Orszag-Tang, du Rotor magnétique et  3D pour MHD Blast sont surtout utilisés pour démontrer la robustesse du code face à des gradients plus important.

Les solveurs disponibles dans le code (CLOVIS) pour le transport des ions sont le solveur de Roe et le solveur HLLD, pour le transport des neutres le solveur HLLC et le solveur de Roe \og f-waves \fg.
La particularité du solveur de Roe \og f-waves \fg, utilisé en HD, est qu'il inclut la gravité directement dans le flux, alors que la gravité est en terme source pour les autres.
Les deux solveurs ont été confrontés sur le cas-test de Brio et Wu. Pour les simuler les autres cas 2D et 3D, nous avons utilisé le solveur HLLD. 

On rappelle que l'objectif de ce stage n'est pas de comparer les solveurs entre eux. Les comparaisons, présentées ici, servent à confirmer la pertinence du choix du cas-test en montrant qu'il est suffisamment discriminant.

\section{Tube à choc de Brio et Wu } 

Ce test permet une première vérification de CLOVIS. En effet, le tube à choc de Brio et Wu est un cas très simple de propagation de différentes ondes de choc, de raréfaction, de discontinuité de contact... C'est un problème identique à celui de Sod en hydrodynamique.  
Il s'agit ici de vérifier la bonne représentation des différentes ondes avec ce cas-test (équivalent à un problème de Riemann particulier). 

\paragraph{Initialisation}

Le cas habituel est initialisé avec comme valeur à gauche  $\rho_l =1 $, $u_l = v_l = 0$, $p_l=1$, $(B_y)_l=1$, à droite $\rho_r=0{,}125$, $u_r=v_r=0$, $p_r=0{,}1$ et $(B_y)_r=-1$, avec également $B_x= 0{,}75$ et $\gamma=2$. Ce cas implique deux ondes de raréfactions rapides, une onde composée lente, une discontinuité de contact et une onde de choc lente. La grille est composée de 800 points avec $x \in [0;1]$ et la séparation en $x=0{,}5$. 

 \begin{figure}[!h]
    \centering
    \includegraphics[scale=0.7]{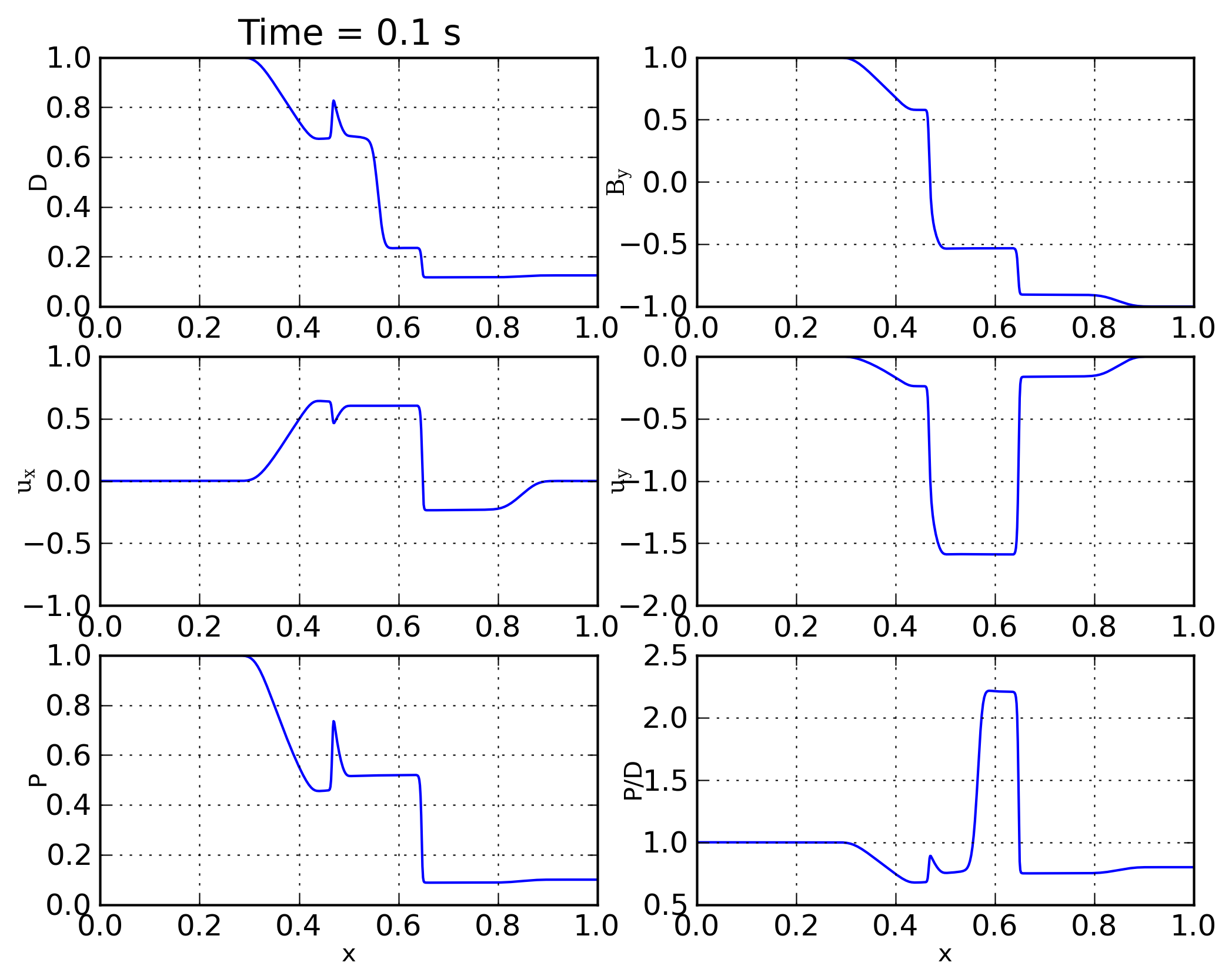}
    \caption{Résultats de la simulation pour le solveur HLLD à l'ordre 1 avec une CFL de 0,5 à $t=0{,}1$ s: en partant d'en haut à gauche, masse volumique des ions $\rho$,  champ magnétique $B_y$,  vitesse du fluide, selon $x$, puis, selon $y$, pression cinétique $p$ et le rapport entre la pression cinétique $p$ et la masse volumique $\rho$. Temps de calcul: environ $10\sim15$ min sur un ordinateur de bureau à 6 c\oe{}urs.}
    \label{Brio}
\end{figure}

\paragraph{Résultats}

Les résultats sont analysés à $t=0{,}1$ s. Nous pouvons voir sur la figure \ref{Brio} de gauche à droite une onde de raréfaction rapide, une onde composée lente,  une discontinuité de contact, une onde de choc lente et encore une onde de raréfaction rapide. Nous pouvons également comparer nos résultats avec ceux de la littérature (figure 2 de l'article de Brio et Wu \cite{brio1988upwind}, sur le site \textit{\og The Athena Code Test Page \fg} \cite{athena_test} ou directement sur l'article d'Athena \cite{stone2008athena}). Ainsi, ce cas-test permet une première validation du CLOVIS.

\paragraph{Comparaison entre les différents solveurs}

Nous avons réduit la CFL à 0,1 pour effectuer cette comparaison, car des oscillations apparaissaient pour le solveur HLLD à l'ordre 2 et des \og undershoots\fg \, ou \og overshoots\fg \, non-négligeables sont présents pour le solveur de Roe à l'ordre 2, dans le cas d'une CFL trop importante. Ceci est un problème bien connu des numériciens.

\begin{figure}[!h]
	\centering
	\begin{tabular}{c}
		\includegraphics[width=0.45\textwidth]{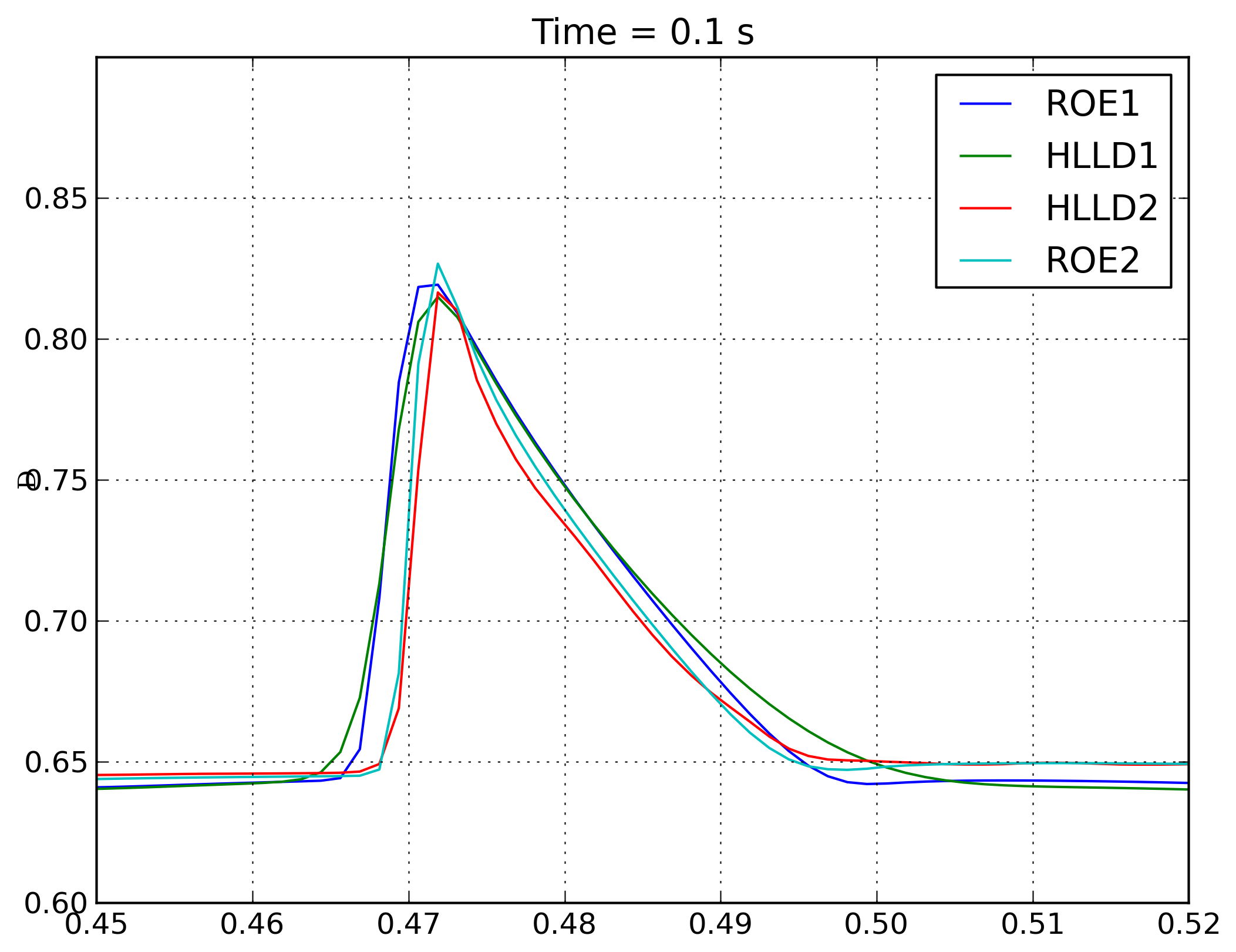}\\[\abovecaptionskip] 
		\small Onde composée
	\end{tabular}
	\begin{tabular}{c}
		\includegraphics[width=0.45\textwidth]{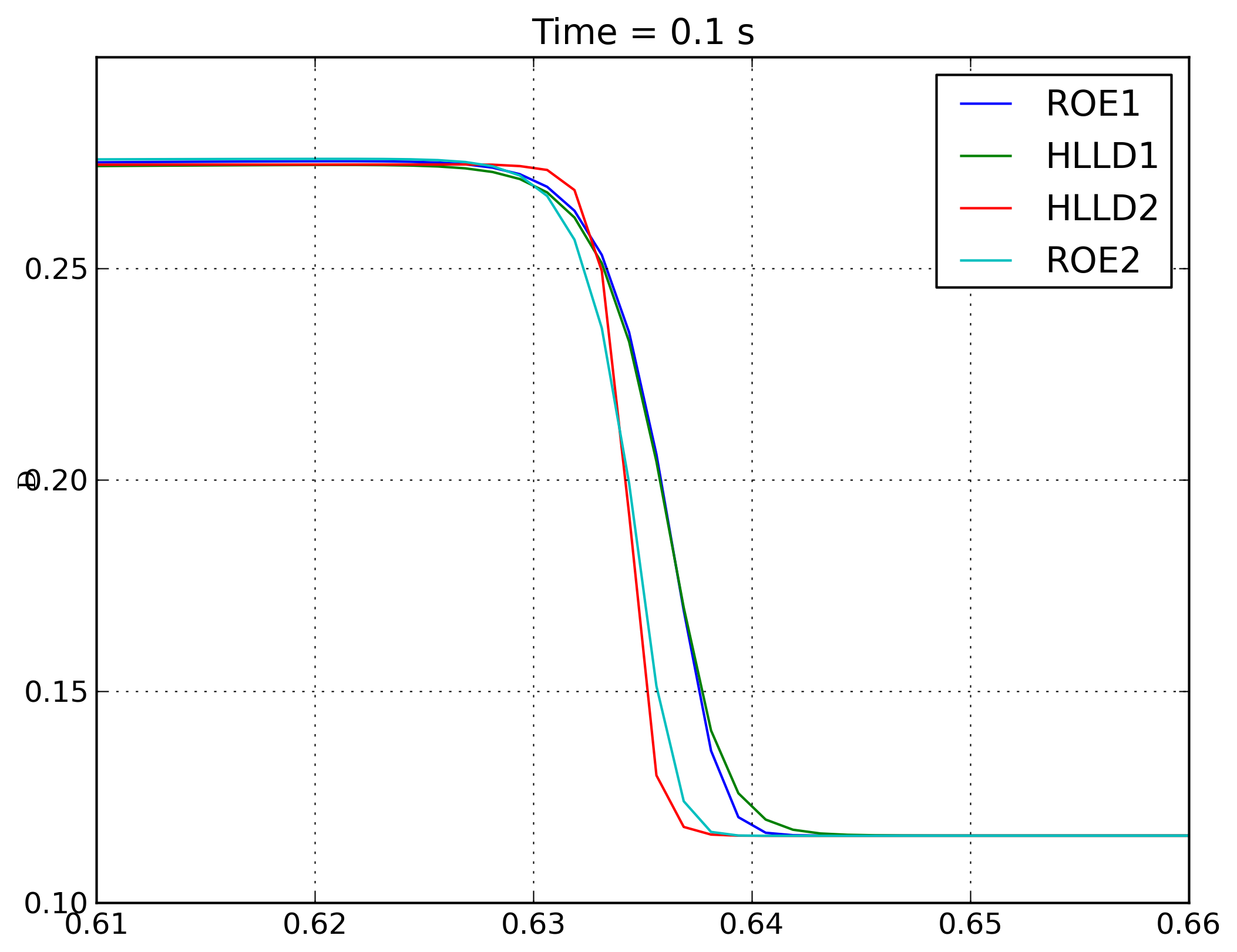}\\[\abovecaptionskip]
		\small Onde de choc lente
	\end{tabular}
	\caption{Zoom sur différentes ondes pour une simulation du cas de Brio et Wu avec une condition CFL de $0{,}1$ pour les solveurs HLLD et Roe à l'ordre 1 et 2. }
	\label{coumpound}
\end{figure}

De façon prévisible, on voit sur les deux figures  \ref{coumpound} que l'ordre 1 d'un solveur est beaucoup plus diffusif que son ordre 2. 

Sur la figure \ref{coumpound} à gauche, l'onde composée est plus amortie  avec le schéma HLLD qu'avec celui de Roe. Nous avons également un léger \og undershoot \fg\, pour le schéma de Roe qui n'est pas présent pour le schéma HLLD (aux ordres respectifs). 

Sur la figure \ref{coumpound} à droite, le schéma de Roe à l'ordre 1 capture de façon plus précise l'onde de choc lente que le schéma HLLD. Ce résultat pourrait s'expliquer par la présence explicite des ondes lentes dans le solveur de Roe, contrairement au solveur HLLD. Néanmoins,  l'ordre 2 de HLLD semble avoir mieux résolu l'onde de choc lente que Roe à l'ordre 2. 

Si nous mettons de côté le dernier point nous trouvons globalement les mêmes différences entre le solveur de Roe et HLLD que dans l'article de Miyoshi \cite{miyoshi2005multi}. 

\section{Vortex d'Orszag-Tang }

Ce test permet d'évaluer la robustesse de CLOVIS, que ce soit sur la résolution d'une propagation d'onde que sur le respect de la contrainte $\nabla \cdot \mathbf{B}=0$.
En effet, il s'agit d'un problème 2D où en partant de conditions initiales continues, on génère des ondes de choc supersoniques qui ensuite interagissent entre elles. 
Il est intéressant de comparer les résultats avec d'autres codes ou schémas numériques existant car ce test reste malgré tout très qualitatif.

\paragraph{Initialisation}

Ce cas est effectué sur un domaine $[0{,}2\pi]\times [0{,}2\pi]$ avec:

\begin{equation}
\begin{array}{cc}
    \mathbf{V} = [ -\sin{y} ; \sin{x}] & \mathbf{B} =  [ -\sin{y} ; \sin{2x}] \\
    \rho = \gamma^2 & p = \gamma \\
    \gamma = 5/3 
\end{array}    
\end{equation}

\begin{figure}[!h]
	\centering
	\begin{tabular}{c}
		\includegraphics[width=0.4\textwidth]{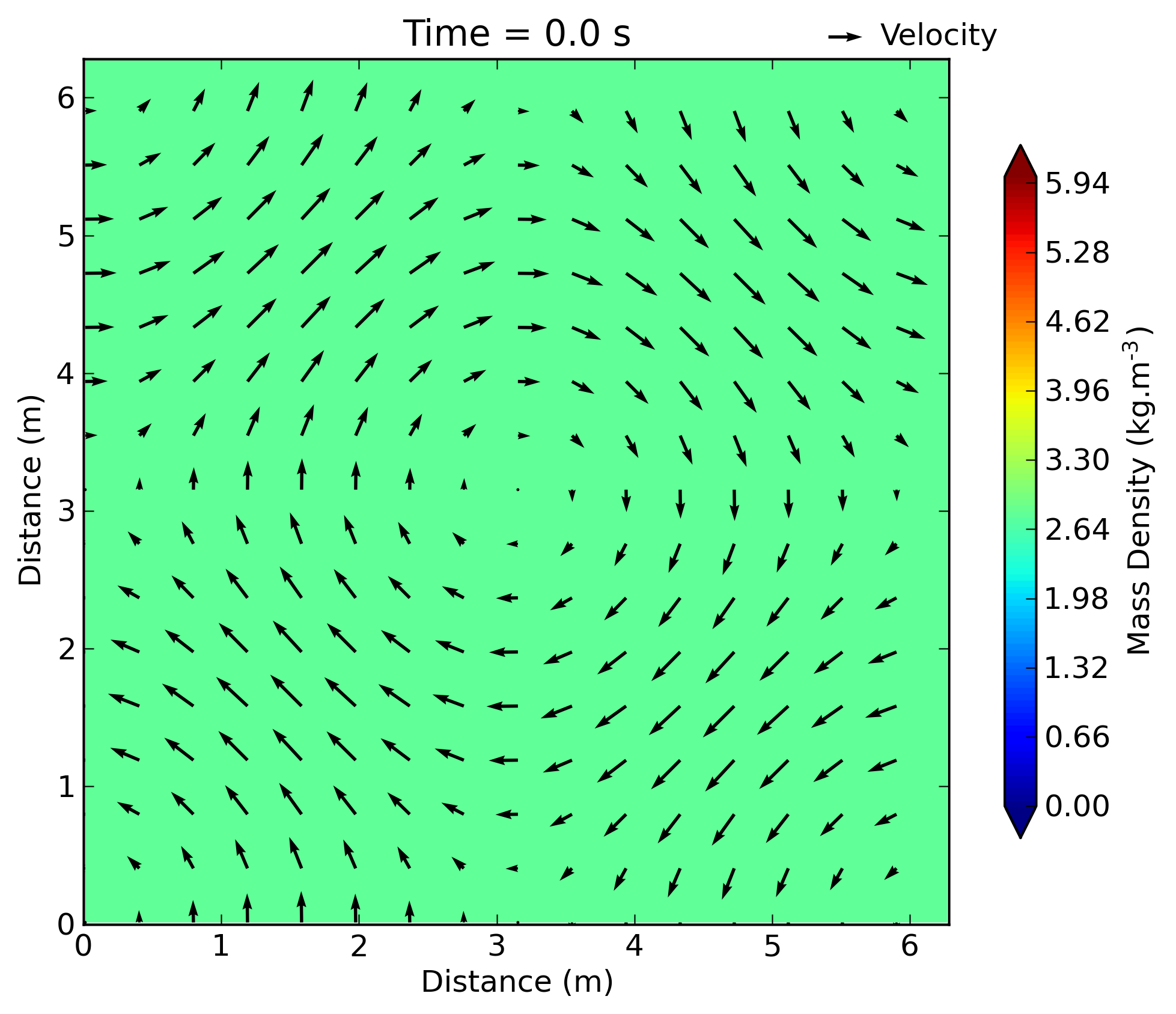}\\[\abovecaptionskip]
		\small à $t=0 $ s
	\end{tabular}
	\begin{tabular}{c}
		\includegraphics[width=0.4\textwidth]{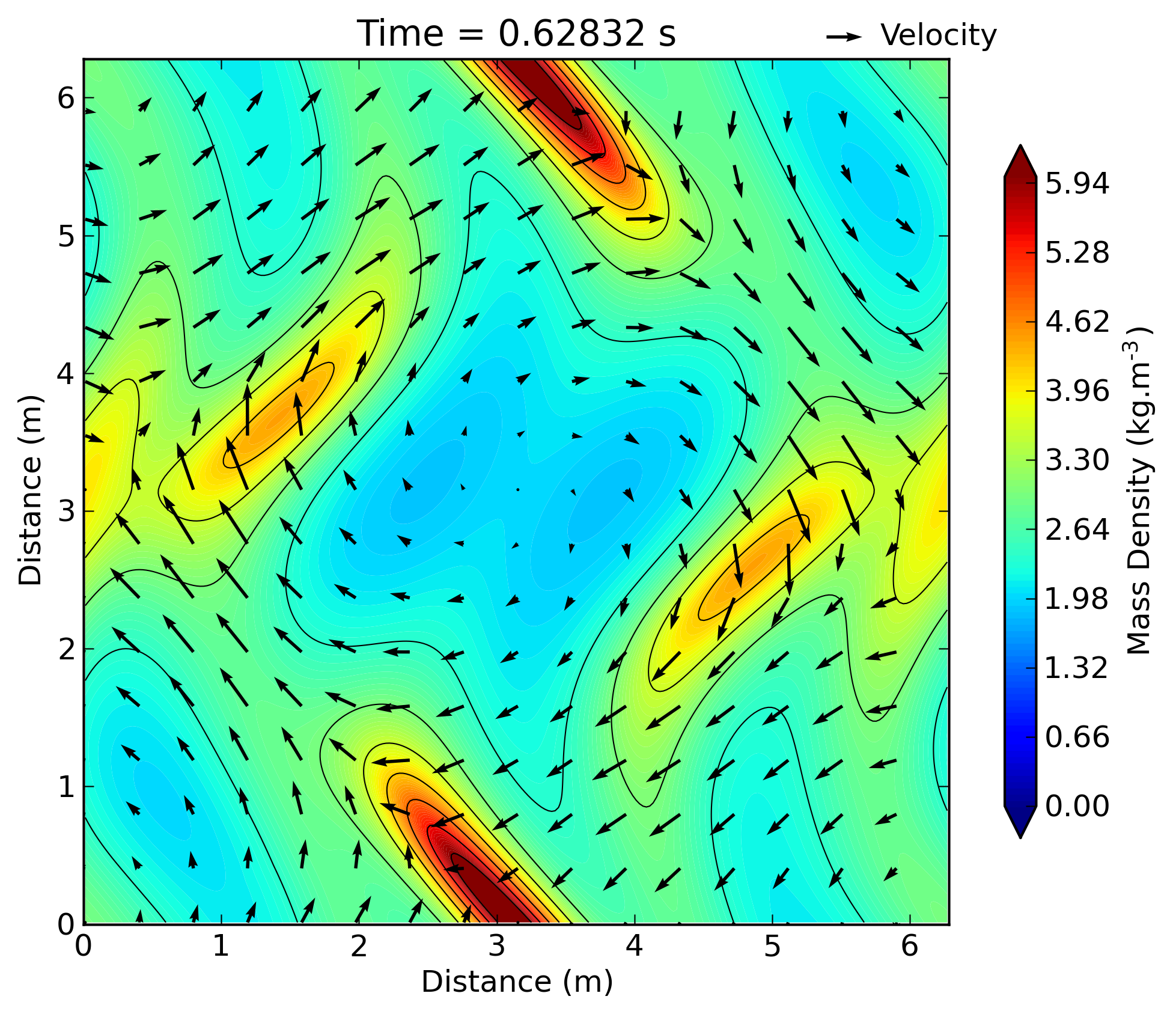}\\[\abovecaptionskip]
		\small à $t=0{,}2\pi $ s
	\end{tabular}
	\begin{tabular}{c}
		\includegraphics[width=0.4\textwidth]{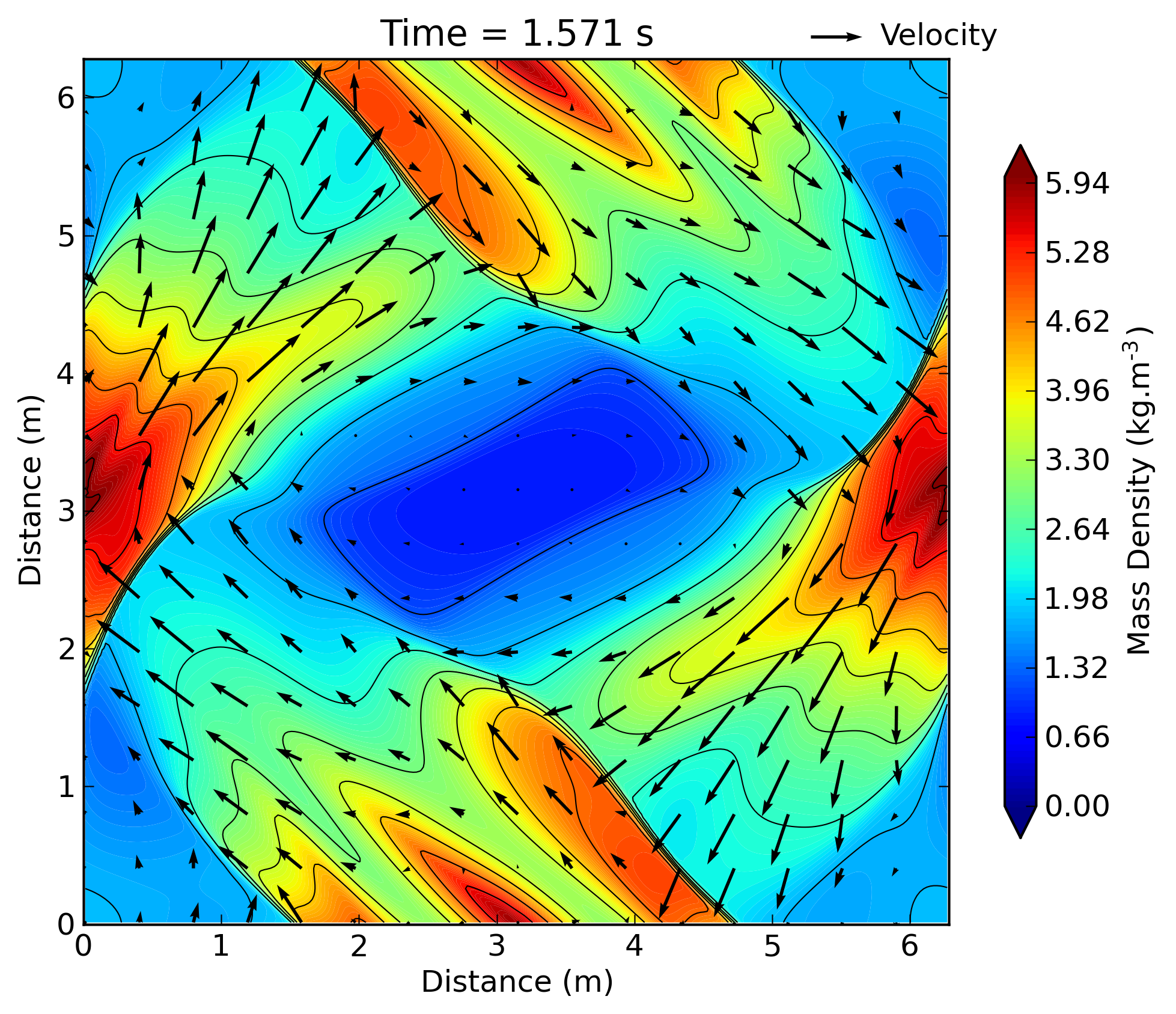}\\[\abovecaptionskip]
		\small à $t=0{,}5\pi $ s
	\end{tabular}
	\begin{tabular}{c}
		\includegraphics[width=0.4\textwidth]{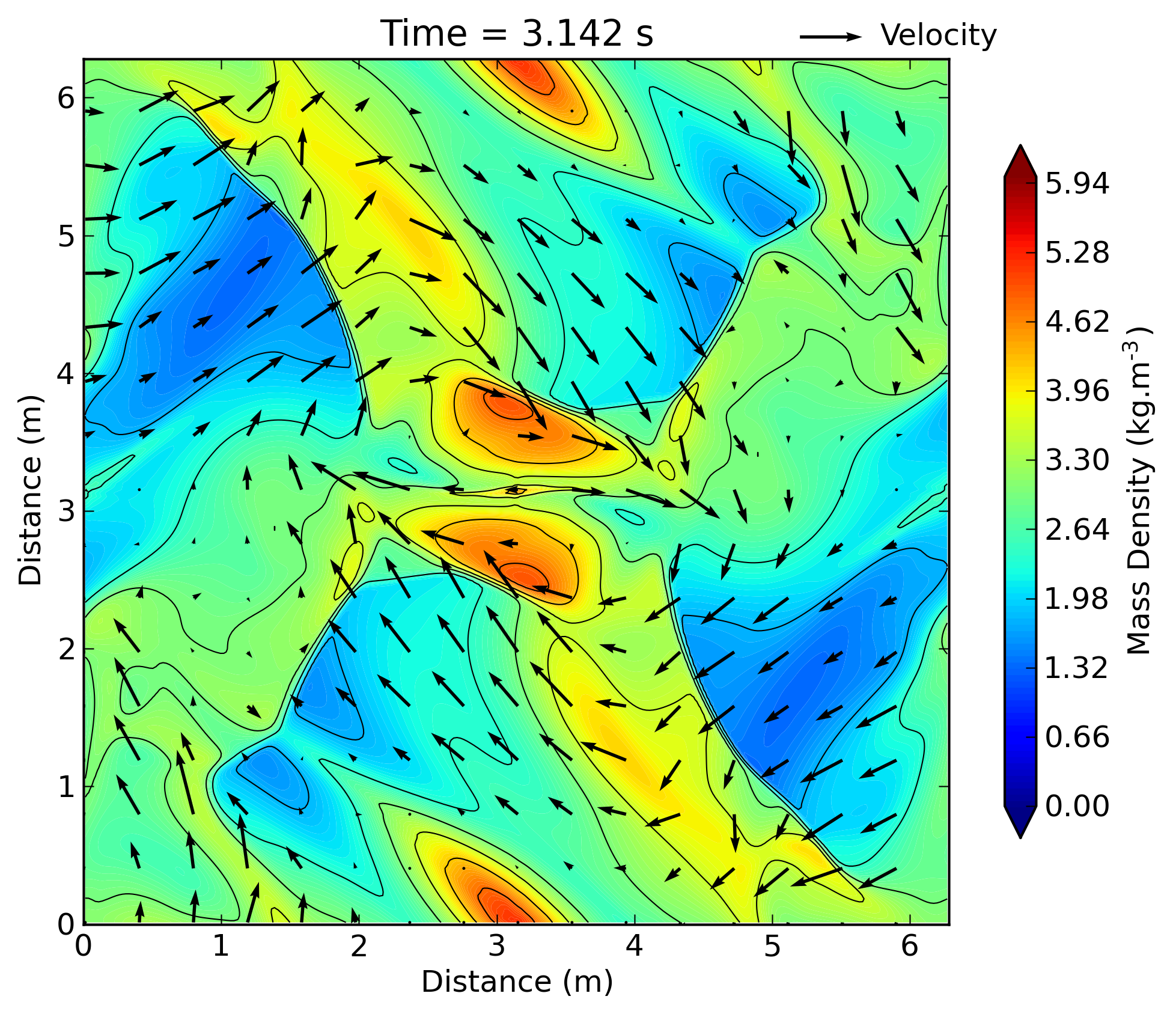}\\[\abovecaptionskip]
		\small à $t=\pi $ s
	
	\end{tabular}
	\caption{Simulation du cas-test d'Orszag-Tang avec un schéma HLLD d'ordre 1 sur une grille de $256\times256$ avec une condition CFL de $0{,}1$. Temps de calcul: environ $30\sim60$ min sur un ordinateur de bureau à 6 c\oe{}urs.}
	\label{Orzsag}
\end{figure}

\paragraph{Résultat} 

Les résultats de ce test sont étudiés à $t=\pi$ s. En comparant qualitativement nos résultats avec ceux de la littérature nous pouvons voir qu'ils concordent bien avec les résultats attendus. Nous pouvons par exemple utiliser les données disponibles sur le site \textit{\og The Athena Code Test Page\fg} \cite{athena_test} ou directement sur l'article d'Athena \cite{stone2008athena}. En effet, les positions des ondes de choc (en haut à gauche sur la deuxième figure \ref{Orzsag} par exemple) sont similaires à celles obtenues par Athena.

\section{MHD Blast}

Ce test permet aussi d'évaluer la robustesse de CLOVIS en décrivant une explosion, modélisée par une sphère de fortes pressions, dans un milieu magnétisé de faibles pressions.
Ce cas est utile car si la condition $\nabla\cdot \mathbf{B}=0$ n'est pas bien respectée, on peut voir apparaître des valeurs non physiques comme par exemple une densité ou une pression négative.

\paragraph{Initialisation}

Ce cas a été effectué sur un domaine 3D $[-0{,}5;0{,}5] \times [-0{,}5;0{,}5]\times [-0{,}5;0{,}5]$ par le code Flash \cite{center2005flash} sur lequel nous avons pris exemple pour notre implémentation. Les valeurs initiales sont une vitesse nulle, $B_x=100/\sqrt{4\pi}$, $B_z=B_y=0$, et dans un cercle de rayon $R=0{,}1$ autour de l'origine une pression $p=1000$ et en dehors du cercle une pression $p=0{,}1$. 

\begin{figure}[!h]
	\centering
	\begin{tabular}{c}
		\includegraphics[width=0.45\textwidth]{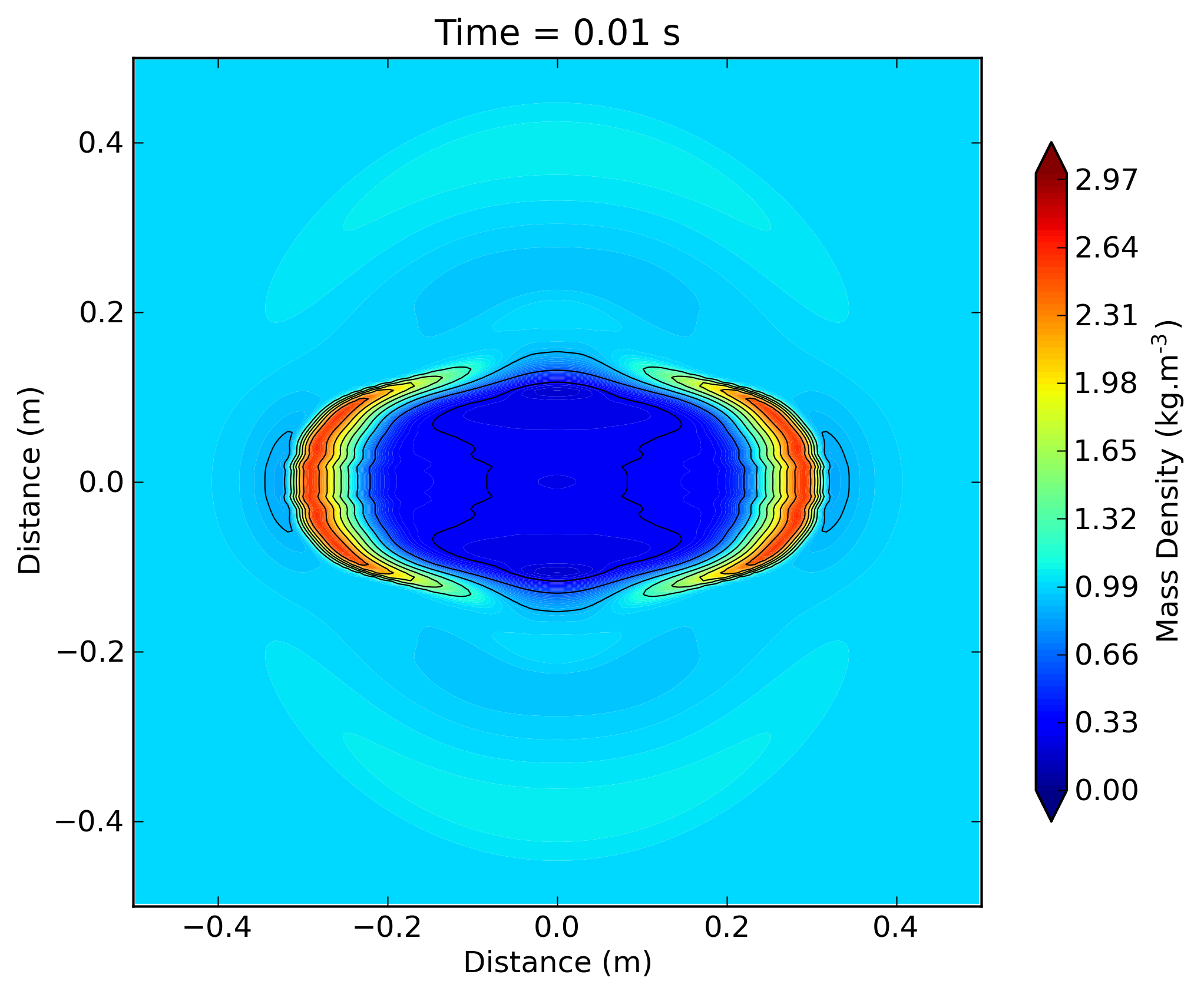}\\[\abovecaptionskip]
		\small Masse volumique des ions
		\label{Blastdens}
	\end{tabular}
	\vspace{1em} 
	\begin{tabular}{c}
		\includegraphics[width=0.45\textwidth]{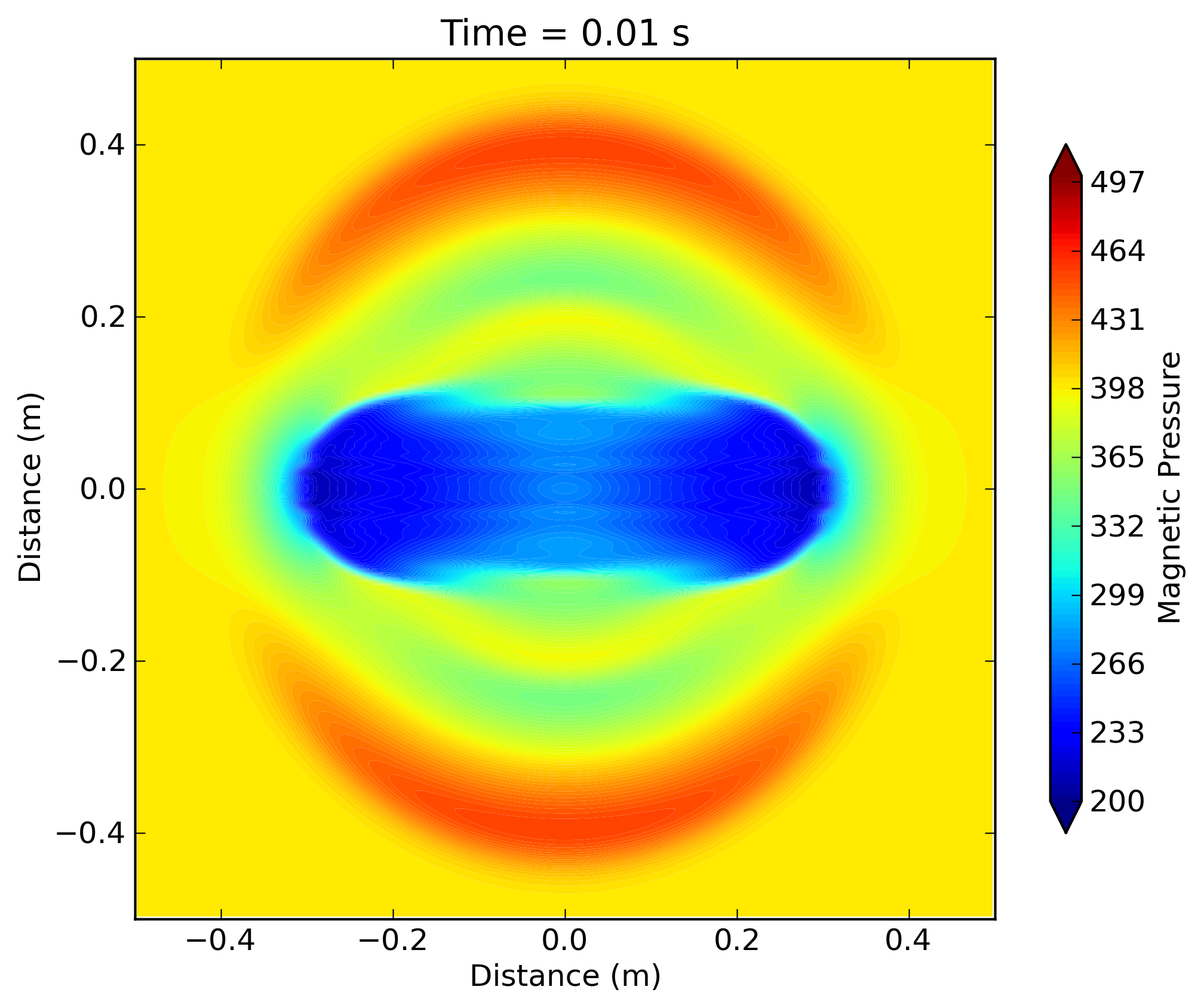}\\[\abovecaptionskip]
		\small Pression magnétique 
		\label{BlastMpress}
	\end{tabular}
	\caption{Simulation du cas-test MHD Blast avec un schéma HLLD d'ordre 1 sur une grille de $200\times200$ avec une condition CFL de $0{,}5$ et à $t=0{,}01$ s. Temps de calcul : environ $4\sim5$ h sur un ordinateur de bureau à 6 c\oe{}urs.}
	\label{Blast}
\end{figure}

\paragraph{Résultats}

Les résultats, observés à $t=0{,}01$ s, montrent une bonne concordance avec ceux du code Flash \cite{center2005flash}. 
Comme attendu, une onde de choc de forme sphérique, se propage à la vitesse de l'onde magnétosonore rapide et le plasma se déplace plus lentement et préférentiellement dans la direction des lignes de champ magnétique.

\section{Rotor magnétique}

Le rotor magnétique nous permet de vérifier que le solveur reproduit correctement l'évolution de fortes ondes d'Alfvén de torsion. C'est un phénomène analogue à la naissance d'un corps stellaire.

\paragraph{Initialisation}

Ce cas est effectué sur un domaine $[-0{,}5;0{,}5]\times[-0{,}5;0{,}5]$ avec comme conditions aux limites des murs absorbants. Les conditions initiales sont \cite{toth2000b}: 

\begin{equation}
    \rho (x,y) = 
    \left\{
    \begin{array}{lc}
         10 & r \leq r_0\\
          1+9 f(r)& r_0 \leq r \leq r_1\\
          1& r_1 \leq r
    \end{array}
    \right.
\end{equation}

\begin{equation}
    u (x,y) = 
    \left\{
    \begin{array}{lc}
         -u_0y/r_0 & r \leq r_0\\
          -u_0 f(r) y/r_0 & r_0 \leq r \leq r_1\\
          0& r_1 \leq r 
    \end{array}
    \right.
\end{equation}
\begin{equation}
    v (x,y) = 
    \left\{
    \begin{array}{lc}
         u_0x/r_0 & r \leq r_0\\
          u_0 f(r) x/r_0& r_0 \leq r \leq r_1\\
          0& r_1 \leq r 
    \end{array}
    \right.
\end{equation}

\begin{equation}
    p(x,y) = 0{,}5
\end{equation}

\begin{equation}
    B_x = \frac{2{,}5}{\sqrt{4\pi}}
\end{equation}

\begin{equation}
    B_y = 0
\end{equation}

où $r=\sqrt{x^2+y^2}$, $r_0=0{,}1$, $r_1=0{,}115$, $f$ est une fonction définie par $f(r) = \frac{r_1-r}{r_1-r_0}$ et $u_0=1$. On a également $w=B_z=0$ et $\gamma = 5/3$. Les résultats sont présentés à $t=0{,}295$ s.

Une autre version peut également être utilisée, avec $B_x=\frac{5}{\sqrt{4\pi}}$, $p = 1$, $u_0=2$ et $\gamma = 1{,}4$ . Ce deuxième cas étant plus contraignant que le précédent, les résultats sont observés après un temps plus long, soit $t=0{,}115$ s. 

\begin{figure}[!h]
	\centering
	\begin{tabular}{c}
		\includegraphics[width=0.45\textwidth]{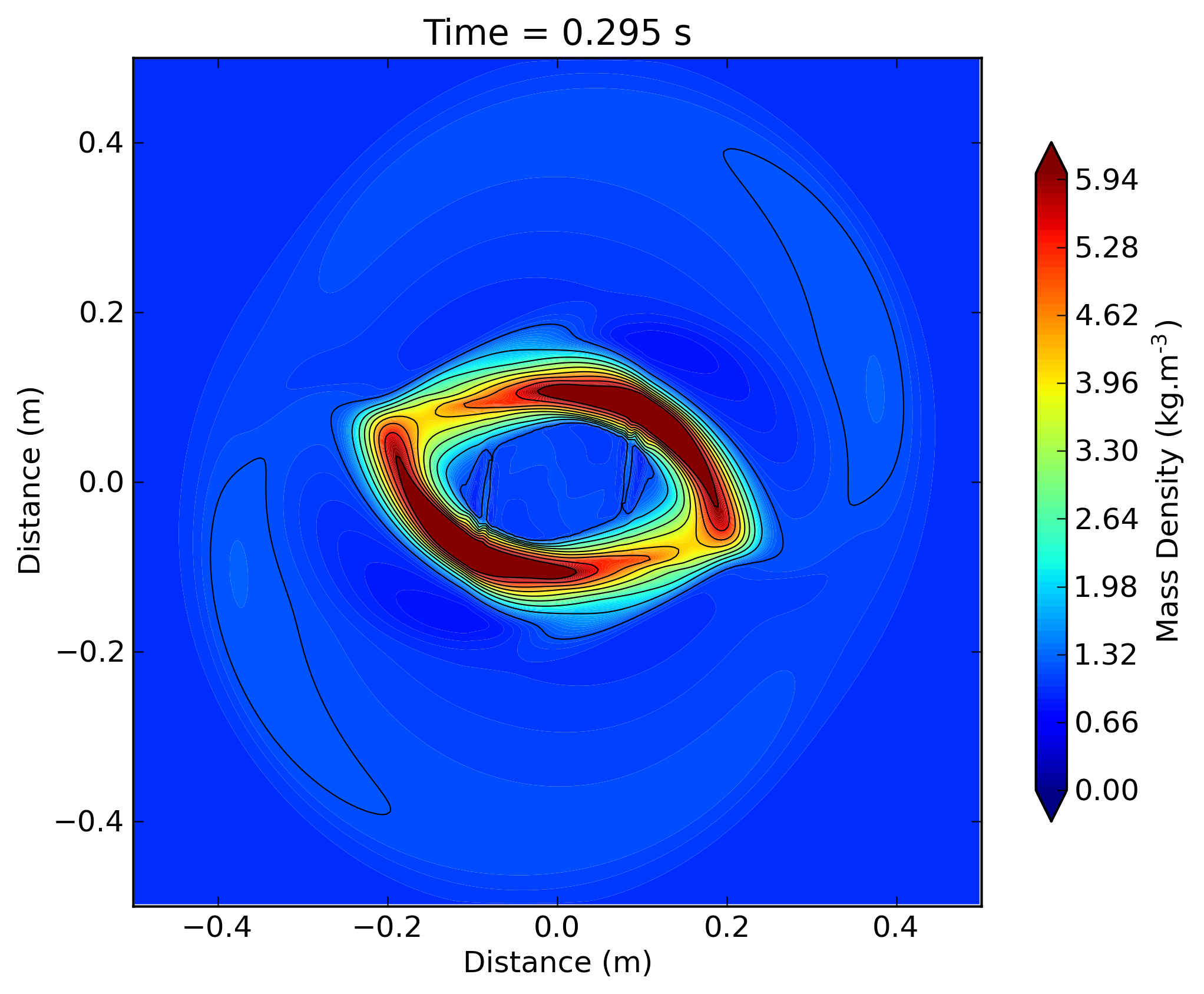}\\[\abovecaptionskip]
		\small Masse volumique des ions
	\end{tabular}
	\begin{tabular}{c}
		\includegraphics[width=0.45\textwidth]{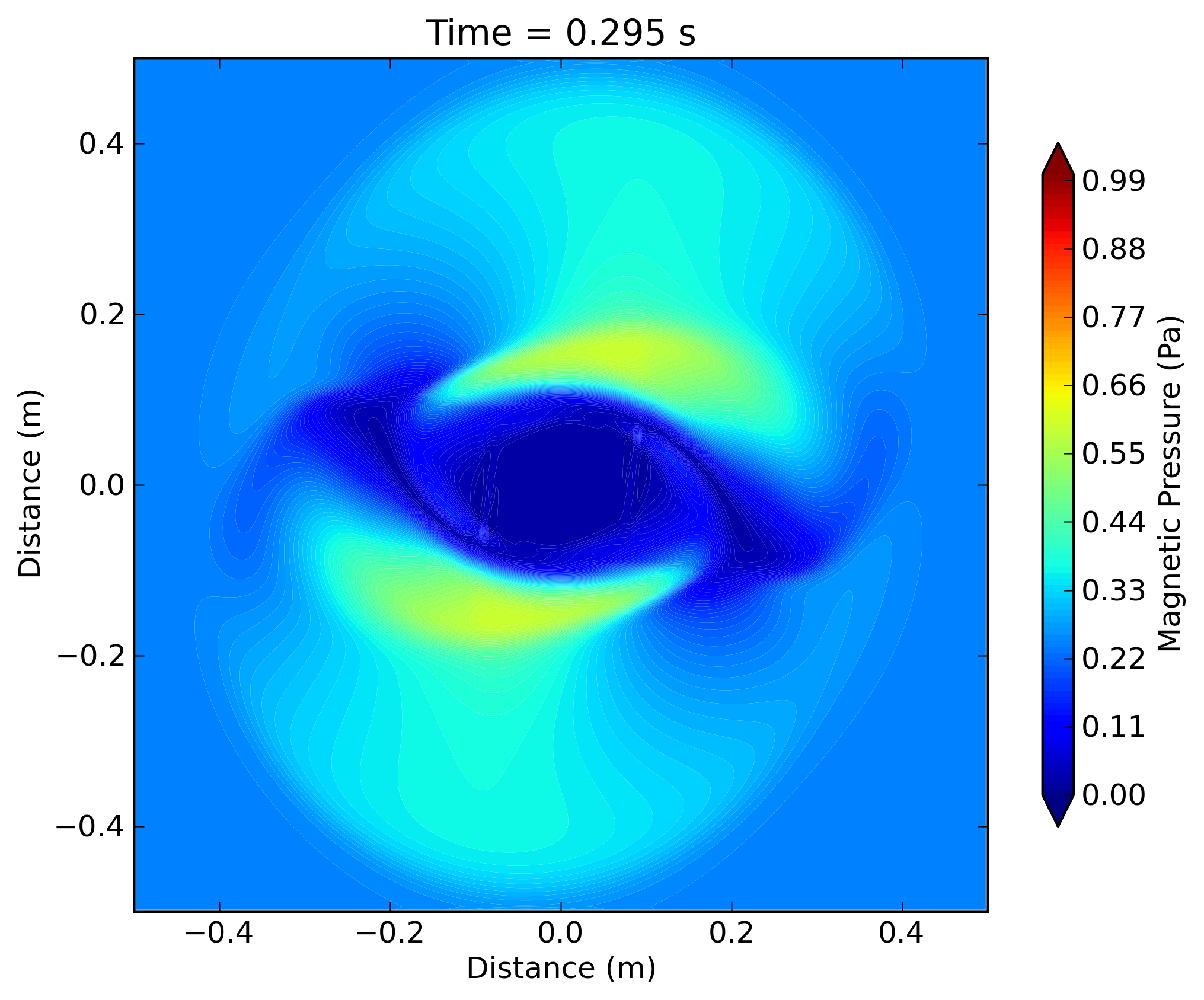}\\[\abovecaptionskip]
		\small Pression magnétique
	\end{tabular}
	\caption{Simulation du premier cas-test du Rotor magnétique avec un schéma HLLD d'ordre 1 sur une grille de $256\times256$ avec une condition CFL de $0{,}1$ et à $t=0{,}295$ s. Temps de calcul: environ $30\sim60$ min sur un ordinateur de bureau à 6 c\oe{}urs. }
	\label{Rotor}
\end{figure}

\paragraph{Résultats}  
 
 Sur la figure \ref{Rotor}, on peut voir que la forme circulaire initiale a été comprimée en un ovale par le champ magnétique. Au cours de ce processus le disque a été ralenti et des ondes d'Alfvén ont été émises dans le milieu ambiant.   
Néanmoins, une plus grande diffusion est observable en comparant avec les résultats de la littérature, notamment sur les contours extérieurs qui devraient être confondus. Cela peut être dû à notre maillage plus grossier et l'ordre de notre schéma. 

\section{Conclusion}

Nous avons sélectionné des cas-tests variés de par leur géométrie 1D, 2D ou 3D, et les ondes MHD présentes.
Le cas-test de Brio et Wu permet simuler à partir d'un problème 1D comprenant une discontinuité initiale, l'évolution de 5 ondes MHD : une onde de raréfaction rapide, une onde composée lente,  une discontinuité de contact, une onde de choc lente et une onde de raréfaction rapide.

Ce premier test a l'avantage d'être simple dans son interprétation physique, avec l'identification des ondes, et il est aussi immédiat de visualiser les effets du solveur (oscillation, \og overshoot \fg).
De même, il permet aussi de comparer aisément les solveurs entre eux, sur les différentes ondes, et d'évaluer leur précision.

Les autres cas-tests, impliquant de plus importantes ondes de choc comme Orszag-Tang, MHD Blast et de torsion comme dans le Rotor magnétique permettent de vérifier la robustesse du code.
En effet, des oscillations amenant à des valeurs non physiques, une perte de symétrie, ou des reconnexions magnétiques (impossible en MHD idéale) peuvent être observées au cours des simulations.

Ces tests ont permis de montrer qu'il est aussi pertinent d'évaluer la précision, la consistance, et la robustesse des solveurs de manière qualitative.
Néanmoins, nous proposons dans le chapitre suivant, une étude spécifique sur un cas-test décrivant la croissance d'une instabilité de Rayleigh-Taylor dont l'analyse théorique nous apportera une approche plus quantitative.

\clearpage
\chapter{Étude de l'Instabilité de Rayleigh-Taylor (IRT)}

Les études analytiques de cette instabilité permettent d'obtenir son taux d'accroissement en phase linéaire et son évolution asymptotique dans la phase non-linéaire.
Le taux d'accroissement de la phase linéaire est bien connu en HD et en MHD, il pourra permettre une confrontation plus quantitative avec le code. 
De plus, des modèles existent pour calculer la croissance asymptotique des structures en phase non-linéaire dans un cas HD.
Cette étude aboutira à proposer les simulations de l'IRT comme un cas-test avec des comparaisons pertinentes.  
De plus, nous avons sélectionné ce test, car il est représentatif de phénomènes physiques observés dans l'ionosphère comme les \og Equatorial spread F\fg.
\section{Cas Hydrodynamique }

\subsection{Mise en évidence physique de l'instabilité}

Un cas simple, pour mettre en évidence cette instabilité, est celui d'un verre d'eau retourné. Soit $g$ la constante gravitationnelle et $l$ la hauteur de liquide, la pression de l'eau $p_{eau}$ à l'interface eau/air vaut (en $x$): 

$$p_{eau} = \rho_{eau}\ g l \approx 10^3 \  \textrm{Pa pour } l=10\  \textrm{cm}$$ 

Or, la pression atmosphérique vaut $10^5 $ Pa, ce qui devrait suffire à maintenir l'eau dans le verre et pourtant l'eau tombe!

\begin{figure}[!h]
	\centering
	\includegraphics[width=0.3\textwidth]{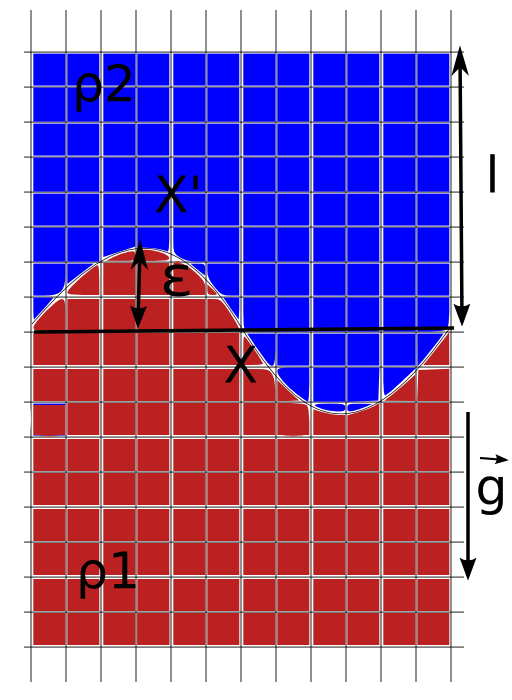}
	\caption{Perturbation de l'interface eau/air }
	\label{RTI_schema}
\end{figure}

Néanmoins, ceci n'est valable que si l'interface est parfaitement horizontale, alors que localement l'interface est sans cesse agitée. Le calcul de la pression ne peut donc pas s'effectuer de manière globale. Il faut considérer les petites perturbations à l'interface. En effet, considérant un point de l'interface perturbée tel que $x'=l-\epsilon$ ($\epsilon >0$) (voir la figure \ref{RTI_schema}), la pression vaut alors: 

$$ p'_{eau} = \rho_{eau}g (l-\epsilon) < p_{eau}$$

Puisque la pression atmosphérique reste la même en tout point de l'interface, 

$$\frac{p_{atm}}{p'_{eau}} > \frac{p_{atm}}{p_{eau}} $$

Ainsi, le gradient de pression exerce une force plus grande sur l'eau en $x'$ qu'en $x$. L'eau aux alentours de $x'$ va être poussée par une bulle d'air alors qu'en $x$ celle-ci va tomber en raison du gradient de pression. 
\subsection{Simulation de l'instabilité de Rayleigh-Taylor (IRT) hydrodynamique}

Nous allons simuler la croissance d'une instabilité de Rayleigh-Taylor hydrodynamique pour présenter les différentes étapes depuis la phase linéaire à la phase non-linéaire où des instabilités secondaires apparaissent. 
Les structures de ces instabilités secondaires sont très variées et dépendent fortement du schéma utilisé.

\begin{figure}[!h]
	\centering
	\begin{tabular}{c}
		\includegraphics[width=0.25\textwidth]{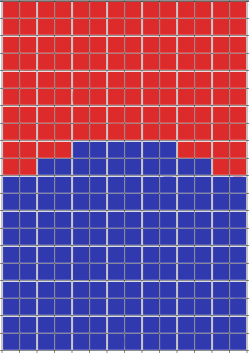}\\[\abovecaptionskip]
	        \small Initialisation avec \\ une perturbation en densité
	   
	\end{tabular}
	\begin{tabular}{c}
		\includegraphics[width=0.25\textwidth]{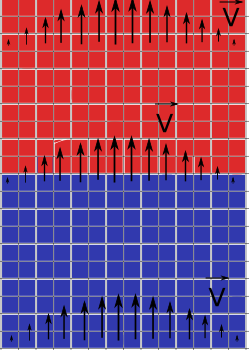}\\[\abovecaptionskip]
		\small Initialisation avec \\ une perturbation en vitesse
	\end{tabular}
\caption{Différentes manières de perturber l'interface pour l'IRT}
\label{RTIinit}
\end{figure}

\paragraph{Initialisation}
Ce cas a été réalisé en partant des conditions initiales proposées par les développeurs d'Athena sur leur site \textit{\og The Athena Code Test Page \fg}  \cite{athena_test}.
Nous avons un domaine rectangulaire défini sur $-0{,}25<x<0{,}25;$ $-0{,}75<y<0{,}75 $ avec des conditions aux limites périodiques, selon $x$ et des murs réfléchissants, selon $y$ ($x$ désigne la coordonnée spatial horizontale croissant de la gauche vers la droite et $y$ désigne la coordonnée spatial verticale croissant du bas vers le haut). 
Pour $y>0$, nous imposons une densité $\rho_2 = 2$ et pour $y<0$, une densité $\rho_1=1$. 
Nous avons également un champ gravitationnel constant, avec $g=0{,}1$. 
La pression suit la condition d'équilibre hydrostatique, $P=P_0 - \rho g y$ avec $P_0=2{,}5$
. 
Pour éviter des erreurs dues à la discrétisation, la perturbation est initialisée au travers de la vitesse du fluide, soit $v = 0{,}01[1+ \cos{4\pi x}][1+\cos{3\pi y}]/4$ pour le mode fondamental de la boite. 
L'initialisation en vitesse est préférable, car elle permet de perturber notre interface avec un maillage plus grossier sans que des erreurs de discrétisation apparaissent. 
Le schéma de gauche \ref{RTIinit} représente parfaitement l'erreur entre la perturbation en densité théorique et celle discrétisée, ce qui pourrait influencer nos résultats pour la phase linéaire de l'instabilité. 
Pour une perturbation en vitesse, illustrée par le schéma de droite \ref{RTIinit}, ce problème n'apparaît pas, mais nous devons toutefois imposer une vitesse $v$ constante selon $y$ ou tout du moins la faire varier de façon continue comme ici pour éviter la présence d'une onde de choc ou de raréfaction au niveau de l'interface. 
 
\begin{figure}[!h]
	\centering
	\begin{tabular}{c}
		\includegraphics[width=0.21\textwidth]{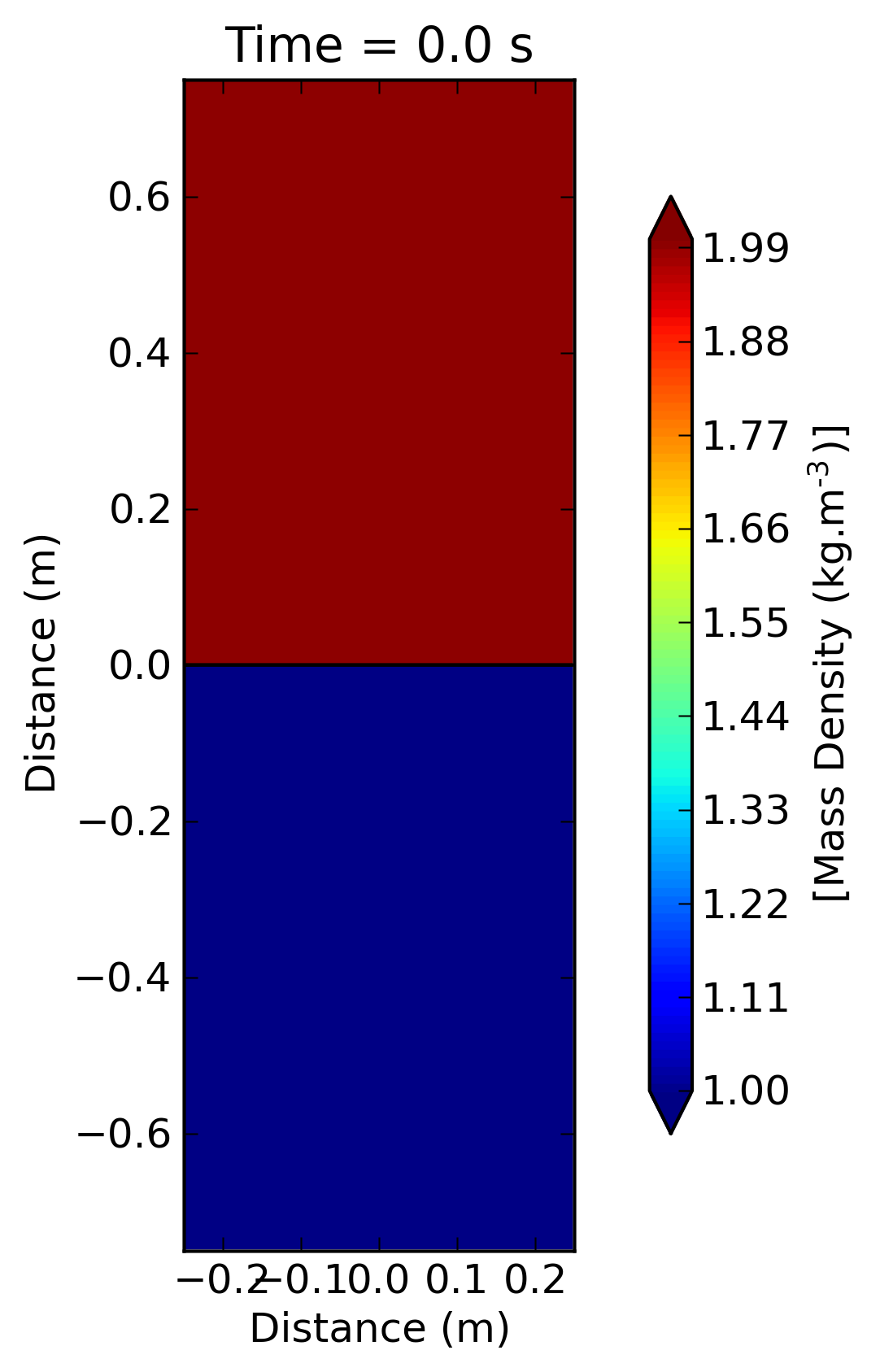}\\[\abovecaptionskip]
	\small à $t=0$ s
	\end{tabular}
	\begin{tabular}{c}
		\includegraphics[width=0.21\textwidth]{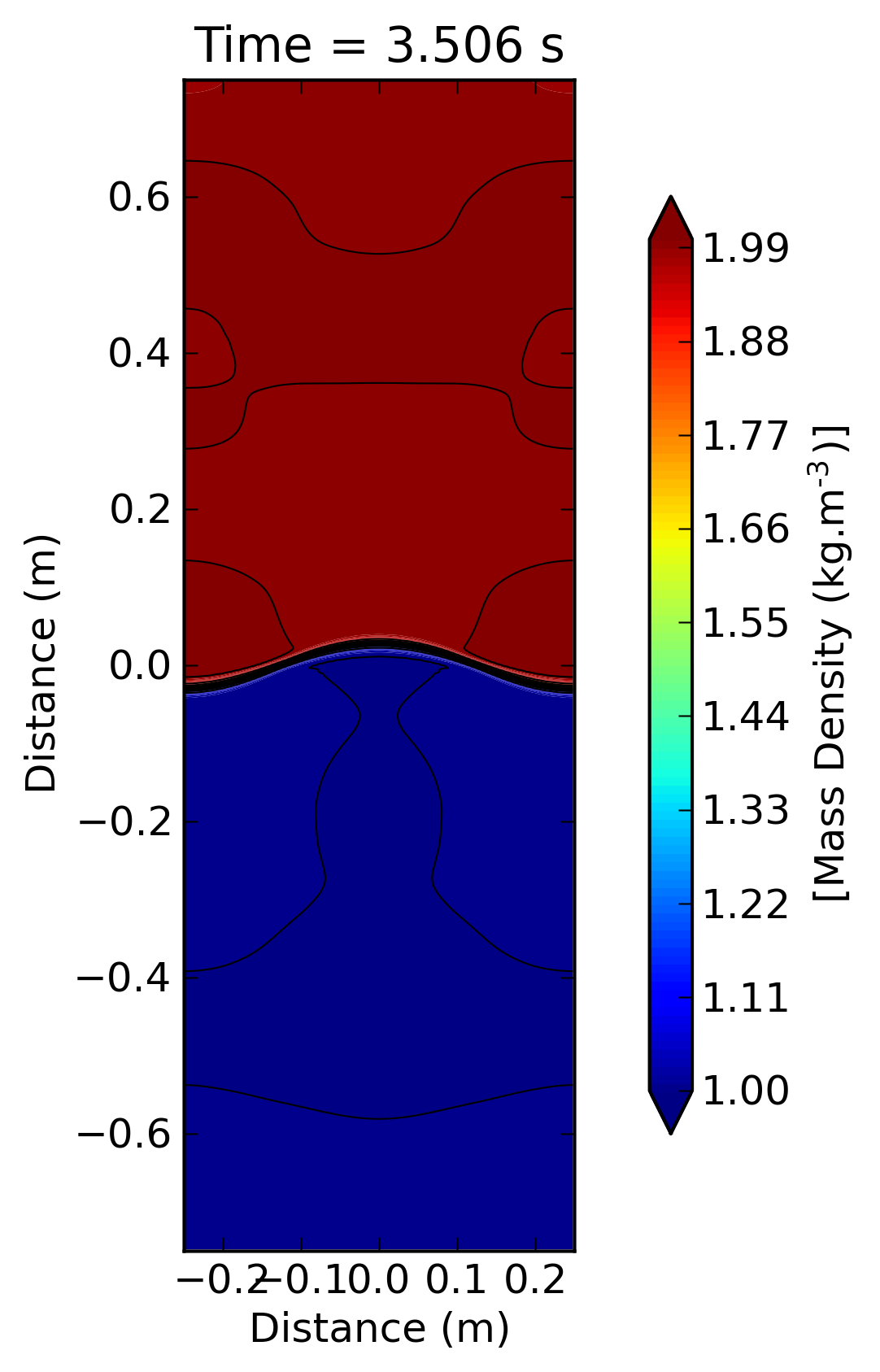}\\[\abovecaptionskip]
	\small à $t=3{,}51$ s
	\end{tabular}
	\begin{tabular}{c}
		\includegraphics[width=0.21\textwidth]{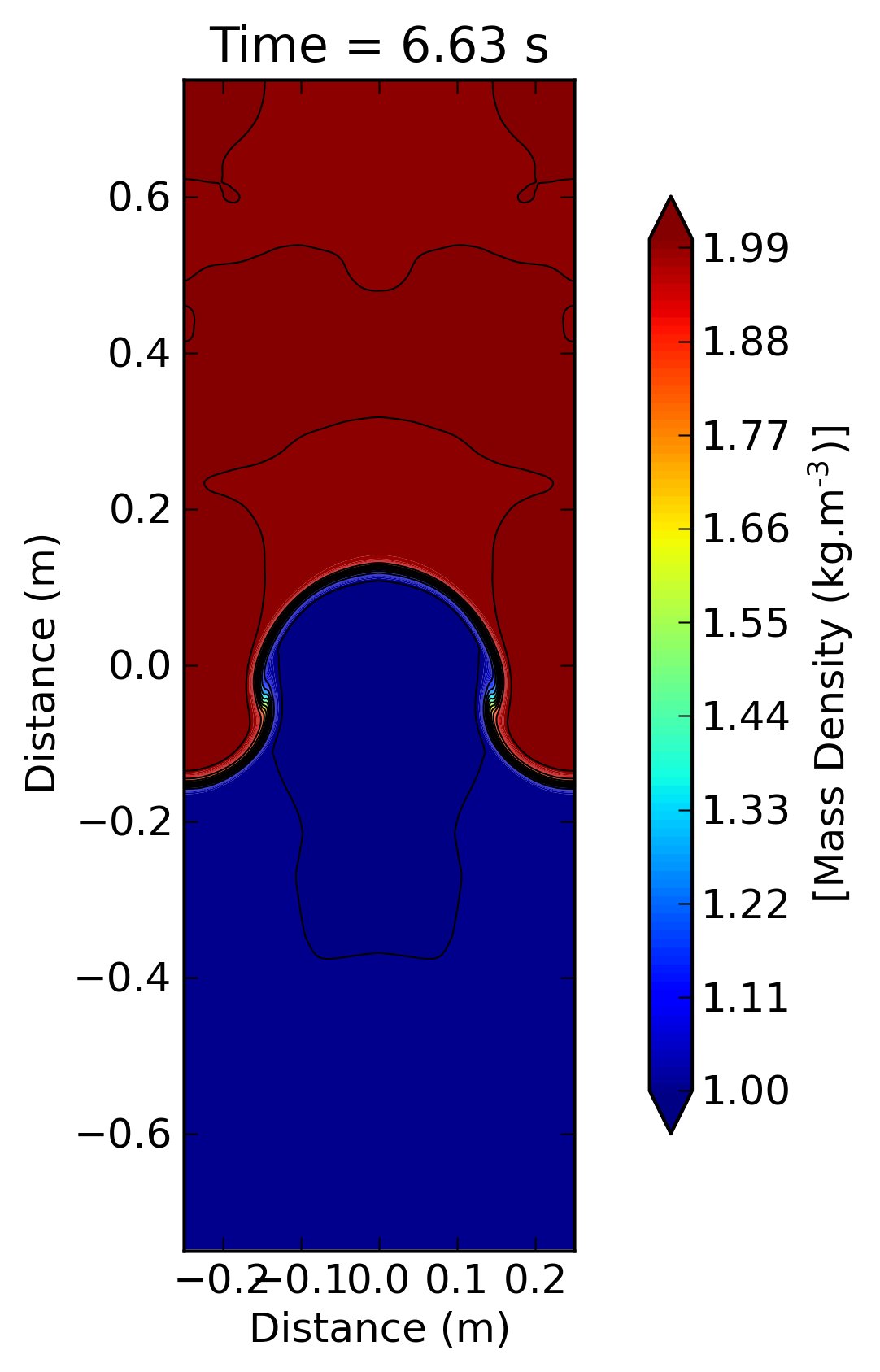}\\[\abovecaptionskip]
	\small à $t=6{,}63$ s
	\end{tabular}
	\begin{tabular}{c}
		\includegraphics[width=0.21\textwidth]{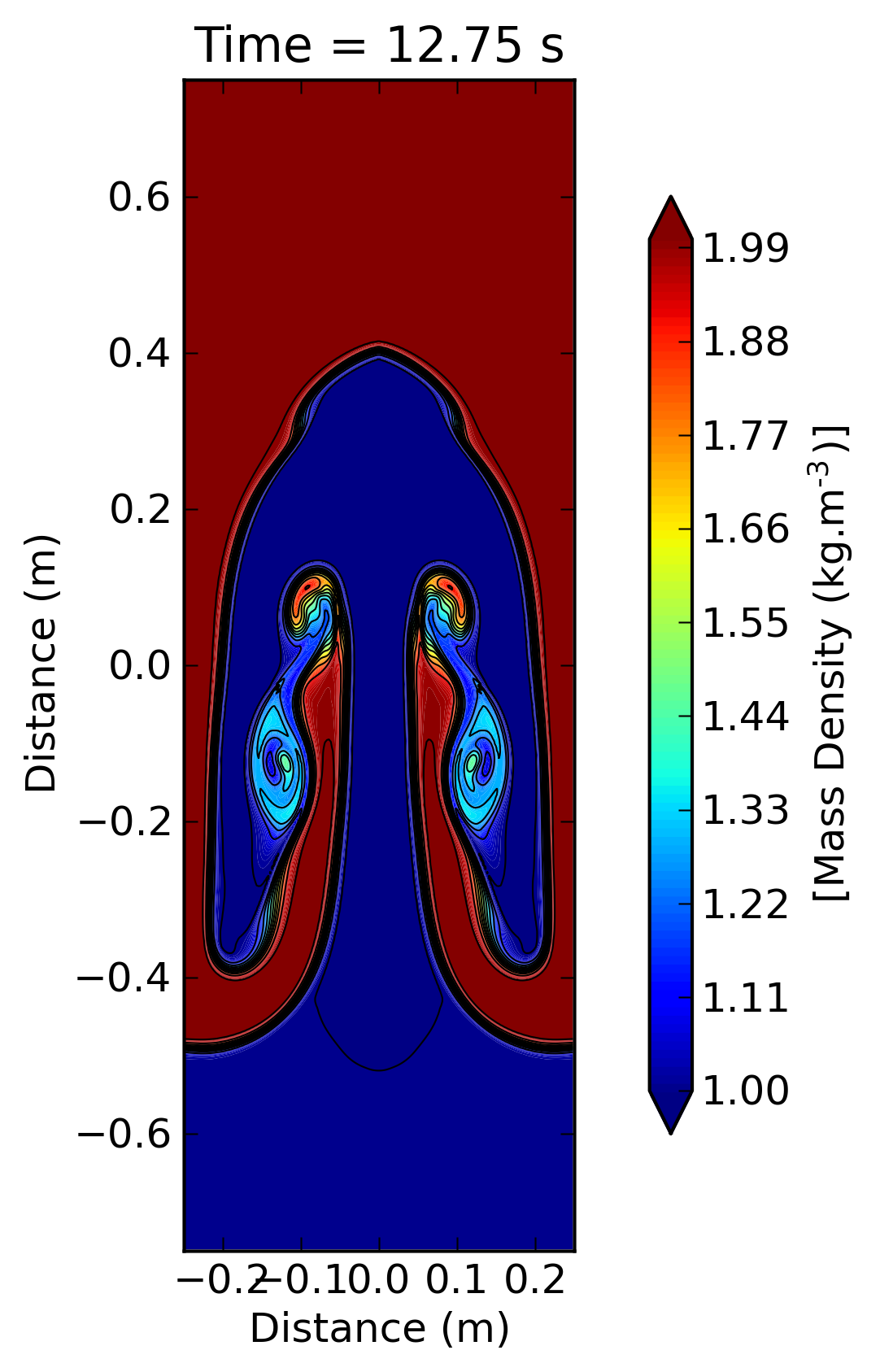}\\[\abovecaptionskip]
	\small à $t=12{,}75$ s
	\end{tabular}
\caption{Simulation du cas-test IRT pour un cas HD avec un solveur de Roe \og f-wave\fg\, avec la gravité incluse dans le flux et une CFL de 0,5,  sur une grille de $200\times600$. Temps de calcul: environ $10$ h sur un n\oe{}ud (28 c\oe{}urs) du supercalculateur \textit{Cobalt}.}
\label{RTIfin}
\end{figure}

\paragraph{Description}

 Nous pouvons voir sur la figure \ref{RTIfin} que la phase linéaire de l'instabilité se termine rapidement pour entrer dans une phase non-linéaire. 
En général, on considère la fin de la phase linéaire quand l'amplitude de l'instabilité atteint la moitié de sa longueur d'onde.
Dans la phase non-linéaire, nous remarquons l'apparition de structures secondaires, notamment des rouleaux créés par l'instabilité de Kelvin-Helmoltz, caractéristique du cisaillement entre les fluides montant et descendant. 
Le nombre d'instabilités de Kelvin-Helmoltz secondaires est très sensible à la manière de perturber l'interface et avec quelle précision le solveur résout la discontinuité de contact.  

En effet, Wendroff et Liska ont effectué une comparaison entre 8 différents schémas pour le test de Rayleigh-Taylor \cite{liska2003comparison} et comme nous pouvons le voir sur la figure 4.8 de leur article, les structures secondaires de Kelvin-Helmoltz  sont très différentes, selon les schémas, aussi bien en nombre qu'en taille. 

Ils ont remarqué que les schémas moins diffusifs, comme \textit{WENO} et \textit{CLAW}, ont plus de structures secondaires que des schémas plus diffusifs, comme \textit{CFLFh}. 
Néanmoins, il ne faut pas extrapoler ce résultat en supposant que plus de structures fines impliquent un meilleur schéma, car des schémas comme \textit{PPM} qui possèdent un \og\, contact steepener \fg (qui permet d'obtenir une discontinuité de contact plus abrupte), font apparaître des marches d'escalier dans les discontinuités de contact et ainsi donnent naissance à des instabilités de Kelvin-Helmoltz causées par des erreurs de discrétisation \cite{athena_test,liska2003comparison}.  

L'IRT a été présentée ici comme un simple cas-test, mais la difficulté à comparer de façon qualitative les résultats entre différents solveurs et la richesse de ce phénomène physique nous ont poussé à étudier de façon beaucoup plus détaillée l'IRT.

\subsection{Analyse linéaire du cas 2D en HD}

Notre configuration initiale est décrite par un champ de vitesse moyenne défini par $\mathbf{V}_0 =  (u_0(x,y),v_0(x,y))$ avec $u_0(x,y) = v_0(x,y) =0$, où le champ gravitationnel est $\mathbf{g} = g \hat{\mathbf{y}}$. Une interface en $y=0$ sépare les fluides de densités $\rho_1$ dans la zone inférieure et $\rho_2$ dans la zone supérieure. 

Nous décrivons la perturbation apportée au système comme un champ de vitesse d'amplitude infiniment petite, $\mathbf{V}' = (u'(x,y),v'(x,y)) $. Nous nous plaçons dans le cas incompressible, ainsi ce champ de vitesse peut être décrit par des lignes de courant, notées $\psi$. 

\begin{equation}
    \mathbf{V}' = (u'(x,y),v'(x,y)) = (\partial_y \psi, -\partial_x \psi )
\end{equation}
Le champ de vitesse est supposé irrotationel ($\nabla  \times \mathbf{V}'= 0$), ainsi les lignes de courant vérifient $\nabla^2\psi = 0$. Comme le système est invariant, selon $x$, on peut chercher une solution sous la forme: 

\begin{equation}
    \psi(x,y,t) = e^{ik(x-ct)}\Psi(y)
\end{equation}

où $k$ est le nombre d'onde spatial, $c$ la vitesse de phase et $\Psi(y)$ une fonction à déterminer. Ainsi, le problème se ramène à la résolution de l'équation: 

\begin{equation}
    (\partial_y^2 - k^2) \psi_{1,2} =0 
    \label{courant_equa}
\end{equation}

où l'indice $1$ désigne le fluide dans la zone inférieure ($-\infty <y<0$) et l'indice $2$ désigne le fluide dans la zone supérieure ($0<y<\infty$). Pour déterminer notre solution complète, il nous faut utiliser les conditions aux limites et à l'interface, ce qui nous permettra de déterminer $c$ et ainsi notre condition de stabilité. 

La première condition est une vitesse nulle à l'infini, car nous travaillons avec une énergie finie. Ainsi, $v'_1 =0$ en $y=-\infty$ et $v'_2 =0$ en $y=+ \infty$, soit pour nos lignes de courant: 

\begin{equation}
    \begin{array}{cc}
    \Psi_1(-\infty) =0     &  \Psi_2(+ \infty) =0  
    \end{array}
\end{equation}

Les trois autres conditions sont fournies par le comportement de l'interface perturbée $y=\eta(x,t)$. 

La \textit{continuité de la composante verticale de la vitesse}, en $y=\eta$, impose que  $v'_1 = v'_2$. Soit en terme de ligne de courant: 

\begin{equation}
    \Psi_1(\eta) = \Psi_2(\eta) 
\end{equation}

Par développement limité en $y=0$, on obtient: 

\begin{equation}
    \Psi_1(0) = \Psi_2(0) + o(y)
\end{equation}

La \textit{condition de surface libre} impose que  le long de la surface $y=\eta(x,t)$, la condition cinétique suivante s'applique: 

\begin{equation}
    \partial_t \eta+ u'\partial_x \eta = v'(\eta)
\end{equation}

Par linéarisation on obtient:

\begin{equation}
    \partial_t \eta = v'(0)
\end{equation}

En utilisant les représentations de mode normal et les lignes de courant, cette condition s'écrit $c\eta = \Psi$.

L'\textit{absence de saut de pression à l'interface} impose (en négligeant la tension superficielle): 

\begin{equation}
    p_2(y=\eta) - p_2(y=\eta) =0 
\end{equation}

Soit, en séparant la pression totale et la pression de la perturbation: 

\begin{equation}
    [p_{02}(\eta) + p'_2] -  [p_{01}(\eta) + p'_1]= 0
\end{equation}

En linéarisant la pression totale avec la pression hydrostatique, on a:  

\begin{equation}
    \begin{array}{cc}
        p_{02} = - \rho_2g \eta +p_0(0) & p_{01} = - \rho_1g \eta +p_0(0) 
    \end{array}
\end{equation}

et donc

\begin{equation}
    p'_2-p'_1 = g \eta (\rho_2 - \rho_1)
\end{equation}

Puis en utilisant la conservation de la quantité de mouvement, $\partial_t u'_i = -\frac{1}{\rho_i}\partial_x p'_i$, avec $i=1,2$, on obtient :

\begin{equation}
    p'_i = \rho_ic \partial_y \Psi 
\end{equation}

Notre condition d'absence de saut devient: 

\begin{equation}
    c(\rho_2 \partial_y \Psi_2 - \rho_1 \partial_y \Psi_1) = g \eta(\rho_2-\rho_1) 
\end{equation}

En utilisant $c \eta = \Psi$, on obtient:

\begin{equation}
    c^2(\rho_2 \partial_y \Psi_2 - \rho_1 \partial_y \Psi_1) = g \Psi(\rho_2-\rho_1) 
    \label{courant_saut}
\end{equation}

A noter que seules les dérivées de $\Psi$ sont indexées car $\Psi_1 = \Psi_2$ en $y=0$. 

L'équation (\ref{courant_equa}) et les conditions à l'infini nous donnent: 
\begin{equation}
    \begin{array}{cc}
         \Psi_1 = A_1 e^{k y}, &  \Psi_2 = A_2 e^{-k y}
    \end{array}
\end{equation}

or comme $\Psi_1 = \Psi_2$ en $y=0$, on a $A_1=A_2=A$ . Enfin, en injectant notre solution dans (\ref{courant_saut}), on obtient : 

\begin{equation}
    c^2 = \frac{g}{k} \frac{\rho_1-\rho_2}{\rho_2+\rho_1}
\end{equation}

Notre solution est instable si $c^2< 0$ ($c$ est imaginaire pur). Nous retrouvons, ainsi, la condition d'instabilité $\rho_2 > \rho_1$ (fluide le plus lourd au-dessus). Et notre solution s'écrit sous la forme: 

\begin{equation}
    \psi(x,y,t) = Ae^{\gamma t}e^{ikx -k\mid y \mid }
\end{equation}

où $\gamma=-ikc$. Ainsi, on obtient le taux d'accroissement donné par $\gamma = \sqrt{gk \frac{\rho_2-\rho_1}{\rho_2+\rho_1}}$ \cite{chandrasekhar1961hydrodynamic,drazin2002introduction}. 

La quantité $\frac{\rho_2-\rho_1}{\rho_2+\rho_1}=A_t$ s'appelle le nombre d'Atwood.

\subsection{Étude de la phase linéaire}

Nous allons tout d'abord choisir un moyen de mettre en évidence proprement l'amplitude théorique et numérique en fonction du temps. Ensuite, nous changerons la longueur d'onde de l'instabilité. Et enfin, nous montrerons l'effet de la résolution du maillage sur la croissance de l'instabilité.

\paragraph{Représentation de l'amplitude théorique et numérique}
Pour pouvoir capturer de façon précise l'instabilité, nous avons réduit le domaine, tel que  $-0{,}25<x<0{,}25$ et  $ -0{,}05<y<0{,}05 $ tout en maintenant une grille de $100\times 300$. Notre perturbation initiale est  $v=0{,}01(1+\cos{4\pi x})/2$. 

Pour mettre en évidence le taux d'accroissement linéaire, nous avons tracé le logarithme de l'amplitude de la position verticale de l'interface. Pour évaluer l'amplitude, nous avons utilisé les trois normes suivantes $L_{inf}$, $L_1$ et $L_2$, qui pour rappel sont définies par: 

\begin{equation}
    \left\{
    \begin{array}{l}
         L_{inf} = \max(\mid y'_i \mid)  \\
         L_1 = \sum \limits_i \mid y'_i \mid \\
         L_2 =  \sqrt{\sum \limits_i \mid y'_i \mid^2} \\
    \end{array}
    \right. \quad i \in [1;100]
\end{equation}

avec $\mathbf{y}' = (y'_i)$ pour $i \in [1;100]$ la position de l'interface en $y$ le long de l'axe $x$. 

\begin{figure}[!h]
    \centering
    \includegraphics[scale=0.5]{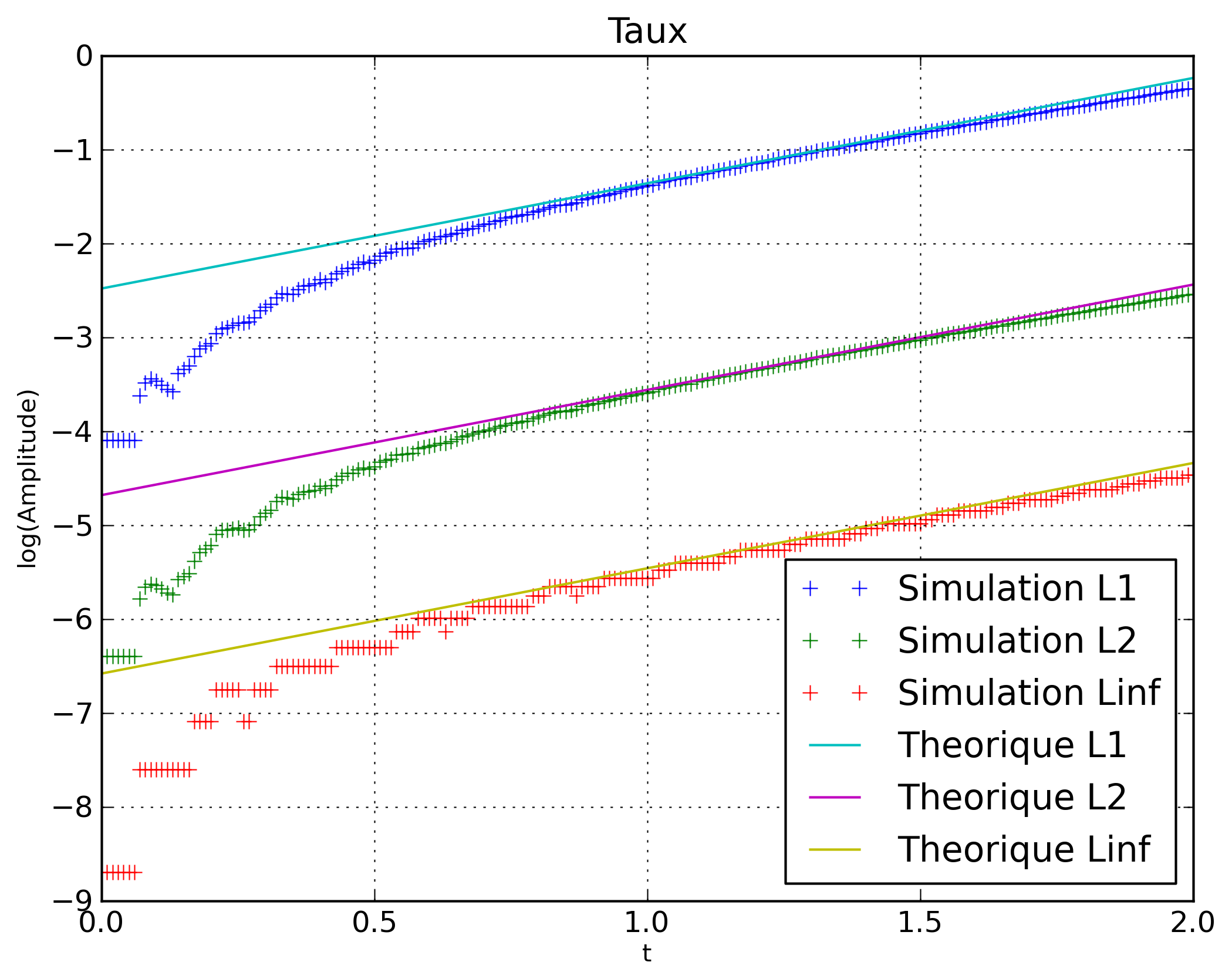}
    \caption{Amplitude de la perturbation en fonction des trois différentes normes pour le cas hydro avec le solveur Roe \og f-waves \fg \, et une CFL de 0,5.}
    \label{TauxRTI}
\end{figure}

Sur la figure \ref{TauxRTI} , nous pouvons distinguer trois phases: 

\begin{itemize}
    \item Une partie de transition due à notre perturbation initiale sur environ $0<t<0{,}7$.  
    \item Une partie qui suit la croissance linéaire analytique sur environ $0{,}7<t<1{,}7$.
    \item Et le début de la partie non-linéaire qui s'écarte de notre courbe théorique pour $t>1{,}7$.  
\end{itemize}

Avec la norme $L_{inf}$, on voit la position du point le plus éloigné de l'interface initiale, ainsi, dans les premiers instants, un nombre considérable de pas de temps est nécessaire pour que ce point se déplace d'un volume discret à un autre. Ceci nous conforte dans notre choix d'une perturbation initiale en vitesse plutôt que sur l'interface elle-même, car une résolution plus importante aurait été nécessaire pour décrire de façon précise une perturbation de faible amplitude. 

 \begin{figure}[!h]
	\centering
	\begin{tabular}{c}
		\includegraphics[width=0.45\textwidth]{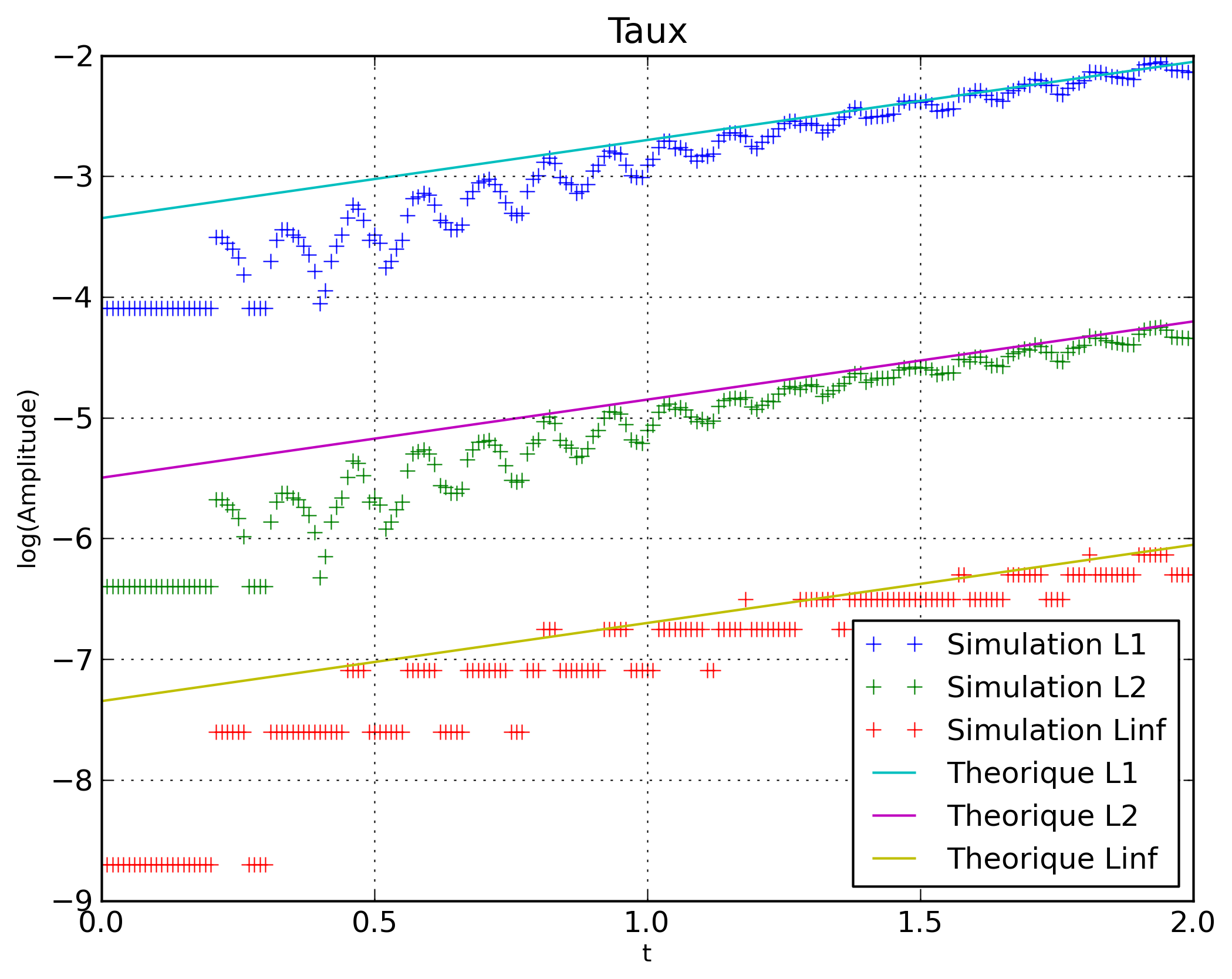}\\[\abovecaptionskip]
		\small $\lambda = L_x$ 
		\label{taux1}
	\end{tabular}
	 \vspace{\floatsep}
	\begin{tabular}{c}
		\includegraphics[width=0.45\textwidth]{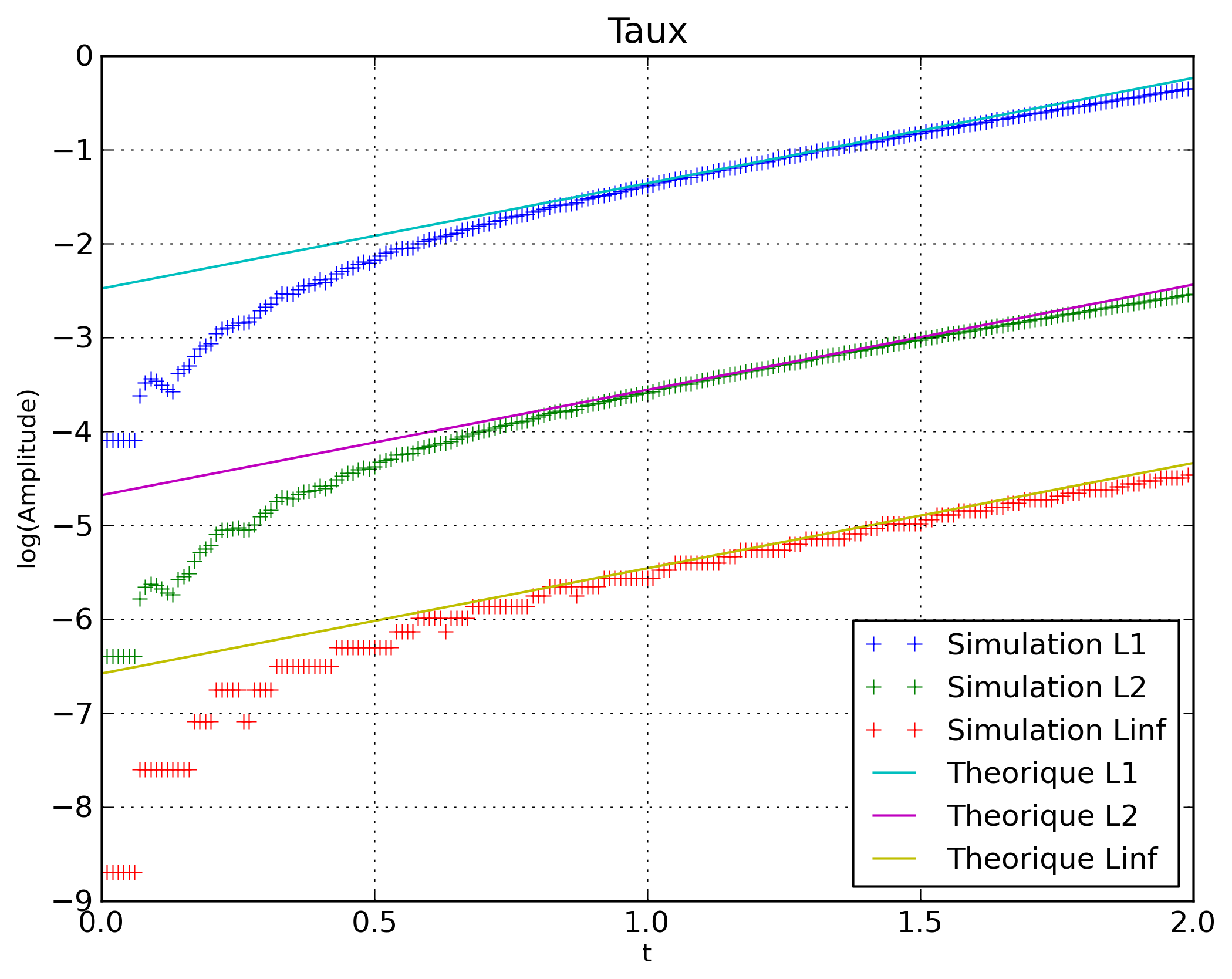}\\[\abovecaptionskip]
		\small $\lambda = L_x/3$ 
		\label{taux3}
	\end{tabular}
	 \vspace{\floatsep}
	\begin{tabular}{c}
		\includegraphics[width=0.45\textwidth]{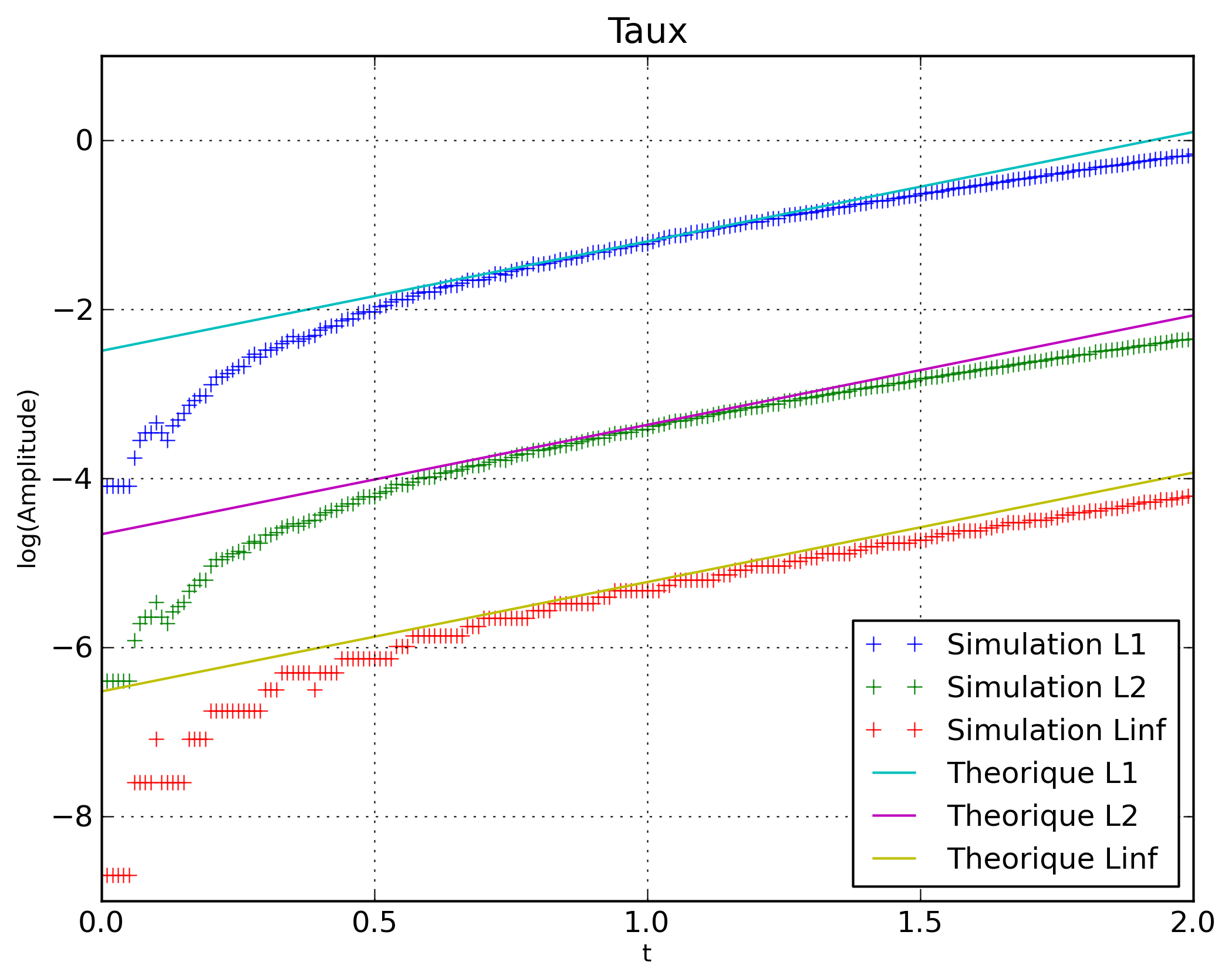}\\[\abovecaptionskip]
		\small $\lambda = L_x/4$
		\label{taux4}
	\end{tabular}
	 \vspace{\floatsep}
	\begin{tabular}{c}
		\includegraphics[width=0.45\textwidth]{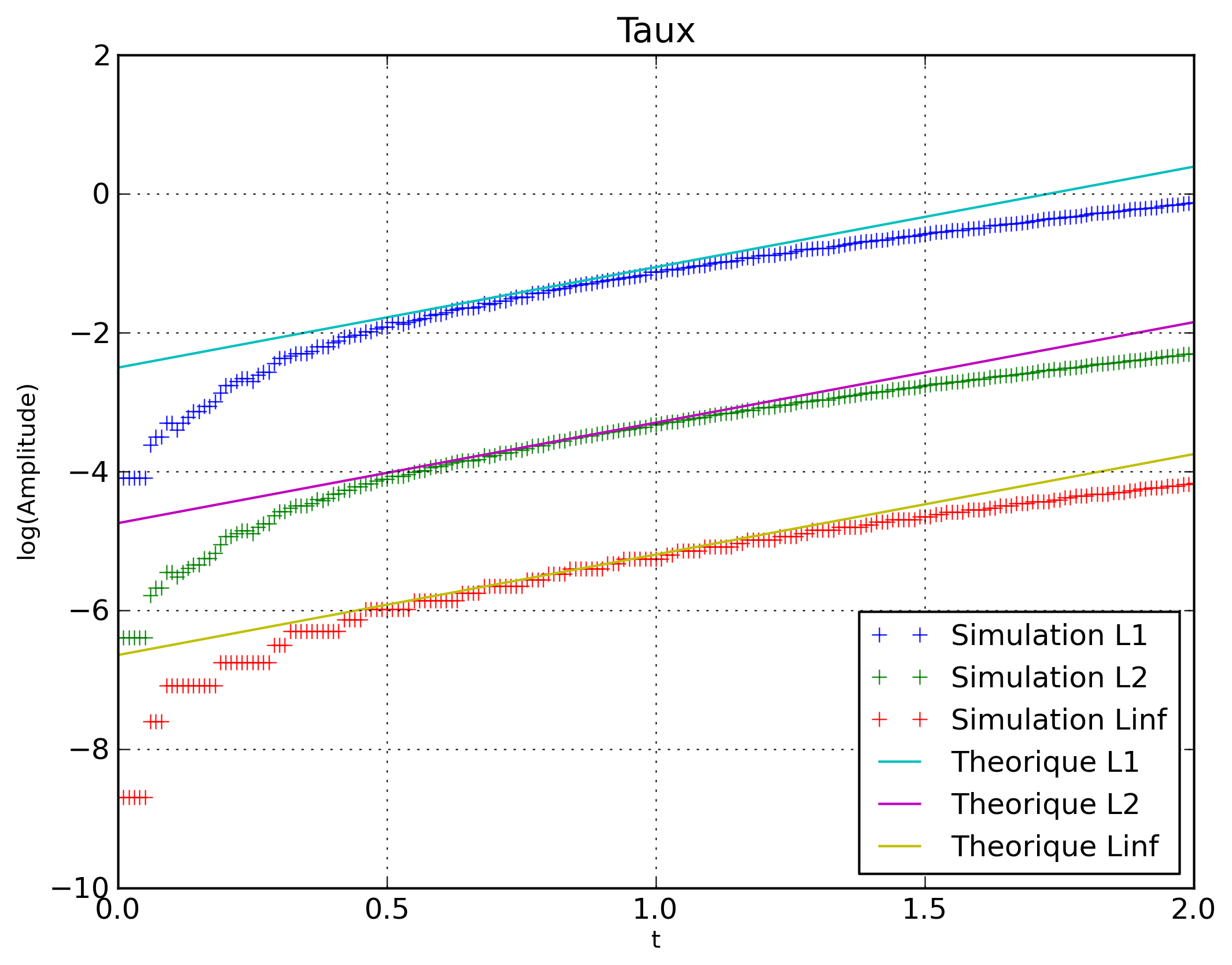}\\[\abovecaptionskip]
		\small $\lambda = L_x/5$
		\label{taux5}
	\end{tabular}
  \caption{Observation du taux d'accroissement pour différentes longueurs d'onde $\lambda$ avec un maillage de  $100\times 300$ ($L_x$ est la largeur du domaine)  }
  \label{RTItauxdiff}
  \end{figure}

\paragraph{Variation de la longueur d'onde}

Les oscillations visibles sur la première figure \ref{RTItauxdiff} sont dues à notre perturbation en vitesse qui fait des aller-retour en se réfléchissant sur parois, provoquant une oscillation de l'interface autour de l'origine, jusqu'à ce que l'énergie de l'instabilité dépasse celle de la perturbation initiale. Néanmoins, notre croissance moyenne suit la croissance théorique de la phase linéaire. 
On peut voir que sur les deuxième et troisième figures \ref{RTItauxdiff}, la phase linéaire de l'évolution de l'instabilité de Rayleigh-Taylor est bien visible ce qui est encourageant pour la validité de notre simulation. Sur la dernière figure \ref{RTItauxdiff}, la phase linéaire est plus courte, puisque, plus on diminue la longueur d'onde, plus on augmente le taux d'accroissement et donc, plus vite on atteint la phase non-linéaire. 

\begin{figure}[!h]
    \centering
    \includegraphics[scale=0.5]{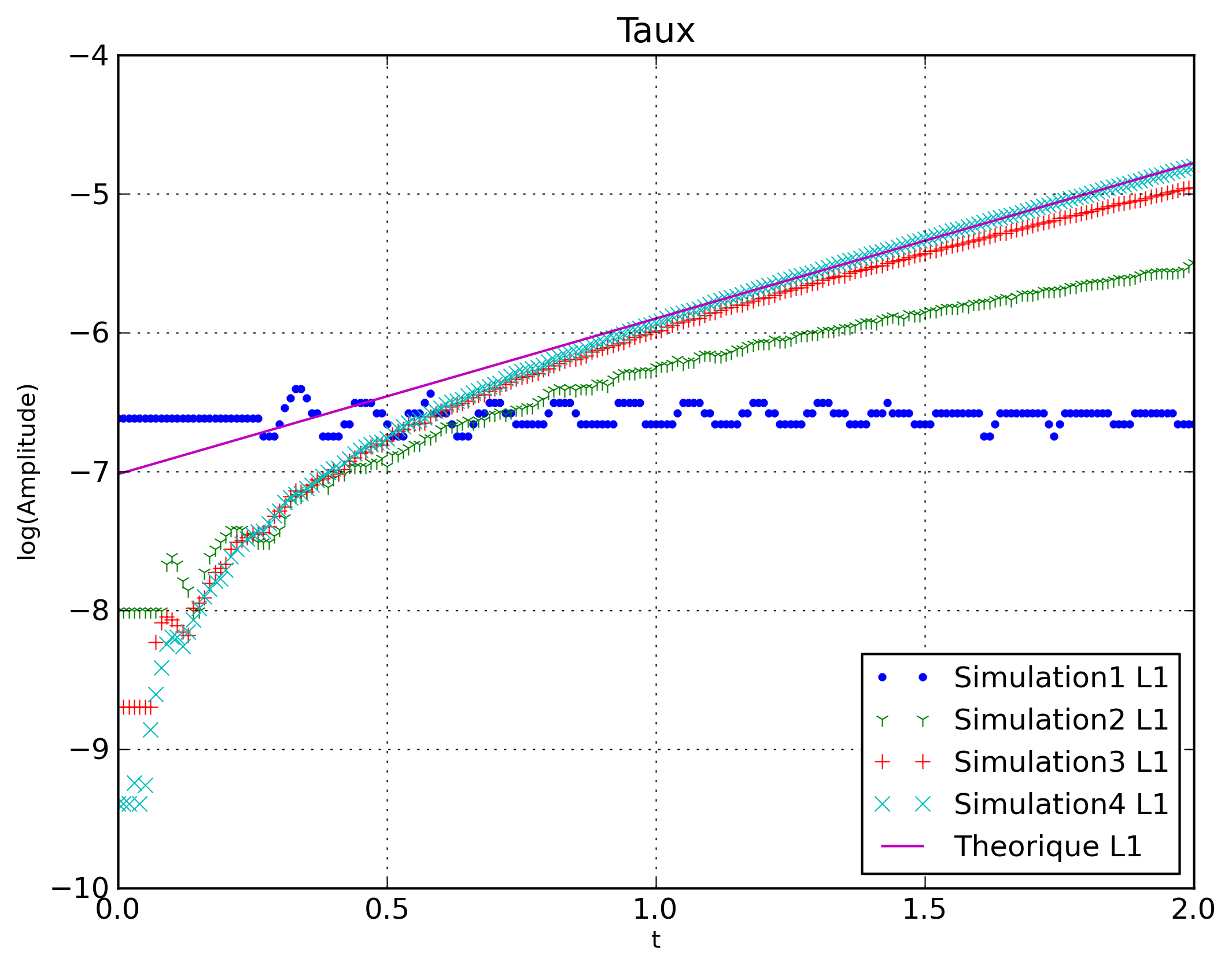}
    \caption{Amplitude de la perturbation avec différents raffinements en maillage pour une longueur d'onde $\lambda = L_x/3 $ avec le solveur de Roe \og f-waves \fg. Les maillages sont respectivement de 1 à 4: $25\times75$, $50\times150$, $100 \times 300$ et $200 \times 600$ }
    \label{Tauxmaillage}
\end{figure}

\paragraph{Convergence en maillage}

\begin{figure}[!h]
	\centering
	\begin{tabular}{c}
		\includegraphics[width=0.45\textwidth]{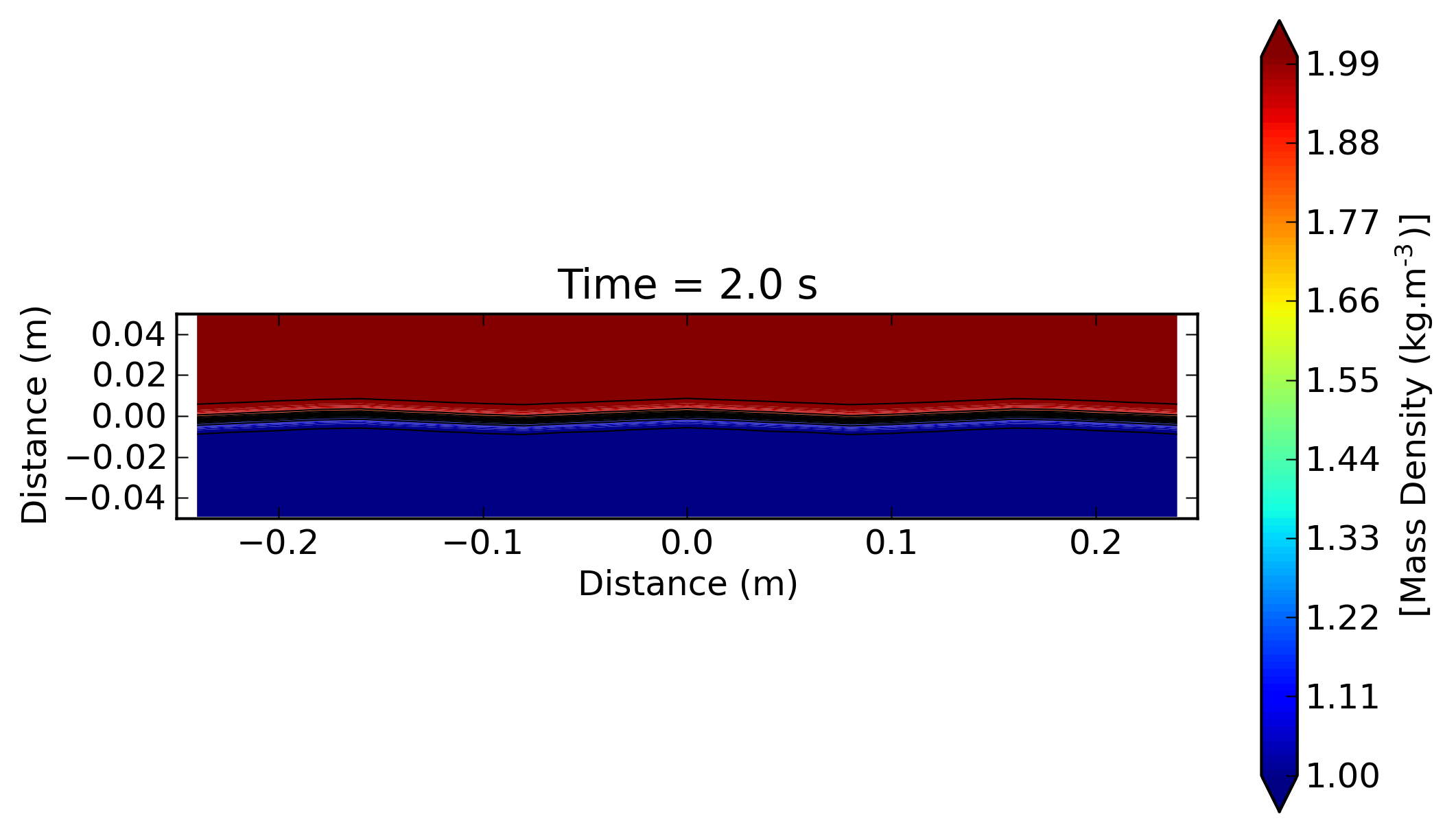}\\[\abovecaptionskip]
		\small Maillage de $25\times 75$
	\end{tabular}
	\begin{tabular}{c}
		\includegraphics[width=0.45\textwidth]{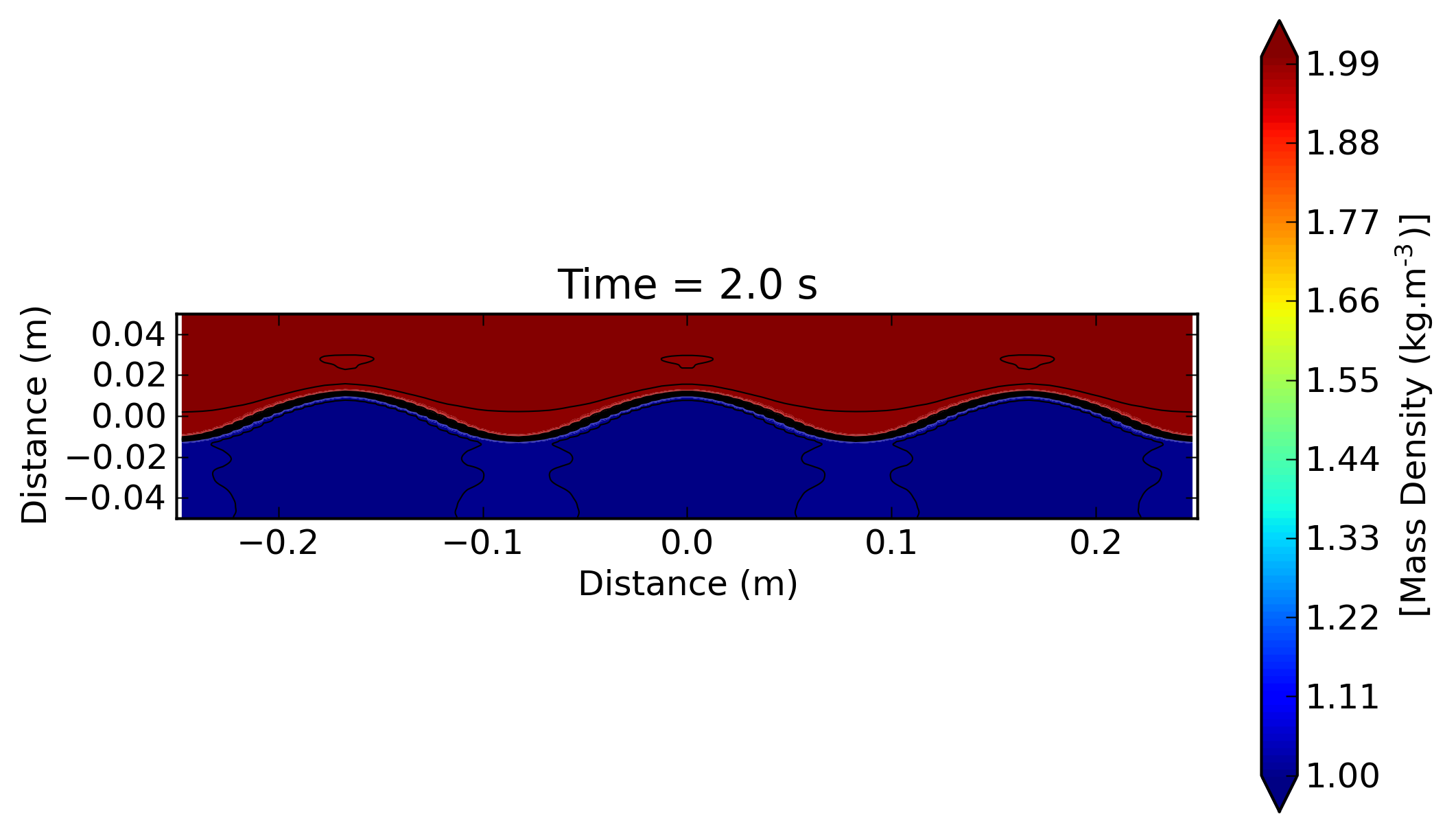}\\[\abovecaptionskip]
		\small Maillage de $100\times 300$ 
	\end{tabular}
\caption{Profil de densité à $t=2$ s pour l'étude de la phase linéaire avec le solveur de Roe \og f-waves \fg. Temps de calcul: environ $4 \sim 5$ h sur un n\oe{}ud (28 c\oe{}urs) du supercalculateur \textit{Cobalt}.}
\label{RTImaillage}
\end{figure}

Sur la figure \ref{Tauxmaillage}, nous voyons que notre schéma converge en maillage pour la phase linéaire de notre instabilité. Nous avons seulement représenté la norme $L_1$ pour plus de lisibilité et de simplicité. En effet, le taux d'accroissement théorique que nous avons calculé précédemment ne prend pas en compte la diffusion numérique. Il y a une certaine \og compétition \fg \, entre la diffusion numérique et la croissance de l'IRT, car le taux d'accroissement est plus faible lorsque nous avons un gradient de densité entre les deux fluides (dû à la diffusion numérique) comparativement à un saut de densité. Ainsi pour le plus petit maillage, la diffusion numérique est telle que l'instabilité n'a pas le temps de croître significativement. Sur la figure \ref{RTImaillage}, on peut voir directement que l'instabilité a été amortie par la diffusion numérique pour le maillage $25\times 75$, contrairement au maillage $100 \times 300$. Pour des maillages plus importants, on voit que la solution converge et que la théorie linéaire se vérifie, ce qui est un signe de validité de notre code. 

\subsection{Phase non-linéaire}

Nous allons maintenant étudier l'évolution des structures dans la  phase non-linéaire. Les structures sont appelées \og bulle \fg\, pour la portion du fluide léger qui \og monte \fg\, dans le fluide lourd et \og jet \fg\, pour la portion de fluide lourd qui  \og descend \fg\, dans le fluide léger.

La perturbation initiale sera mono-mode, pour éviter les interactions entre les structures.
Nous avons préparé nos simulations pour obtenir des tailles de structure telles que $\lambda /2<h<L_x/2$, où $h$ est la hauteur de la structure par rapport à l'interface initiale.

Un autre test consisterait à initialiser une perturbation en multi-modes pour observer spécifiquement les fusions de  bulles ou de jets pour former finalement des structures plus grandes. 

\subsubsection{Nombre d'Atwood égal à 1 ($A_t=1$)}

\begin{figure}[!h]
	\centering
	\begin{tabular}{c}
	\includegraphics[width=0.21\textwidth]{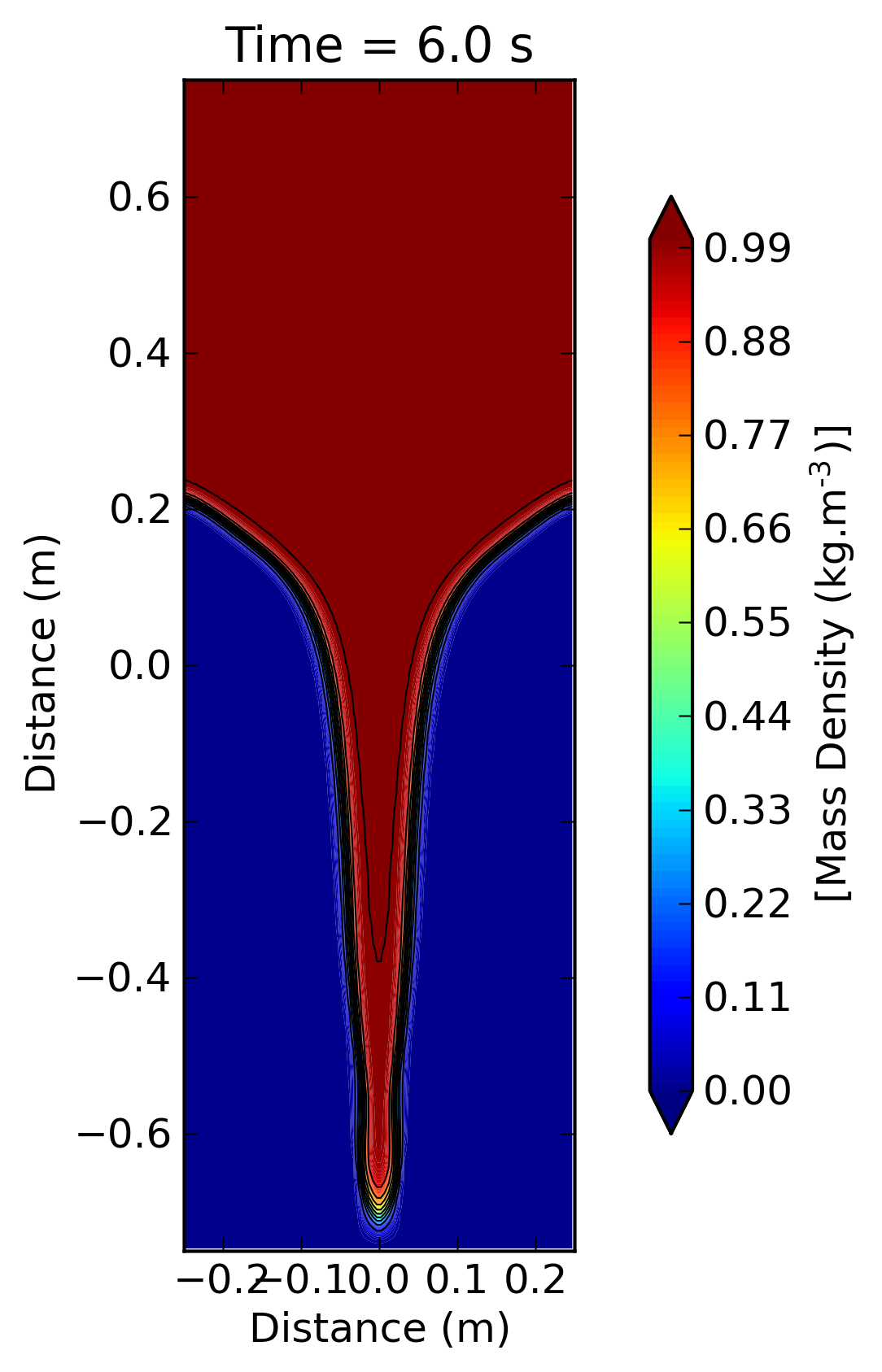}\\[\abovecaptionskip]
	\small Roe 'f-wave' ordre 2  \\
	et une CFL de 0,5
	\end{tabular}
	\begin{tabular}{c}
	\includegraphics[width=0.21\textwidth]{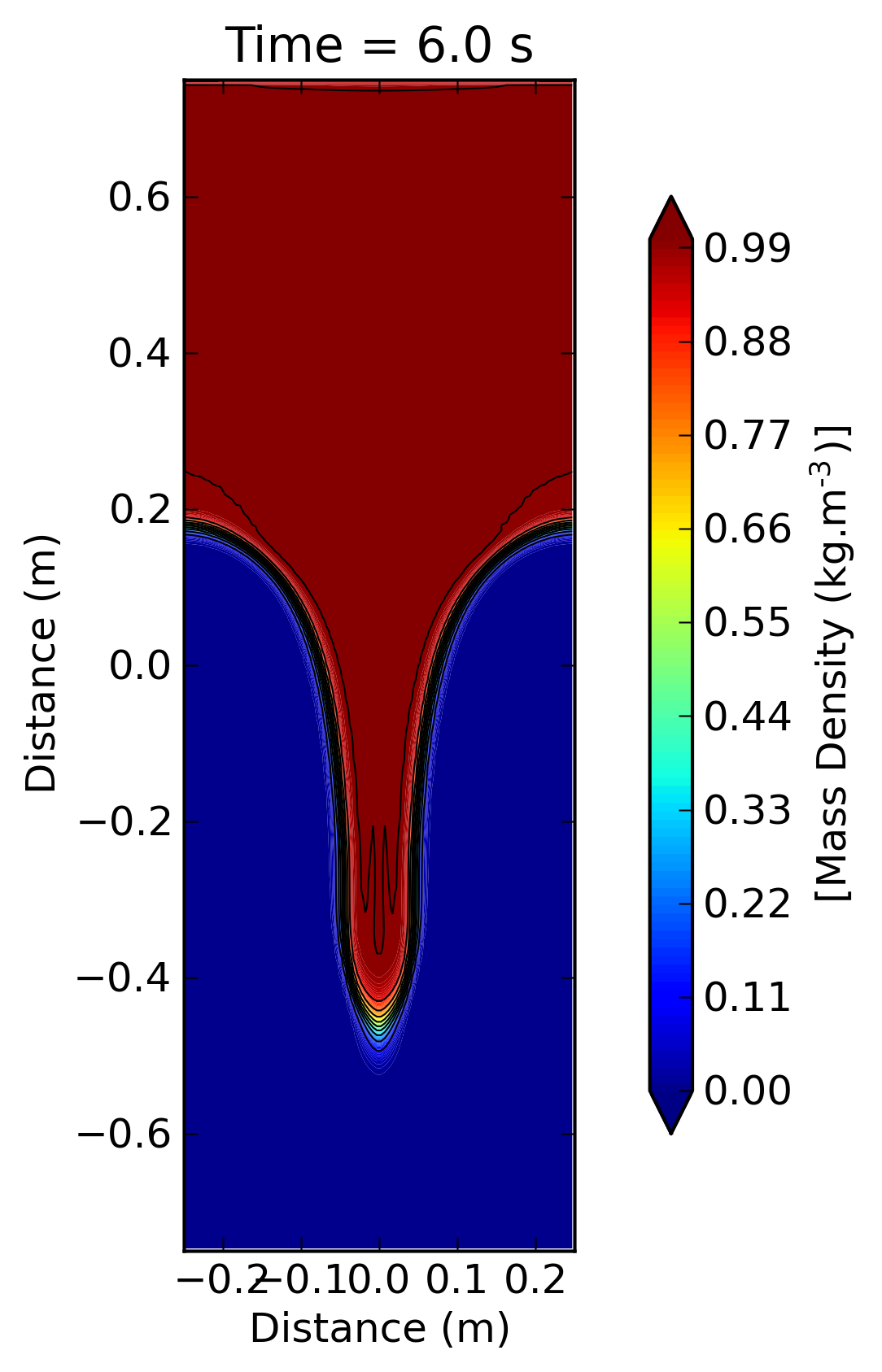}\\[\abovecaptionskip]
		\small Roe ordre 2 \\
		et une CFL de 0,2
	\end{tabular}
	\begin{tabular}{c}
	\includegraphics[width=0.21\textwidth]{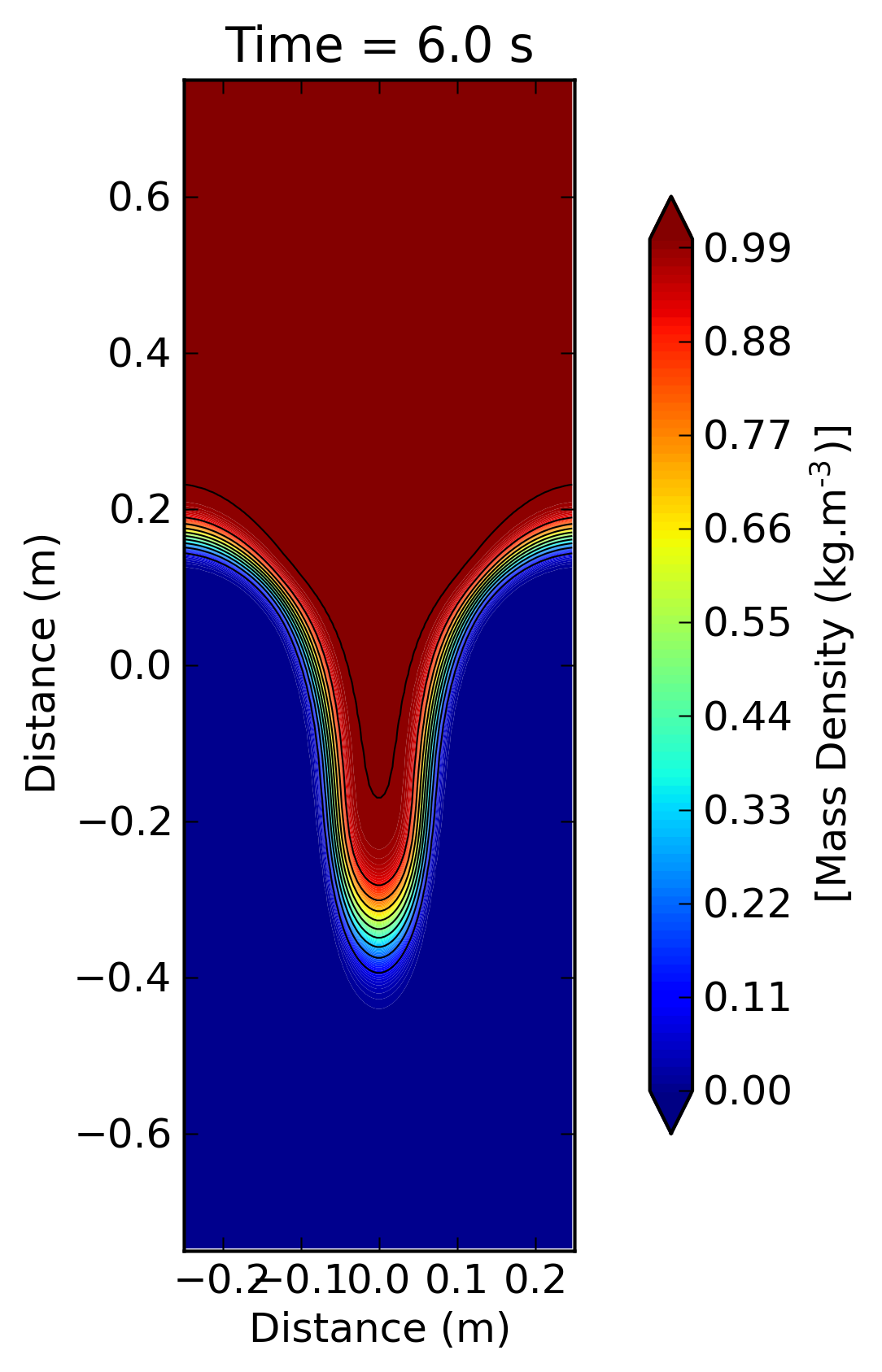}\\[\abovecaptionskip]
	\small Roe 'f-wave' ordre 1 \\
	et une CFL de 0,5
	\end{tabular}
	\vspace{1em} 
	\begin{tabular}{c}
	\includegraphics[width=0.21\textwidth]{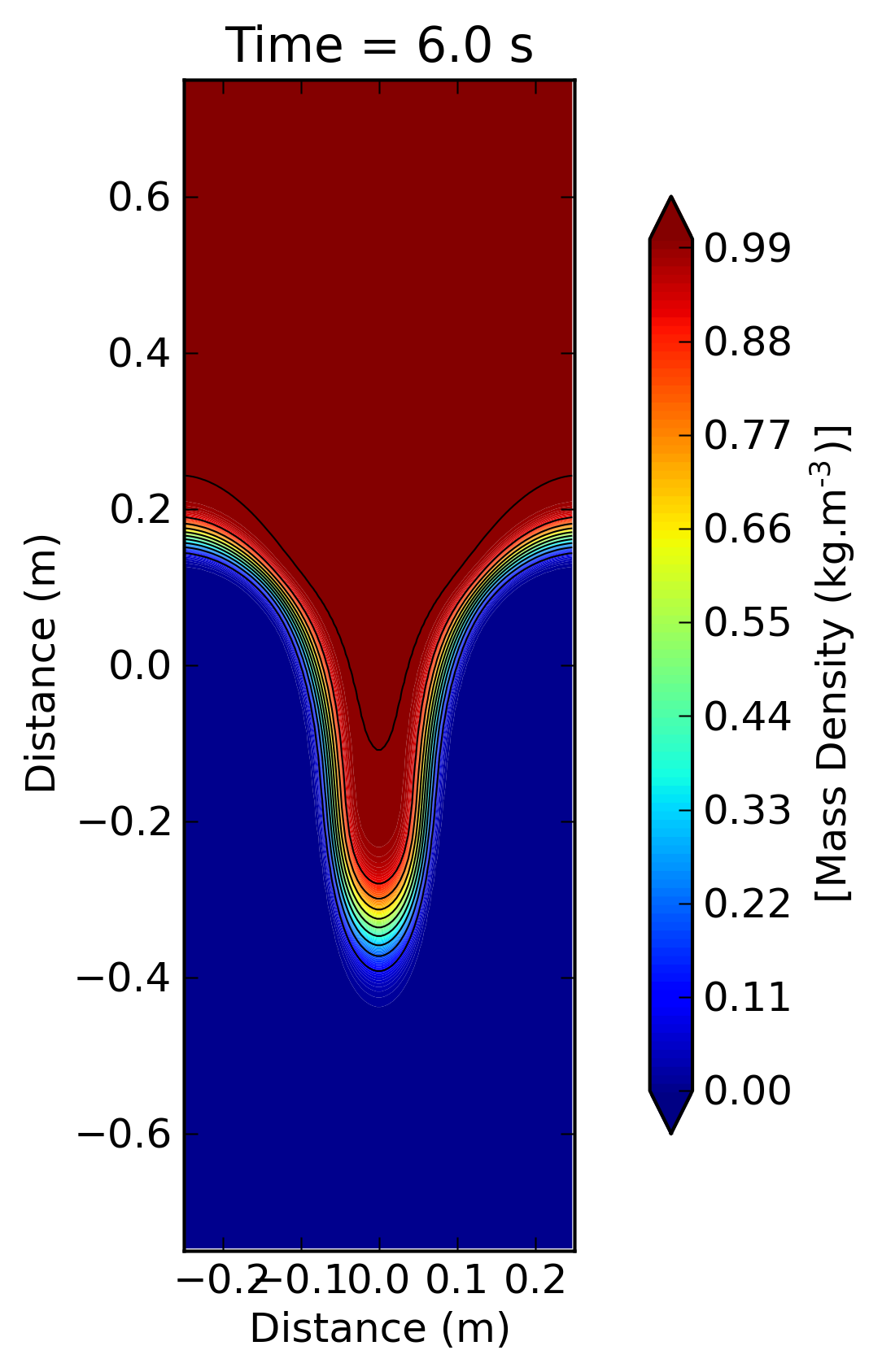}\\[\abovecaptionskip]
	\small HLLC ordre 1 \\
	et une CFL de 0,5
	\end{tabular}
\caption{Phase non linéaire de notre instabilité de Rayleigh-Taylor avec différents solveurs pour un maillage de $100\times 300$ ($A_t=1$). Temps de calcul: environ $6 \sim 7$ h sur un n\oe{}ud (28 c\oe{}urs) du supercalculateur \textit{Cobalt}.}
\label{RTInonlin}
\end{figure}

Lorsque $A=1$, cela équivaut à une interface entre un fluide et le vide, ainsi nous avons modifié les densités des deux fluides tel que $\rho_1 =0{,}001$ et $\rho_2 = 1$. Les instabilités de Kelvin-Helmoltz deviennent alors inexistantes, comme on peut le voir sur les figures \ref{RTInonlin}. 
 
Il existe des modèles analytiques pour déterminer la vitesse de croissance des bulles et des jets \cite{mikaelian1998analytic}. Ces modèles donnent pour les bulles:

\begin{equation}
    v_B(t) = \sqrt{\frac{\lambda g}{6 \pi}}
\end{equation}

soit: 

\begin{equation}
    h_B(t) = \sqrt{\frac{\lambda g}{6 \pi}}t
\end{equation}

et pour les jets: 
\begin{equation}
    v_J(t) = gt
\end{equation}

soit: 
\begin{equation}
    h_J(t) = \frac{1}{2}gt^2
\end{equation}
\medskip

\begin{figure}[!h]
	\centering
	\begin{tabular}{c}
	\includegraphics[width=0.4\textwidth]{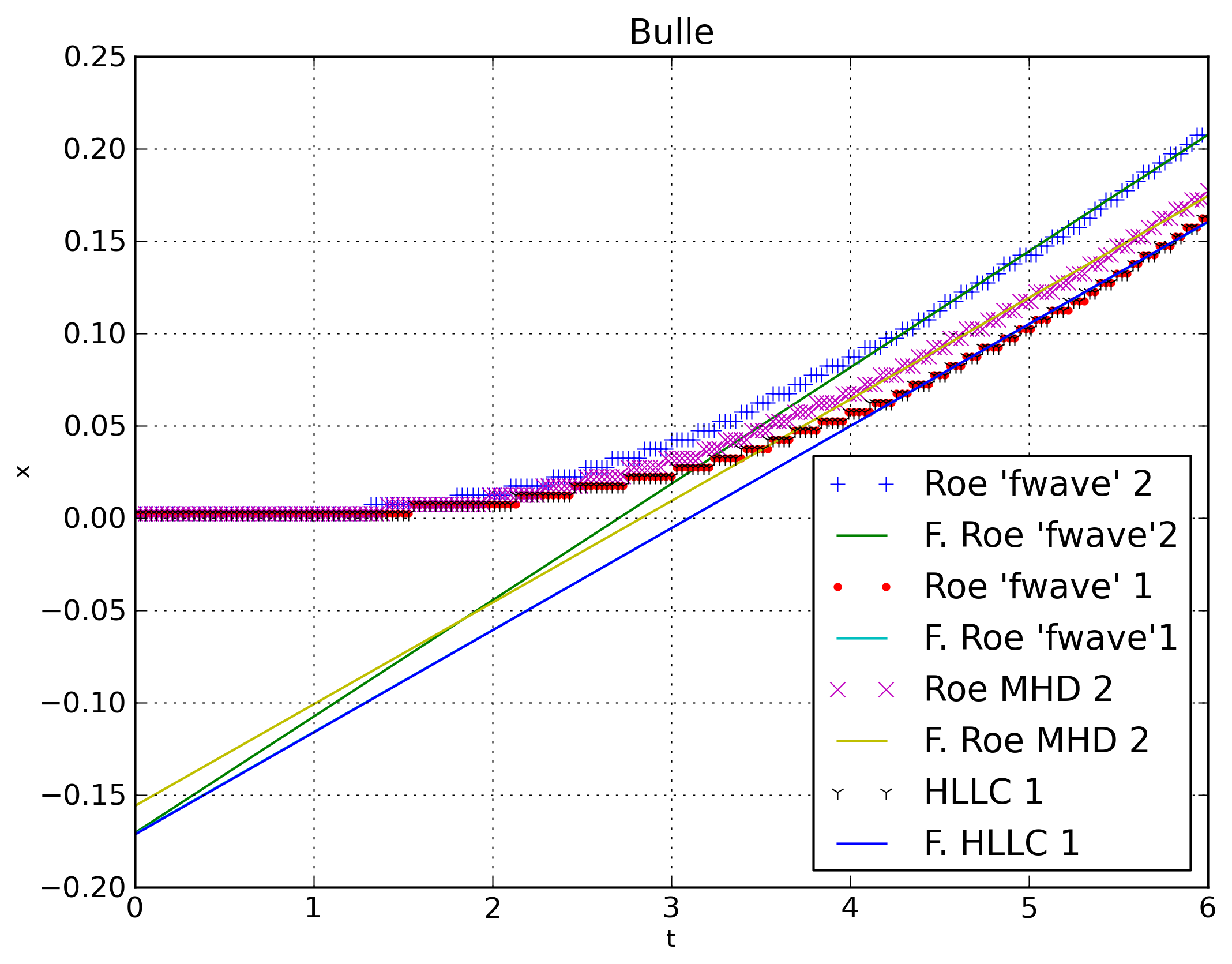}\\[\abovecaptionskip]
	\small Bulle
	\end{tabular}
	\vspace{1em} 
	\begin{tabular}{c}
	\includegraphics[width=0.4\textwidth]{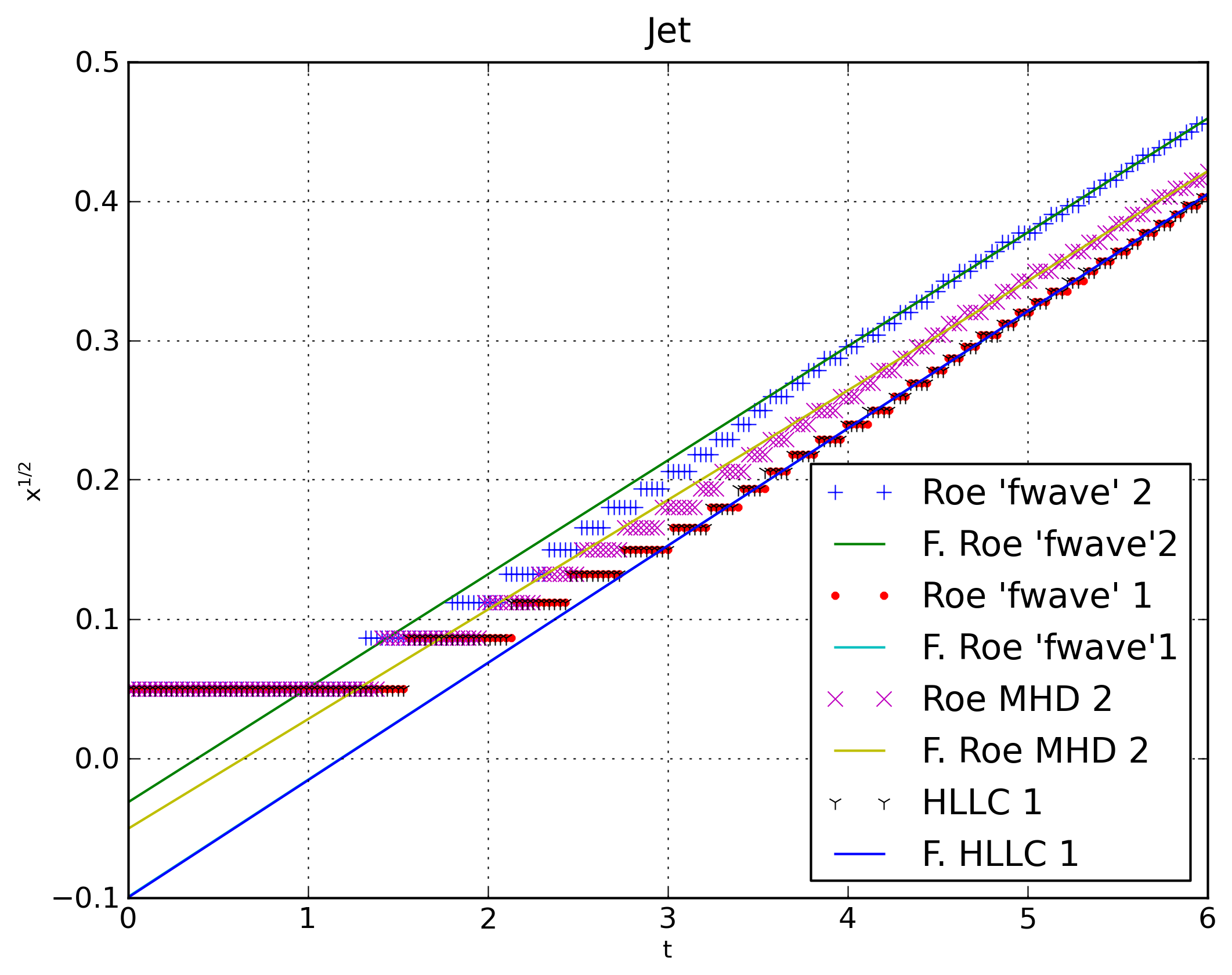}\\[\abovecaptionskip]
		\small Jet
	\end{tabular}
\caption{Évolution de la hauteur de la bulle et de la longueur du jet avec différents solveurs: Roe \og f-waves\fg\, à l'ordre 2, Roe \og f-waves\fg \, à l'ordre 1, Roe à l'ordre 2 et HLLC à l'ordre 1  }
\label{RTInonlinfit}
\end{figure}

Pour comparer de façon quantitative, nous avons effectué une régression linéaire sur la hauteur de nos bulles et la racine de la longueur de nos jets, comme on peut le voir sur la figure \ref{RTInonlinfit}. Nous avons mis les valeurs des pentes de nos régressions linéaires dans deux tableaux \ref{tabbulle} pour les bulles et \ref{tabjet}  pour les jets. Une première remarque est que les solveurs d'ordre 1 comme HLLC et Roe \og f-wave\fg donnent des résultats proches.

L'évolution de la bulle semble suivre le modèle théorique comme on peut le voir dans le tableau  \ref{tabbulle}.
Pour les trois solveurs, Roe ordre 2, Roe \og f-waves \fg\, ordre 1 et HLLC, on obtient une différence de l'ordre de 7\% avec la valeur théorique. 
En revanche, le solveur \og f-wave \fg obtient une valeur supérieure d'environ 14\%. Cette différence peut s'expliquer par la forme en \og ogive \fg\,  de la bulle comparée à une forme en \og tête d'un champignon \fg\, pour les autres solveurs (voir figure \ref{RTInonlin}).
Une autre possibilité est que nous n'avons pas encore atteint totalement le régime non-linéaire puisque la hauteur de la bulle est inférieure à la moitié de la longueur d'onde de la perturbation.

\begin{table}[!h]
    \centering
 \begin{tabular}{|l|C{2.5cm}|C{2.5cm}|C{2.5cm}|C{2.5cm}|}
     \hline
      & Roe \og f-waves \fg\, à l'ordre 2 & Roe \og f-waves \fg\, à l'ordre 1 & Roe à l'ordre 2 & HLLC à l'ordre 1  \\
     \hline
     Théorique ($  \sqrt{\frac{\lambda g}{6 \pi}}\approx 0{,}0515) $& 0,0617 & 0,055& 0,055& 0,055 \\ 
     Erreur (en \%) & 14,5 & 7,2 & 6,8 & 7,38 \\
     \hline
\end{tabular}
    \caption{Tableau des différentes valeurs de notre régression linéaire pour les bulles avec $A_t = 1 $}
    \label{tabbulle}
\end{table}

Pour l'évolution du jet, on remarque que la valeur obtenue avec Roe \og f waves \fg\, à l'ordre 2 est très proche de la valeur théorique avec seulement 7\% d'erreur, alors que Roe d'ordre 2 est à 35\% et les solveurs d'ordre 1 sont aux environs de 50\%.
La figure \ref{RTInonlin}, montre l'effet de la diffusion numérique sur l'interface entre les fluides, notamment pour les solveurs d'ordre 1.
L'interface est plus nette dans la zone de la bulle que dans la zone jet. La forme du jet étant plus fine, la diffusion numérique devient bidimensionnelle, longitudinale et transverse.
\begin{table}[!h]
    \centering
 \begin{tabular}{|l|C{2.5cm}|C{2.5cm}|C{2.5cm}|C{2.5cm}|}
     \hline
      & Roe \og f-waves \fg\, à l'ordre 2 & Roe \og f-waves\fg\, à l'ordre 1 & Roe à l'ordre 2 & HLLC à l'ordre 1  \\
     \hline
     Théorique ($  \sqrt{\frac{ g}{2}}\approx 0{,}223) $& 0,215 & 0,15& 0,188 &0,16  \\
     $\frac{1}{2}$ & 0,46 & 0,23 & 0,33 & 0,24  \\
     Erreur (en \%) & 7,3 & 52 & 35 & 51 \\ 
      \hline
\end{tabular}
    \caption{Tableau des différentes valeurs de notre régression linéaire pour les jets avec $A_t = 1 $}
    \label{tabjet}
\end{table}

\subsubsection{Nombre d'Atwood différent de 1}

Nous avons deux modèles pour décrire l'évolution de la bulle et du jet dans la phase non-linéaire. 
\subparagraph{Modèle 1 \cite{goncharov2002analytical}:} 

Pour les bulles, on a: 

\begin{equation}
    v_B(t) = \sqrt{\frac{2A_t}{1+A_t} \times \frac{\lambda g}{6 \pi}}
\end{equation}

et pas de valeur pour les jets. 

\subparagraph{Modèle 2 \cite{SergeBouquet}:} 

pour les bulles, on a: 

\begin{equation}
    v_B(t) = \sqrt{\frac{8A_t}{3+5A_t} \times \frac{\lambda g}{6 \pi}}
\end{equation}

et pour les jets: 

\begin{equation}
    h_J(t) = \frac{1}{2}gt^2
\end{equation}

Nous avons utilisé la même méthode que précédemment pour un nombre d'Atwood différent de 1 (Rappel $A_t=\frac{\rho_2-\rho_1}{\rho_2+\rho_1}$). Les résultats utilisés sont ceux du solveur Roe \og f-waves \fg\, présentés sur la figure \ref{RTIfin}. 

Pour les bulles, les valeurs de croissance sont proches de celles théoriques avec un accord un peu plus satisfaisant (10\%) avec le modèle 2 \cite{SergeBouquet}.
Pour le jet, nous obtenons une évolution par simulation numérique plus lente d'un facteur 10 par rapport au modèle 2.
Cette différence est sûrement due à l'instabilité de Kelvin-Helmoltz dont l'effet n'est pas pris en compte dans les modèles asymptotiques. En effet, cette instabilité n'est plus négligeable avec un nombre d'Atwood différent de 1.
L'évolution du jet semble plus impactée par l'instabilité secondaire que l'évolution de la bulle. 

\begin{table}[!h]
    \centering
 \begin{tabular}{|l|C{2.5cm}|C{2.5cm}|}
     \hline
      Erreur & Jet & Bulle  \\
     \hline
     Modèle 1 (en \%)& none & 17 \\ 
     Modèle 2 (en \%)& 95& 10 \\
     \hline
     Valeur mesuré & 0,049 & 0,042\\
     \hline
\end{tabular}
    \caption{Tableau des différentes valeurs de notre régression linéaire pour les bulles et les jets avec $At\neq 1 $}
    \label{tabatwood}
\end{table}

\section{Cas Magnétohydrodynamique}

\subsection{Résultats analytiques}

Le taux d'accroissement théorique de la partie linéaire suit la formule \cite{chandrasekhar1961hydrodynamic, jun1995numerical}: 

\begin{equation}
    \gamma^2 = gk \frac{(\rho_2-\rho_1)}{(\rho_2  +\rho_1)} - \frac{2 (\mathbf{B}\cdot \mathbf{k})^2}{(\rho_2 +\rho_1)}
\end{equation}

où $\mathbf{k}$ est le vecteur d'onde de l'instabilité (avec $k=\|\mathbf{k}\|$). Avec un champ magnétique $\mathbf{B}$ nul ou perpendiculaire à l'interface, nous retrouvons le taux d'accroissement purement hydrodynamique. Le champ magnétique réduit significativement le taux de croissance pour les instabilités de faible longueur d'onde, on peut ainsi définir une longueur d'onde de coupure $\lambda_c$ qui désigne la longueur d'onde minimale à partir de laquelle une instabilité  peut croître (toute instabilité plus petite sera amortie par le champ magnétique).

\begin{equation}
    \lambda_c = \frac{4\pi (\mathbf{B}\cdot \mathbf{n})^2}{g(\rho_2-\rho_1)}
\end{equation}

où $\mathbf{n}$ est le vecteur unitaire dirigé dans la même direction que le vecteur d'onde $\mathbf{k}$. On peut facilement montrer que la longueur d'onde ayant le taux de croissance maximal est $\lambda_{max } = 2 \lambda_c$. On peut également définir une valeur de coupure $B_c$ pour un champ magnétique parallèle à l'interface de telle sorte qu'aucune instabilité ne puisse croître dans un domaine de largeur $L_x$:

\begin{equation}
    B_c = \left[ \frac{gL_x(\rho_2-\rho_1)}{4\pi}\right]^{\frac{1}{2}} 
\end{equation}

Dans le cas d'un champ magnétique perpendiculaire à l'interface, le taux de croissance n'est pas modifié, en revanche le champ magnétique joue toujours sur les instabilités secondaires de Kelvin-Helmoltz. Nous n'avons pas de modèle asymptotique de bulles ou de jets pour la phase non-linéaire dans le cas MHD \cite{jun1995numerical,levy2012etude}. 

\subsection{Simulation}

\paragraph{Champ magnétique de coupure}

\begin{figure}[!h]
    \centering
    \includegraphics[scale=0.5]{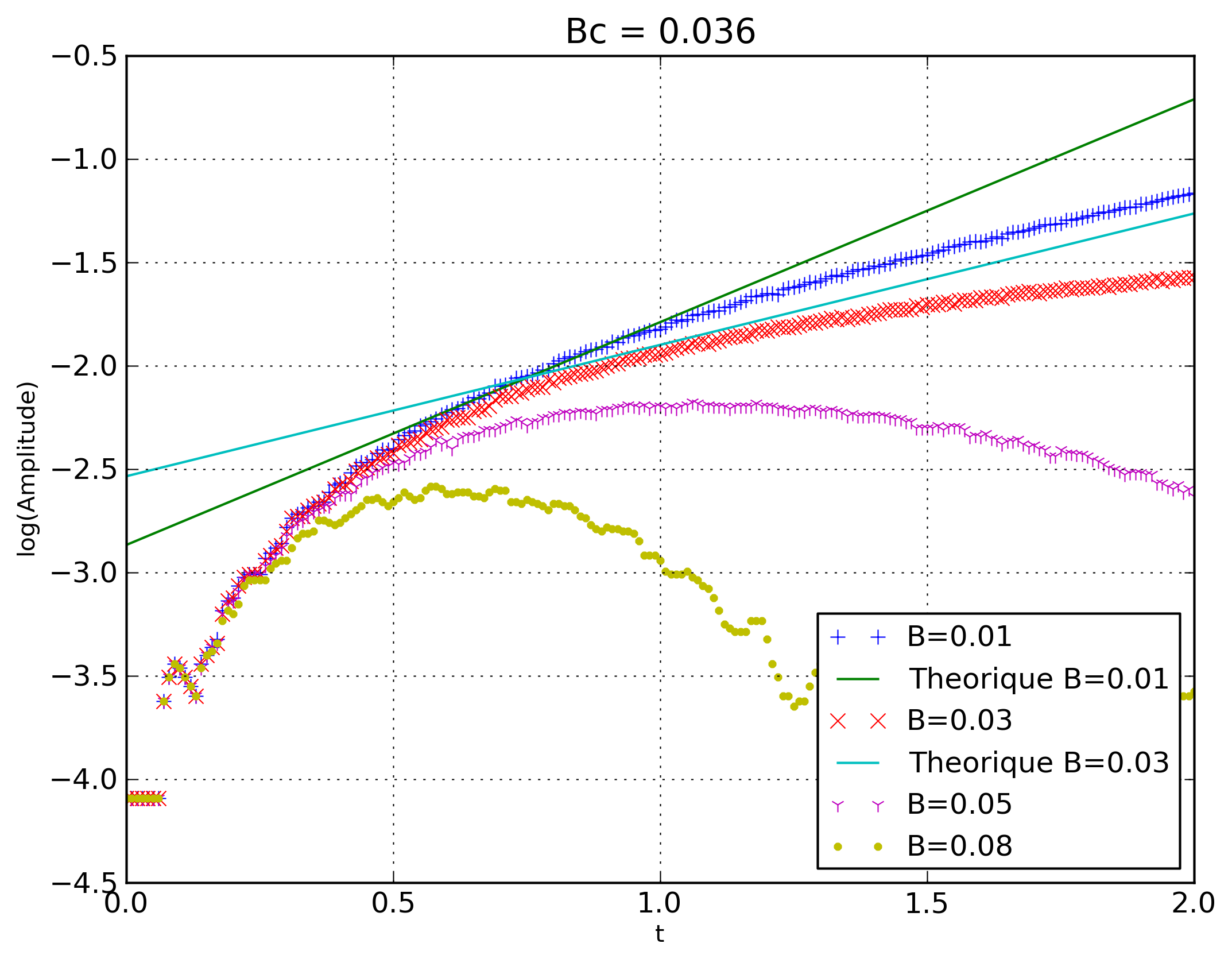}
    \caption{Amplitude de la perturbation ($\lambda=L_x/3$) en fonction du temps avec différentes valeur de $B$ pour le solveur de Roe MHD à l'ordre 2 et une CFL de 0,2.}
    \label{Bvarie}
\end{figure}

Pour étudier le champ magnétique de coupure, nous avons comparé différents taux de croissance linéaires  dus à différentes valeurs du champ magnétique parallèle à l'interface. Nous avons initialisé notre perturbation en vitesse avec une longueur d'onde $\lambda = L_x/3 \approx 0{,}167$, ainsi $B_c \approx 0{,}036$ avec $\rho_1 = 1$ et $\rho_2 = 2$. On peut voir sur la figure  \ref{Bvarie} que nous  avons bien le résultat attendu, pour un champ magnétique en dessous de la valeur de coupure, tel que $B=0{,}03$, l'instabilité croit, alors que pour un champ plus grand, comme par exemple $B=0{,}08$, la perturbation est amortie. 

\paragraph{Longueur d'onde de coupure}

Pour observer la longueur d'onde de coupure et la longueur d'onde présentant le taux d'accroissement maximal, nous avons fait des simulations avec les dix premiers modes propres, notés $m$, de notre boite, tels que $\lambda_m = L_x/m$ (avec toujours $L_x =0{,}5$). Nous avons fixé notre champ magnétique parallèle à l'interface, tel que $B=0{,}02$ et ainsi $\lambda_c \approx 0{,}05 $, ce qui correspond à $m=10$, et $\lambda_{max} \approx 0{,}10$, ce qui correspond à $m=5$.

\begin{figure}[!h]
	\centering
	\begin{tabular}{c}
	\includegraphics[width=0.45\textwidth]{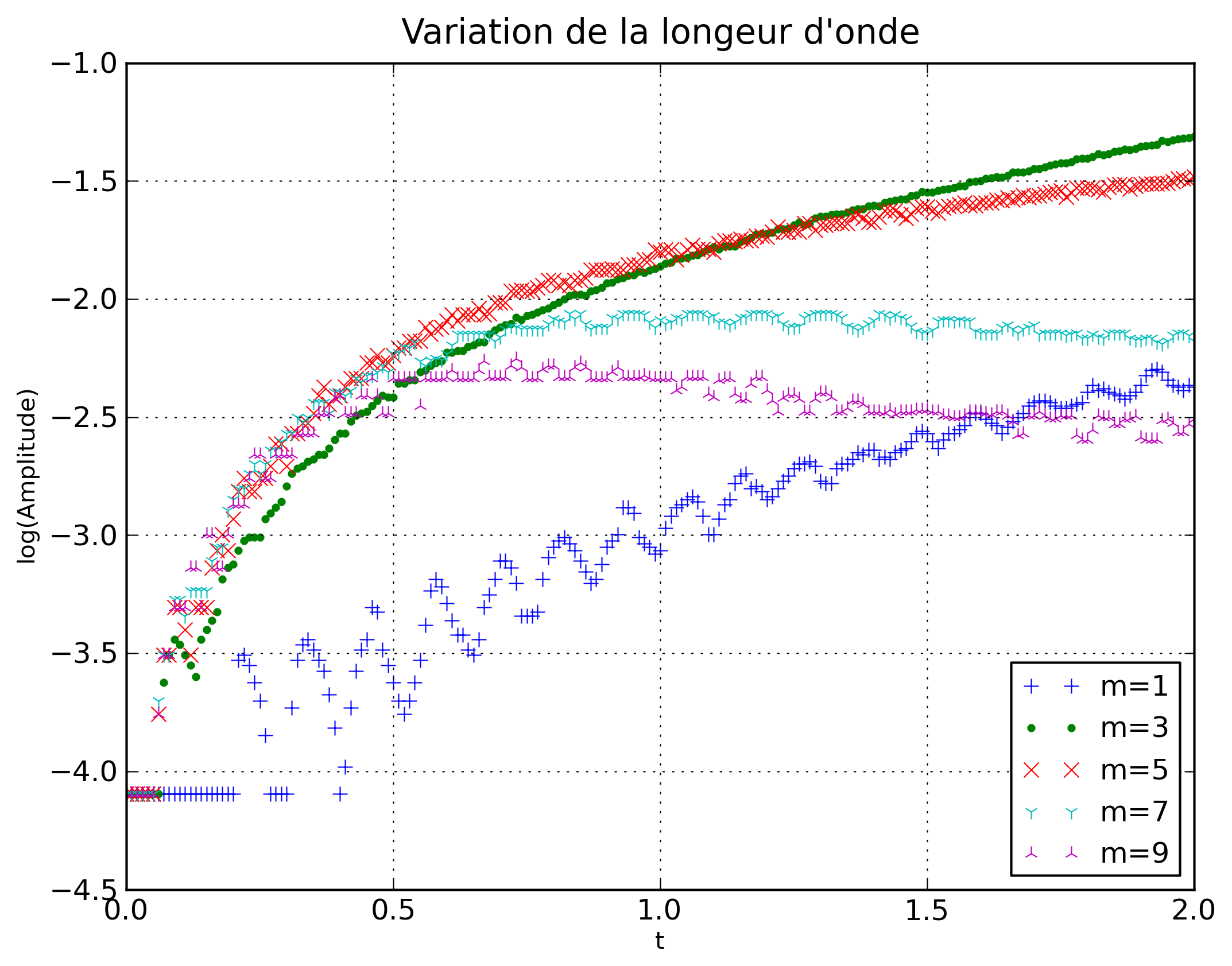}\\[\abovecaptionskip]
	\small Variation de la norme L1 pour différents modes 
	\end{tabular}
	\begin{tabular}{c}
	\includegraphics[width=0.45\textwidth]{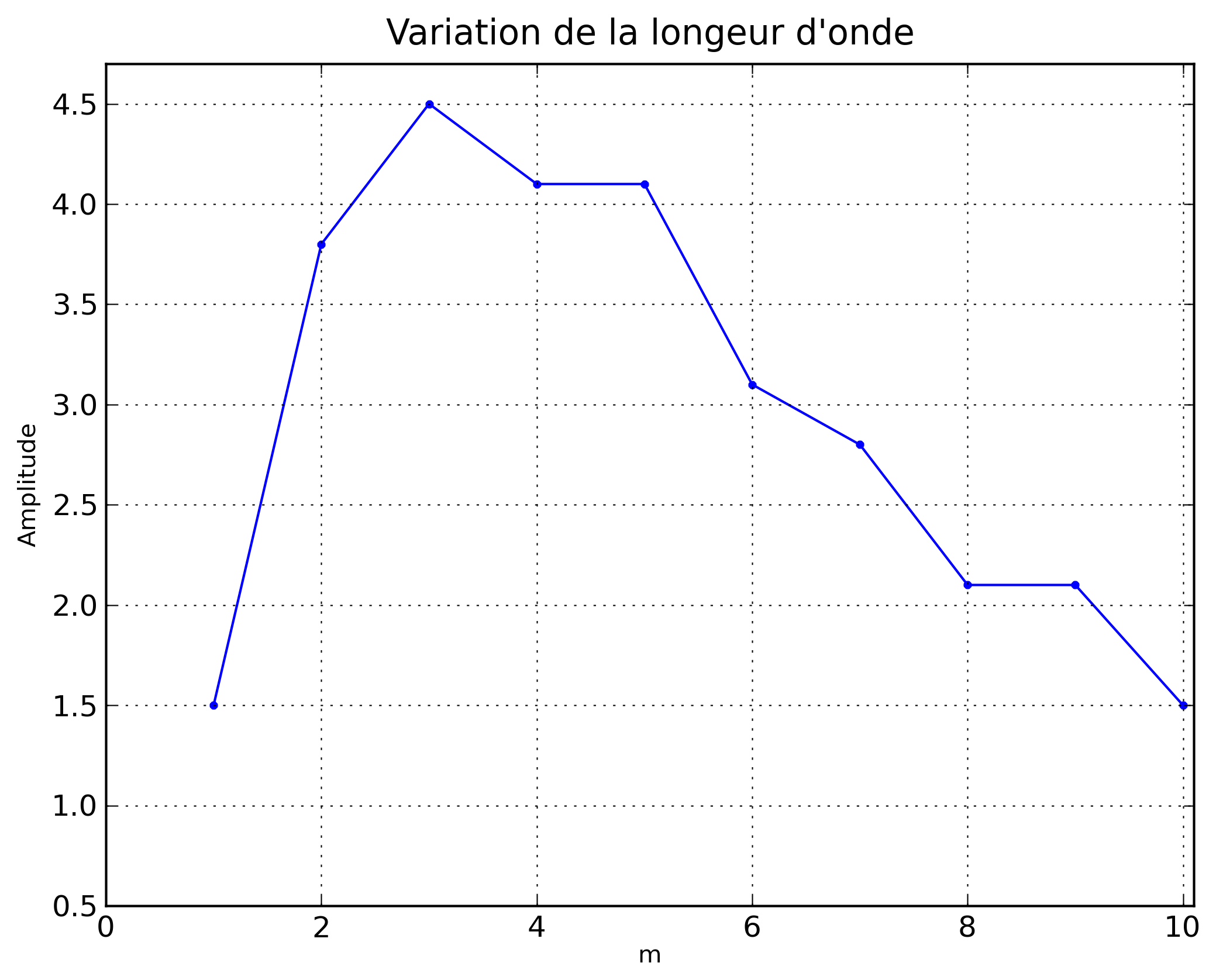}\\[\abovecaptionskip]
	\small Amplitude finale pour les différents modes
	\end{tabular}
	\caption{Amplitude de la perturbation avec différentes longueurs d'onde  pour la phase linéaire avec un maillage de $100\times 300$ avec le solveur de  Roe MHD.  }
	\label{RTIM_mvarie}
\end{figure}

Sur la figure de droite \ref{RTIM_mvarie}, on remarque que le maximum n'est pas atteint pour le mode prévu, mais pour $m=3$. Cela peut s'expliquer par le fait que pour un mode plus petit, la phase non-linéaire est atteinte plus rapidement, ce qui ralentit la croissance absolue de notre instabilité. En effet, sur la figure de gauche \ref{RTIM_mvarie}, on peut remarquer que le mode $m=5$ a une amplitude (pour la norme L1) supérieure à celle du mode $m=3$ jusqu'à $t=1{,}2$ s environ. 
De même, la longueur d'onde de coupure, prévue pour  $m=10$, est déjà un peu visible sur les modes  $m=7$ et  $m=9$, dont l'amplitude décroit après la première phase transitoire.

La diffusion numérique peut également jouer un rôle non négligeable quand nous diminuons notre longueur d'onde, puisque l'instabilité est moins bien résolue (pour $m=1$ la longueur d'onde est résolue sur 100 points, alors que pour $m=10$ la longueur d'onde est résolue sur 10 points). Cela pourrait en partie expliquer pourquoi, on peut déjà voir un amortissement pour le mode $m=7$ sur la figure de gauche \ref{RTIM_mvarie}. 

\paragraph{Phase non-linéaire}

\begin{figure}[!h]
    \centering
    \includegraphics[scale=0.5]{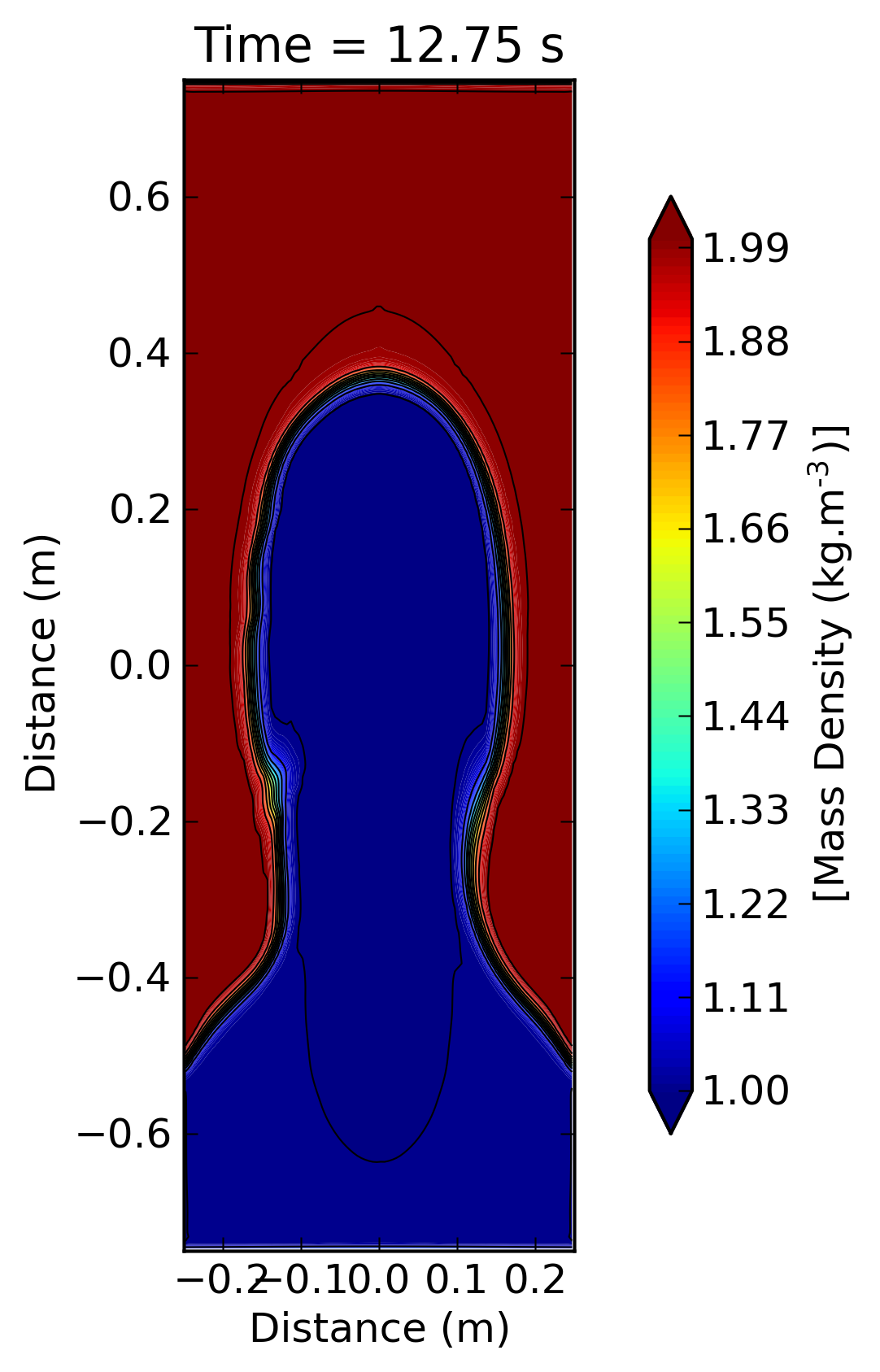}
    \caption{ Phase non-linéaire de l'IRT avec un champ magnétique perpendiculaire à l'interface $B_y=0{,}04$ pour le solveur de Roe MHD à l'ordre 2 et une CFL de 0,2. Temps de calcul: environ $6 \sim 7$ h sur un n\oe{}ud (28 c\oe{}urs) du supercalculateur \textit{Cobalt}.}
    \label{Bvperp}
\end{figure}

On peut voir sur la figure \ref{Bvperp} qu'aucune instabilité de Kelvin-Helmoltz n'a pu se développer. 
En effet, on rappelle que le taux d'accroissement de Kelvin-Helmoltz (non présenté ici) dépend aussi du champ magnétique. 
Ici, l'instabilité de Kelvin-Helmoltz est parallèle aux lignes de champ magnétique qui empêche sa croissance.

\newpage

\section{Synthèse}


L'étude de l'instabilité de Rayleigh Taylor a permis de mettre en évidence les grandeurs théoriques pertinentes à utiliser pour comparer aux simulations numériques à savoir les vitesses HD de bulle et de jet dans la phase non-linéaire et le taux d'accroissement de la phase linéaire HD et MHD.


Après avoir redémontré le taux d'accroissement théorique de la phase linéaire hydrodynamique, nous avons confronté l'amplitude théorique à l'amplitude numérique et ainsi nous avons déterminé une plage de validité, entre une phase transitoire et le début de la phase non-linéaire.
L'étude de la convergence au maillage a permis de montrer que l'évolution de l'amplitude de la perturbation coïncidait au régime linéaire sur une durée plus longue avec le raffinement du maillage.

En MHD, nous avons de plus mis en évidence d'autres propriétés, comme l'existence d'une longueur d'onde de coupure et d'un champ magnétique de coupure au-delà desquels le taux d'accroissement de l'instabilité s'annule. Les simulations ont réussi à reproduire ces phénomènes.

La comparaison entre la croissance asymptotique des structures en phase non-linéaire et les résultats numériques est satisfaisante, notamment pour les bulles.
La comparaison avec les modèles de jet est intéressante car les écarts sont liés la précision des solveurs et des méthodes utilisées.

Ce cas-test se démarque des autres présentés dans le chapitre précédent par la présence de la gravité. Il permet ainsi de vérifier la qualité d'implémentation de celle-ci, puisque l'on a pu observer que le solveur de Roe \og f-waves\fg\, avec la gravité incluse dans le terme de flux donnent des résultats plus précis que le solveur de Roe avec la gravité implémentée en terme source.


\clearpage
\newpage

\chapter{Conclusion}

Les solveurs de Riemann sont des outils puissants et robustes  adaptés aussi bien à une description hydrodynamique (HD)
pour le transport d'un fluide neutre, qu'à une description magnétohydrodynamique (MHD) qui concerne le transport des fluides chargés.
Deux types de solveurs de Riemann ont été présentés: le solveur de Roe et les solveurs de type HLL avec HLLC pour l'HD et HLLD pour la MHD.

Une liste de cas-tests, parmi ceux utilisés dans la littérature scientifique, a été constituée pour vérifier la qualité des solveurs.
Les simulations et les comparaisons réalisées entre les solveurs disponibles dans le code ont permis de justifier de la pertinence de chaque cas-test.

Le cas-test de Brio and Wu permet de comparer la précision des solveurs pour traiter la propagation d'onde en une dimension.
Il est alors aisé, d'observer la diffusion numérique des différents schémas numériques comme par exemple entre les schémas d'ordre 1 et les schémas d'ordre 2.

Les cas-tests 2D (Orszag-Tang, le Rotor magnétique)  et 3D (\og Blast wave\fg ) permettent de monter la robustesse des solveurs.
En effet, les solveurs d'ordre 2 ont dû faire l'objet d'attentions plus particulières comme la réduction de la contrainte CFL, ou le choix d'un limiteur adapté pour simuler ces conditions contraignantes.
Par exemple le cas-test d'Orszag-Tang décrit, à partir de conditions initiales continues, l'apparition d'ondes de choc, la propagation, et ensuite leurs interactions.
Dans le cas-test Blast, on simule l'explosion d'une sphère chaude, c'est-à-dire de pression élevée, dans un milieu froid, c'est-à-dire de pression faible, où un champ magnétique uniforme s'applique.
Le cas-test du rotor, similaire au phénomène de formation stellaire, simule la mise en mouvement circulaire d'un plasma qui comprime la matière et génère une onde d'Alfvén torsionnelle.

Même si la majorité des cas-tests utilisés en MHD sont qualitatifs, ils permettent néanmoins de montrer que les solveurs sont capables d'obtenir une évolution spatiale et temporelle cohérente avec les solveurs d'autres codes.
En revanche, l'étude de l'instabilité de Rayleigh-Taylor (IRT) a permis d'obtenir des comparaisons plus quantitatives. 
Le taux d'accroissement théorique hydrodynamique a été comparé graphiquement au taux mesuré dans la simulation lors de deux études, une sur la variation de la longueur d'onde de la perturbation et une sur la convergence en maillage. 
Nous retrouvons la phase transitoire, la phase linéaire et la phase non-linéaire. La plage temporelle d'existence de la phase linéaire varie suivant les conditions: généralement plus le taux est élevé, plus le régime non-linéaire apparait vite. 
De plus, pour les simulations MHD, nous nous sommes assurés que les solveurs restituaient bien la dépendance avec champ magnétique et avec la longueur d'onde. Nous avons ainsi observé l'amortissement de l'instabilité une fois le champ magnétique de coupure ou  la longueur d'onde de coupure dépassée. 

Lors de la phase non-linéaire hydrodynamique, l'évolution des bulles et des jets a pu être comparée aux modèles asymptotiques. 
Pour l'évolution des bulles, on obtient avec les différents solveurs des résultats proches avec  environ 10\% d'erreur. 
Pour l'évolution du jet, seul le solveur de Roe \og f-waves\fg \, obtient des résultats dont l'erreur est inférieur à 10\%. En effet, nous avons remarqué que la diffusion numérique de l'interface était plus marquée sur le jet que sur la bulle. La sensibilité de ce test le rend pertinent, car il discrimine nettement les solveurs entre eux. Des modèles asymptotiques de bulles ou de jets, dans des conditions MHD, permettraient d'améliorer ce cas-test.


Finalement, ces cas-tests constituent une base diversifiée qui permet de vérifier aussi bien la précision que la robustesse des solveurs utilisés pour résoudre les équations de la MHD idéale.
\clearpage
\newpage

\bibliography{biblio}

\newpage

\appendix

\chapter{Loi d'Ohm tri-fluides}
\label{OHm3}
    Les équations de conservation de la quantité de mouvement pour les ions et les électrons sont: 

\begin{equation}
\begin{split}
\partial_t \mathbf{V}_{e,i} + \nabla . \mathbf{V}_{e,i} \otimes \mathbf{V}_{e,i} = - \frac{1}{n_{e,i}m_{e,i}}\nabla p_{e,i} + \frac{q_{e,i}}{m_{e,i}} (\mathbf{E} + \mathbf{V}_{e,i} \times \mathbf{B})\\
- \nu_{in,en} (\mathbf{V}_{e,i} - \mathbf{V}_n)\nu_{ei,ie} (\mathbf{V}_{e,i} -\mathbf{V}_{i,e} )  
\end{split}
\end{equation}

Avec $\mathbf{V}_e$ ($\mathbf{V}_i$) la vitesse des électrons (respectivement des ions), $\mathbf{V}_n$ la vitesse fluide des  neutres et $\nu_{\alpha \beta}$ la fréquence de collision  des particules $\alpha$ sur les particules $\beta$. 

Nous travaillons dans l'hypothèse de quasi-neutralité $n=n_e=n_i$ et $ m_i \| \mathbf{V}_i\|  \gg m_e \| \mathbf{V}_e \|$. Nous introduisons également le courant $\mathbf{J}  = en(\mathbf{V}_i - \mathbf{V}_e)$ ce qui donne donc $\mathbf{V}_i = \mathbf{V}$ et $\mathbf{V}_e  = \mathbf{V}-\frac{1}{ne}\mathbf{J}$. Les équations de conservation de la quantité de mouvement deviennent: 

\begin{equation}
\begin{split}
\partial_t \mathbf{V} -\frac{1}{ne} \partial_t \mathbf{J} + \nabla \cdot [ (\mathbf{V} - \frac{1}{ne} \mathbf{J} ) \otimes (\mathbf{V} - \frac{1}{ne} \mathbf{J} )] = 
- \frac{1}{nm_e} \nabla p_e -\frac{e}{m_e} \mathbf{E}\\ - \frac{e}{m_e} (\mathbf{V}-\frac{1}{ne}\mathbf{J}) \times \mathbf{B} - \nu_{en} ( \mathbf{V} - \frac{1}{ne} \mathbf{J} - \mathbf{V}_n) + \frac{\nu_{ei}}{ne} \mathbf{J} 
\label{eneutre}
\end{split}
\end{equation}

\begin{equation}
\partial_t \mathbf{V} + \nabla \cdot  (\mathbf{V}  \otimes \mathbf{V}  ) = - \frac{1}{nm_i} \nabla p_i +\frac{e}{m_i} \mathbf{E} + \frac{e}{m_i} \mathbf{V} \times \mathbf{B} - \nu_{in} ( \mathbf{V} - \mathbf{V}_n) - \frac{\nu_{ie}}{ne} \mathbf{J} 
\label{ineutre}
\end{equation}

En négligeant le terme de dérivée temporelle (terme de gauche) et le gradient de pression, puis, en faisant disparaître  
la vitesse des neutres $\mathbf{V}_n$, à l'aide de (\ref{eneutre}) et (\ref{ineutre}), la loi d'Ohm tri-fluides s'écrit \cite{song2001three}: 

\begin{equation}
en(n_e \nu_{en} + n_i \nu_{in})(\mathbf{E} + \mathbf{V} \times \mathbf{B}) = m_i \nu_{in} (\mathbf{J} \times \mathbf{B}) + \frac{m_i m_e}{e}  ( \nu_{in} \nu_{en} + \nu_{ei} \nu_{in} + \nu_{ie}\nu_{en}) \mathbf{J} 
\end{equation} 

Soit en se souvenant que $ m_e \nu_{ei} = m_i  \nu_{ie}$: 

\begin{equation}
\mathbf{J} + \sigma_H \mathbf{J} \times \mathbf{B} = \sigma_P (\mathbf{E} + \mathbf{V} \times \mathbf{B}) 
\end{equation}

avec: 

\begin{equation}
\sigma_P = \frac{e^2n(m_e \nu_{en}+m_i \nu_{in})}{m_e [m_e\nu_{en}\nu_{ei} + m_i\nu_{in}(\nu_{en}+\nu_{ei})]} 
\end{equation}
et 
\begin{equation}
\sigma_H = \frac{e m_i \nu_{in}}{ m_e [m_e\nu_{en}\nu_{ei} + m_i\nu_{in}(\nu_{en}+\nu_{ei})]}
\end{equation}

\newpage

\chapter{Précision sur les notations de la loi d'Ohm}
\label{notationprésicion}

Voici une petite précision sur la forme de la loi d'Ohm habituellement utilisée pour l'étude des phénomènes physiques de l'ionosphère. 
En effet, la communauté préfère utiliser les conductivités, $\sigma_0$, $\sigma_P$ de Pedersen et $\sigma_H$ de Hall contrairement à l'étude MHD où il est plus facile d'utiliser les résistivités \cite{song2001three}. 
Voici les équations permettant de passer de notre loi d'Ohm généralisée (\ref{OhmHall}) à (\ref{OHMcond}),

\begin{equation}
    \mathbf{J} = \sigma_0 \mathbf{E}_\parallel + \sigma_p (\mathbf{E}_\perp +\mathbf{V}\times \mathbf{B} ) + \sigma_H \mathbf{b} \times ( \mathbf{E}_\perp + \mathbf{V} \times \mathbf{B}) 
    \label{OHMcond}
\end{equation}

où

\begin{equation}
    \sigma_0 = \frac{1}{\eta_0}
\end{equation}

\begin{equation}
    \sigma_p =  \frac{\sigma_0 }{1 + \frac{\alpha_H^2 \Omega_{ce}^2}{\nu_{ei}^2}}
\end{equation}

\begin{equation}
    \sigma_H =  \frac{\sigma_0 }{\frac{\nu_{ei}} {\alpha_H \Omega_{ce}}+ \frac{\alpha_H \Omega_{ce}}{\nu_{ei}}}
\end{equation}

avec l'approximation: 
\begin{equation}
    \frac{\eta_H B}{\eta_0} \approx \frac{\alpha_H \Omega_{ce}}{\nu_{ei}}
\end{equation}

On peut remarquer que $\sigma_H \neq \frac{1}{\eta_H}$ contrairement à ce que l'on pourrait croire. Nous avons également la forme matricielle de la loi d'Ohm: 

\begin{equation}
\mathbf{J} = \mathbf{\sigma}(\mathbf{E} + \mathbf{V} \times \mathbf{B}   )
\end{equation}

\begin{equation}
    \mathbf{\sigma} = 
    \begin{pmatrix}
    \sigma_p & -\sigma_H & 0 \\
    \sigma_H & \sigma_p & 0 \\ 
    0 & 0& \sigma_0 
    \end{pmatrix}
\end{equation}

\newpage

\chapter{Différentes implémentations de la MHD non-idéale}
\label{MHDNONIDEAL}

L'étude de l'ionosphère et plus particulièrement des \og equatorial spread F \fg\, nécessitent de prendre en compte davantage de termes dans la loi d'Ohm que pour le cas de la MHD idéale.
Nous allons donc maintenant décrire différentes manières de résoudre ces nouvelles équations. 

\section{Différentes formes d'un système MHD}

\subsection{MHD idéale}

Nous rappelons ici le cas de la MHD idéale, avec la loi d'Ohm: 

\begin{equation}
    \mathbf{E} + \mathbf{V} \times \mathbf{B} = 0 
\end{equation}

ce qui nous donne: 

\begin{equation}
    \partial_t \mathbf{B} + \nabla \cdot (\mathbf{B} \otimes \mathbf{V} - \mathbf{V} \otimes \mathbf{B}) = 0 
\end{equation}

\subsection{MHD résistive}

La loi d'Ohm en MHD résistive est: 

\begin{equation}
    \mathbf{E} + \mathbf{V} \times \mathbf{B} = \eta \mathbf{J}
\end{equation}

ce qui nous donne: 

\begin{equation}
    \partial_t \mathbf{B}  + \nabla \times (-\mathbf{V} \times \mathbf{B} + \eta \mathbf{J}) = 0
\end{equation}

en utilisant la relation de Maxwell-Ampère (\ref{Maxwelllaw}) pour remplacer $\mathbf{J}$, on obtient: 

\begin{equation}
\partial \mathbf{B} +\nabla \cdot (\mathbf{B} \otimes \mathbf{V} - \mathbf{V} \otimes \mathbf{B}) = \eta_0\Delta \mathbf{B}
\end{equation}

avec $\eta_0$ la résistance du terme diffusif. 

\subsection{MHD résistive avec effet Hall} 

La loi d'Ohm incluant l'effet Hall est: 

\begin{equation}
    \mathbf{E} + \mathbf{V} \times \mathbf{B} = \eta_0 \mathbf{J} +\eta_H \mathbf{J} \times \mathbf{B}
    \label{OhmHall}
\end{equation}

ce qui donne de façon équivalente: 

\begin{equation}
\partial \mathbf{B} +\nabla \cdot (\mathbf{B} \otimes \mathbf{V} - \mathbf{V} \otimes \mathbf{B}-\eta_H \mathbf{B}\otimes \mathbf{B}+ \eta_H B^2\mathbf{I}) = \eta_0 \Delta \mathbf{B}
\end{equation}

avec $\eta_H$ la résistance de Hall.

\section{Les différentes implémentations}

\subsection{Problématique}

Rappelons les équations de la MHD (termes de gauche) auxquelles nous avons ajouté les termes résistifs et de Hall (termes de droite). 

\begin{equation}
 \left\{
    \begin{array}{l}
      \partial_t \rho + \nabla \cdot (\rho \mathbf{V})  =0  \\
        \partial_t (\rho \mathbf{V}) +  \rho \mathbf{V}  \nabla \cdot \mathbf{V} + p_t - \mathbf{J} \times  \mathbf{B} =0 \\
        \partial_t \mathbf{B} + \nabla \times (\mathbf{V} \times \mathbf{B}) = -\nabla \times (\eta_0 \mathbf{J}) - \nabla \times (\eta_H \mathbf{J} \times \mathbf{B}) \\ 
        \partial_t \mathcal{E}_t + \nabla \cdot [\mathbf{V}\cdot (p_t+\mathcal{E}_t) - \mathbf{V} \cdot (\mathbf{B} \otimes \mathbf{B})  ] = - \nabla \cdot (\eta_0 \mathbf{J} \times \mathbf{B}) -\nabla \cdot ( (\eta_H \mathbf{J} \times \mathbf{B})\times \mathbf{B})
    \end{array}
    \right.
\end{equation}

L'implémentation de ces termes additionnels soulève plusieurs difficultés. D'une part, l'apparition d'une dérivée spatiale seconde en $\mathbf{B}$ et d'autre part l'apparition de nouvelles contraintes sur le pas de temps. 

Premièrement, la dérivée seconde de $\mathbf{B}$ est un problème, car les solveurs de Riemann sont conçus pour résoudre des problèmes hyperboliques, ainsi ces nouveaux termes paraboliques doivent être traité spécifiquement. 

Deuxièmement, les nouvelles contraintes en temps peuvent poser problèmes, car celles-ci peuvent grandement allonger le temps de calcul de notre code. Ainsi, dans un cas où $\Delta t_{MHD}$ (pas de temps maximal imposé par la condition CFL pour les termes de la MHD idéale) est largement supérieur à $\Delta t_{par}$ (pas de temps maximal imposé par la condition CFL pour les termes paraboliques: résistivité et Hall), le temps de calcul deviendrait énorme comparé au cas de MHD idéale. Ici nous avons mis les expressions des différents pas de temps imposés par la contrainte CFL respectivement pour la MHD idéale, le terme de Hall et le terme résistif et également leurs ordres de grandeur dans le cas de l'ionosphère (aux alentours de 100 km d'altitude):

\begin{equation}
    \Delta t_{MHD} = \frac{\Delta x}{\lambda_{max}} \sim 1-10 s 
\end{equation}

\begin{equation}
    \Delta t_{Hall} = K\frac{\Delta x^2en\mu_0}{B} \sim 10^{-2} s 
\end{equation}

\begin{equation}
    \Delta t_{R} = K\frac{\Delta x^2\mu_0}{ \eta_0} \sim 10^{3} s 
\end{equation}

avec $ \lambda_{max}$ le maximum des valeurs propres de la Jacobienne $\mathbf{A}$ définie dans le chapitre \ref{Riemansolveur}, $K$ une constante variant autour de 1 selon l'ordre du schéma et le nombre de dimensions spatiales étudié, $\Delta x$ notre pas d'espace, $e$ la charge d'un électron et $n$ la densité du plasma.
 On retrouve bien pour les termes paraboliques $\Delta t_{par} \sim \Delta x^2$, ce qui deviendrait contraignant pour des maillages fins.  

\subsection{Méthode de \og splitting \fg}

Une des implémentations consiste à séparer les termes pour obtenir un système d'équations différentielles hyperboliques et un système d'équations différentielles paraboliques. C'est le \og splitting \fg.
Ainsi, nous retrouvons le système de la MHD idéale que nous pouvons résoudre avec les solveurs de Riemann.
La résolution de l'équation parabolique du champ magnétique utilise souvent la méthode des Différences Finies soit implicite comme FLASH \cite{center2005flash} 
ou explicite comme AMRVAC \cite{porth2014mpi}.
D'autres méthodes, plus sophistiquées, diminuent la contrainte sur le pas en temps comme la méthode \og Super Time Stepping \fg\, ou  \og Runge et Kutta Legendre \fg.
Certaines de ces méthodes permettent d'inclure les termes de Hall comme par exemple \og Super Time Stepping \fg\, que nous détaillons ci-dessous.

\subsubsection{Super Time Stepping (STS)}

Le \og Super Time Stepping \fg (STS) est une méthode numérique qui permet de contourner le problème du pas de temps sans avoir recours à un schéma implicite \cite{alexiades1996super}. 
Cette méthode est très bien adaptée pour résoudre des systèmes paraboliques à moindre coût.

Pour un schéma explicite classique, les équations à résoudre pour chaque pas de temps $\Delta t$ sont de la forme: 

\begin{equation}
\begin{array}{ccc}
    \mathbf{U}^{n+1} = \mathbf{U}^n -\Delta t \mathbf{A} \mathbf{U}^n,  & n \in \mathbb{N}, & \mathbf{U}^0 = \mathbf{U}_0 
\end{array}
\end{equation}

où $\mathbf{A}$ est une matrice symétrique positive de taille $m\times m $ correspondant au schéma explicite, $\mathbf{U}_0$ est un vecteur appartenant à $\mathbb{R}^m$ représentant les conditions initiales. Dans ce cas, la stabilité de ce schéma sera assurée si: 

\begin{equation}
    \Delta t < \Delta t_{CFL} = \frac{2}{\lambda_{max}}
\end{equation}

où $\lambda_{max}$ est la plus grande valeur propre de $\mathbf{A}$, ici pour nos termes paraboliques $\lambda_{max} \propto \Delta x^2$. Ceci est la condition de stabilité de  \og Courant-Friedrichs-Lewy \fg\, (CFL) qui peut être très contraignante pour nos termes paraboliques.
Pour réduire l'impact de cette contrainte sur le temps de calcul, on utilise l'astuce suivante qui est de rechercher la stabilité sur un pas de temps plus grand $\Delta T$ composé lui-même de $N$ sous pas de temps $\tau_j,$ $j \in [1,N]$ (ceci est semblable à la méthode de Runge Kutta). Notre schéma devient alors: 

\begin{equation}
    \begin{array}{cll}
\mathbf{U}^{n+1} = \left[ \prod \limits_{j=1}^N (\mathbf{I}-\tau_j \mathbf{A})\right]  \mathbf{U}^n ,        & n \in \mathbb{N}, & \mathbf{U}^0 = \mathbf{U}_0 
    \end{array}
\end{equation}

On obtient alors la limite suivante: 

\begin{equation}
    \Delta T \underset{\nu \rightarrow 0 }{\longrightarrow} N^2\Delta t_{CFL}
\end{equation}

Notre nouveau schéma est donc  asymptotiquement $N$ fois plus rapide qu'un schéma explicite classique. Néanmoins, il faut être prudent, car bien que le cas $N$ grand et $\nu$ petit accélère grandement nos calculs, le coût à payer est une perte de précision. Il y a donc un compromis à faire entre la précision et le temps de calcul. 

Cette méthode est utilisée dans le code \textit{ATHENA} \cite{bai2012non} pour le terme de résistif et le terme de Hall et également dans le code \textit{PLUTO} \cite{pluto2018user,mignone2007pluto,mignone2011pluto} pour le terme résistif uniquement. Une étude plus complète du \og Super Time Stepping\fg\,  est faite dans l'article de  Alexiades  et al. \cite{alexiades1996super}.  

\subsubsection{Runge-Kutta-Legendre à l'ordre 2 (RKL2)}

Runge-Kutta-Legendre est une méthode différences finies similaire à STS qui impose une stabilité sur un pas de temps plus grand $\Delta T$ grâce à $N$ étapes récursives. 
Les détails de la méthode RKL2 sont disponibles dans l'article de Meyer et al. \cite{meyer2014stabilized}, et pour la méthode RKL1, ils sont disponibles dans l'article \cite{meyer2012second}.\\
La condition de stabilité que doit suivre notre super pas de temps est:

\begin{equation}
    \Delta T < \Delta t_{CFL} \frac{N^2 +N-2}{4}
\end{equation}
 
 Pour passer de l'état $\mathbf{U}^n$ à l'état $\mathbf{U}^{n+1}$ Nous devons suivre les étapes de récurrence suivantes. 
 
 \begin{equation}
 \left\{
  \begin{array}{l}
           \mathbf{Y}_0 = \mathbf{U}_0   \\
        \mathbf{Y}_1 = \mathbf{Y}_0 + \mu_1 \Delta T \mathbf{A} \mathbf{Y}_0 \\ 
        \mathbf{Y}_j = \mu_j \mathbf{Y}_{j-1} + \nu_j \mathbf{Y}_{j-2} +(1-\mu_j-\nu_j) \mathbf{Y}_0 +  \Tilde{\mu}_j \Delta T \mathbf{A} \mathbf{Y}_{j-1} + \Tilde{\gamma}_j \Delta T \mathbf{A} \mathbf{Y}_0; \quad j \in [2,N] \\ \mathbf{U}^{n+1} = \mathbf{Y}_N
  \end{array}
  \right.
 \end{equation}
 
 avec: 
 
 \begin{equation}
     \begin{array}{ll}
          b_j = \frac{j^2+j-2}{2j(j+1)},& j \in [2,N]  \\
          b_0=b_1=b_2= \frac{1}{3},& a_j = 1-b_j \\
          \Tilde{\mu_1} = \frac{4}{3(N^2+N-2)}, & \Tilde{\mu}_j = \frac{4(2j-1)}{j(N^2+N-2)}\frac{b_j}{b_{j-1}}\\ 
          \mu_j = \frac{(2j-1)}{j} \frac{b_j}{b_{j-1}}, & \nu_j = -\frac{j-1}{j} \frac{b_j}{b_{j-2}} \\ 
          \Tilde{\gamma}_j = a_j - \Tilde{\mu}_j & 
     \end{array}
 \end{equation}
 
 RKL2 possède plusieurs avantages comparativement à STS: 
 \begin{itemize}
     \item Tout d'abord, il s'agit d'un schéma d'ordre 2 en temps. 
     \item L'absence d'un paramètre libre tel que $\nu$ pour STS. 
     \item Le schéma parvient à maintenir une précision accrue à moindre coût. 
 \end{itemize}
 
 \medskip 
 
 Ce schéma a été implémenté dans le code \textit{PLUTO} \cite{pluto2018user} pour le terme résitif et a été comparé avec la méthode STS dans l'article de Vaidya et al. \cite{vaidya2017scalable}.
 
 \subsubsection{Solveur HLL}
 
 Une autre méthode, présente dans le code \textit{GIZMO} \cite{hopkins2016anisotropic}, consiste à utiliser un solveur HLL sur $\mathbf{B}$.
Ensuite, une correction de l'énergie doit être apportée.

\subsection{L'effet Hall dans le terme de flux}

Le terme de Hall n'étant pas un terme de diffusion, il peut être inclut dans le terme de flux et résolue par le solveur MHD.
Les codes comme \textit{PLUTO} \cite{lesur2014thanatology} et \textit{AMRVAC} \cite{porth2014mpi} utilise cette méthode appliquée à un solveur HLL.
Les solveurs de Roe ou HLLD, plus complexes, ne semble pas être de bons candidats car l'ajout des termes de Hall doit modifier significativement toutes les vitesses caractéristiques.\\
Les équations de la MHD résitive incluant le terme de Hall sont de la forme: 

\begin{equation}
 \left\{
    \begin{array}{l}
      \partial_t \rho + \nabla \cdot (\rho \mathbf{V})  =0  \\
        \partial_t (\rho \mathbf{V}) + \nabla \cdot ( \rho \mathbf{V} \otimes \mathbf{V} + p_t \mathbf{I} - \mathbf{B} \otimes \mathbf{B} ) = 0 \\
        \partial_t \mathbf{B} + \nabla [\mathbf{B} \otimes (\mathbf{V}-\mathbf{V}_H) - (\mathbf{V}-\mathbf{V}_H) \otimes \mathbf{B} ]= -\nabla \times (\eta_0 \mathbf{J}) \\ 
        \partial_t \mathcal{E}_t + \nabla \cdot [\mathbf{V} \cdot (p_t+\mathcal{E}_t) - \mathbf{V}\cdot (\mathbf{B} \otimes \mathbf{B}) -\mathbf{V}_H \cdot (\mathbf{I}B^2 - \mathbf{B} \otimes \mathbf{B}   ] = \nabla\cdot (\eta_0 \mathbf{J} \times \mathbf{B})
    \end{array}
    \right.
\end{equation}

Cela donne en ajoutant le courant de Hall dans le terme de flux:

\begin{equation}
    \mathbf{F}(\mathbf{U})= 
    \begin{pmatrix}
    \rho u \\
    \rho u^2 + p_t - \frac{B_x^2}{\mu_0} \\ 
    \rho uv - \frac{B_x B_y}{\mu_0} \\ 
     \rho wu - \frac{B_x B_z}{\mu_0} \\ 
     B_y(u+u_H) - B_x(v+v_H) \\ 
     B_z(u+u_H) - B_x(w+w_H) \\ 
     (\mathcal{E}_t+p_t)u + u_H\frac{B^2}{\mu_0} - \frac{B_x}{\mu_0}[(\mathbf{V}+\mathbf{V}_H)\cdot \mathbf{B}]
    \end{pmatrix}
\end{equation}

avec: 

\begin{equation}
    \mathbf{V}_H = \eta_H \mathbf{J} = 
    \begin{pmatrix}
    u_H\\
    v_H\\
    w_H\\
    \end{pmatrix}
\end{equation}
 
Ce type de méthode calcule $\mathbf{V}_H$ à l'aide d'un schéma explicite et/ou implicite pour ensuite l'implémenter dans notre solveur de Riemann comme paramètre extérieur.  

Le problème de ce type de schéma est l'apparition d'une nouvelle onde, mode siffleur, qui peut être plus rapide que notre onde magnétosonore rapide. 
La vitesse de cette nouvelle onde amenée par l'effet Hall est inversement proportionnelle à sa plus petite longueur d'onde qui est ici 2 fois $\Delta x$.
 Ainsi une réduction drastique de notre pas de temps, lors de simulations à fortes résolutions, peut devenir contraignante.  

\begin{equation}
    c_w = \| v \| +c_f + \frac{\eta_H \|\mathbf{B}\| \pi}{\Delta x } 
\end{equation}

\begin{equation}
    \Delta t < \frac{\Delta x}{c_w}
\end{equation} 

En effet, on voit que pour des $\Delta x$ très petits, on a $\Delta t \propto \Delta x^2$, ce qui peut devenir un facteur limitant dans nos calculs. 

\paragraph{Méthode HLL}

Ce type de méthode a été utilisé par \textit{ARMVAC} et \textit{PLUTO} et s'inspire de la méthode décrite par Toth \cite{toth2008hall} qui l'a mise en place pour schéma de type TVD (un autre type de solveur de Riemann) \cite{Toro}. 
Les vitesses caractéristiques du schéma HLL doivent prendre en compte la vitesse de l'onde \og whistler \fg. Ainsi elles deviennent: 

\begin{equation}
    S^\pm = v \pm max \left( c_f, \frac{\eta_H \|\mathbf{B}\| \pi}{\Delta x }  \right)
\end{equation}

et pour les flux 

\begin{equation}
    \mathbf{F}_{\mid i + 1/2}^* = \left\{
    \begin{array}{lr}
        \mathbf{F}^L & si \quad S^L >0;  \\
         \mathbf{F}^R & si \quad S^R <0; \\ 
         \frac{S^R S^L(\mathbf{U}^R-\mathbf{U}^L) + S^R \mathbf{F}^L - S^L \mathbf{F}^R}{S^R - S^L} & sinon,
    \end{array}
    \right.
\end{equation}
 
 \section{Conclusion}
 
La MHD non-idéale repose sur une loi d'Ohm généralisée possédant davantage de termes que la loi d'Ohm de la MHD idéale. 
Nous nous sommes intéressés plus précisément, aux termes résitif et de Hall, et surtout aux méthodes de résolution proposées dans la littérature.
La résolution de la MHD non-idéale avec uniquement les termes résistifs basée sur une méthode de différences finies explicite est simple à mettre en place.
 Des méthodes, plus complexes, permettent de diminuer la contrainte sur le pas en temps comme les méthodes Super Time Stepping, Runge Kutta Legendre sans pour autant devenir implicite. 
Bien que l'ajout du terme de Hall apporte la possibilité de décrire des phénomènes physiques plus intéressants, sa résolution est moins triviale que la résistivité.
Le seul solveur de Riemann utilisé est HLL qui est connu pour être numériquement très diffusif. Néanmoins, des méthodes en différences finies comme Super Time Stepping peuvent être utilisées pour traiter aussi le terme de Hall.

\end{document}